\documentclass[longauth]{aa}

\usepackage{graphicx}
\usepackage{natbib}
\usepackage{scalerel}
\usepackage{mathtools}
\usepackage{url}
\usepackage[table]{xcolor}
\usepackage{xcolor}

\usepackage{amsmath}    
\usepackage{bm}

\usepackage{multirow}
\usepackage{tabularx}

\usepackage{txfonts}
\usepackage[pdfencoding=auto,psdextra]{hyperref}
\hypersetup{
    colorlinks=true,
    linkcolor=blue,
    filecolor=magenta,      
    urlcolor=blue,
    citecolor=blue
}
\makeatletter
\renewcommand*\aa@pageof{, page \thepage{} of \pageref*{LastPage}}
\makeatother

\usepackage[utf8]{inputenc}

\usepackage[switch, modulo]{lineno}
\renewcommand{\linenumbers}[0]{}

\usepackage{euclid}

\usepackage{cleveref}
\usepackage{rotating}
\usepackage{ragged2e}
\usepackage{array}

\begin{document}

%
%

\title{\Euclid: Optimising tomographic redshift binning for 3$\times$2pt power spectrum constraints on dark energy\thanks{This paper is published on behalf of the Euclid Consortium.}}    

   

\newcommand{\orcid}[1]{} 
\author{J.~H.~W.~Wong\orcid{0000-0001-7133-7741}\thanks{\email{jonathanhw.wong@gmail.com}}\inst{\ref{aff1}}
\and M.~L.~Brown\orcid{0000-0002-0370-8077}\inst{\ref{aff1}}
\and C.~A.~J.~Duncan\orcid{0009-0003-3573-0791}\inst{\ref{aff1}}
\and A.~Amara\inst{\ref{aff2}}
\and S.~Andreon\orcid{0000-0002-2041-8784}\inst{\ref{aff3}}
\and C.~Baccigalupi\orcid{0000-0002-8211-1630}\inst{\ref{aff4},\ref{aff5},\ref{aff6},\ref{aff7}}
\and M.~Baldi\orcid{0000-0003-4145-1943}\inst{\ref{aff8},\ref{aff9},\ref{aff10}}
\and S.~Bardelli\orcid{0000-0002-8900-0298}\inst{\ref{aff9}}
\and D.~Bonino\orcid{0000-0002-3336-9977}\inst{\ref{aff11}}
\and E.~Branchini\orcid{0000-0002-0808-6908}\inst{\ref{aff12},\ref{aff13},\ref{aff3}}
\and M.~Brescia\orcid{0000-0001-9506-5680}\inst{\ref{aff14},\ref{aff15},\ref{aff16}}
\and J.~Brinchmann\orcid{0000-0003-4359-8797}\inst{\ref{aff17},\ref{aff18}}
\and A.~Caillat\inst{\ref{aff19}}
\and S.~Camera\orcid{0000-0003-3399-3574}\inst{\ref{aff20},\ref{aff21},\ref{aff11}}
\and V.~Capobianco\orcid{0000-0002-3309-7692}\inst{\ref{aff11}}
\and C.~Carbone\orcid{0000-0003-0125-3563}\inst{\ref{aff22}}
\and J.~Carretero\orcid{0000-0002-3130-0204}\inst{\ref{aff23},\ref{aff24}}
\and S.~Casas\orcid{0000-0002-4751-5138}\inst{\ref{aff25},\ref{aff26}}
\and M.~Castellano\orcid{0000-0001-9875-8263}\inst{\ref{aff27}}
\and G.~Castignani\orcid{0000-0001-6831-0687}\inst{\ref{aff9}}
\and S.~Cavuoti\orcid{0000-0002-3787-4196}\inst{\ref{aff15},\ref{aff16}}
\and A.~Cimatti\inst{\ref{aff28}}
\and C.~Colodro-Conde\inst{\ref{aff29}}
\and G.~Congedo\orcid{0000-0003-2508-0046}\inst{\ref{aff30}}
\and C.~J.~Conselice\orcid{0000-0003-1949-7638}\inst{\ref{aff1}}
\and L.~Conversi\orcid{0000-0002-6710-8476}\inst{\ref{aff31},\ref{aff32}}
\and Y.~Copin\orcid{0000-0002-5317-7518}\inst{\ref{aff33}}
\and F.~Courbin\orcid{0000-0003-0758-6510}\inst{\ref{aff34},\ref{aff35}}
\and H.~M.~Courtois\orcid{0000-0003-0509-1776}\inst{\ref{aff36}}
\and A.~Da~Silva\orcid{0000-0002-6385-1609}\inst{\ref{aff37},\ref{aff38}}
\and H.~Degaudenzi\orcid{0000-0002-5887-6799}\inst{\ref{aff39}}
\and G.~De~Lucia\orcid{0000-0002-6220-9104}\inst{\ref{aff5}}
\and A.~M.~Di~Giorgio\orcid{0000-0002-4767-2360}\inst{\ref{aff40}}
\and J.~Dinis\orcid{0000-0001-5075-1601}\inst{\ref{aff37},\ref{aff38}}
\and F.~Dubath\orcid{0000-0002-6533-2810}\inst{\ref{aff39}}
\and X.~Dupac\inst{\ref{aff32}}
\and S.~Dusini\orcid{0000-0002-1128-0664}\inst{\ref{aff41}}
\and M.~Farina\orcid{0000-0002-3089-7846}\inst{\ref{aff40}}
\and S.~Farrens\orcid{0000-0002-9594-9387}\inst{\ref{aff42}}
\and F.~Faustini\orcid{0000-0001-6274-5145}\inst{\ref{aff43},\ref{aff27}}
\and S.~Ferriol\inst{\ref{aff33}}
\and M.~Frailis\orcid{0000-0002-7400-2135}\inst{\ref{aff5}}
\and E.~Franceschi\orcid{0000-0002-0585-6591}\inst{\ref{aff9}}
\and S.~Galeotta\orcid{0000-0002-3748-5115}\inst{\ref{aff5}}
\and K.~George\orcid{0000-0002-1734-8455}\inst{\ref{aff44}}
\and W.~Gillard\orcid{0000-0003-4744-9748}\inst{\ref{aff45}}
\and B.~Gillis\orcid{0000-0002-4478-1270}\inst{\ref{aff30}}
\and C.~Giocoli\orcid{0000-0002-9590-7961}\inst{\ref{aff9},\ref{aff10}}
\and A.~Grazian\orcid{0000-0002-5688-0663}\inst{\ref{aff46}}
\and F.~Grupp\inst{\ref{aff47},\ref{aff44}}
\and L.~Guzzo\orcid{0000-0001-8264-5192}\inst{\ref{aff48},\ref{aff3}}
\and S.~V.~H.~Haugan\orcid{0000-0001-9648-7260}\inst{\ref{aff49}}
\and W.~Holmes\inst{\ref{aff50}}
\and I.~Hook\orcid{0000-0002-2960-978X}\inst{\ref{aff51}}
\and F.~Hormuth\inst{\ref{aff52}}
\and A.~Hornstrup\orcid{0000-0002-3363-0936}\inst{\ref{aff53},\ref{aff54}}
\and S.~Ili\'c\orcid{0000-0003-4285-9086}\inst{\ref{aff55},\ref{aff56}}
\and K.~Jahnke\orcid{0000-0003-3804-2137}\inst{\ref{aff57}}
\and M.~Jhabvala\inst{\ref{aff58}}
\and E.~Keih\"anen\orcid{0000-0003-1804-7715}\inst{\ref{aff59}}
\and S.~Kermiche\orcid{0000-0002-0302-5735}\inst{\ref{aff45}}
\and A.~Kiessling\orcid{0000-0002-2590-1273}\inst{\ref{aff50}}
\and B.~Kubik\orcid{0009-0006-5823-4880}\inst{\ref{aff33}}
\and M.~Kunz\orcid{0000-0002-3052-7394}\inst{\ref{aff60}}
\and H.~Kurki-Suonio\orcid{0000-0002-4618-3063}\inst{\ref{aff61},\ref{aff62}}
\and S.~Ligori\orcid{0000-0003-4172-4606}\inst{\ref{aff11}}
\and P.~B.~Lilje\orcid{0000-0003-4324-7794}\inst{\ref{aff49}}
\and V.~Lindholm\orcid{0000-0003-2317-5471}\inst{\ref{aff61},\ref{aff62}}
\and I.~Lloro\orcid{0000-0001-5966-1434}\inst{\ref{aff63}}
\and G.~Mainetti\orcid{0000-0003-2384-2377}\inst{\ref{aff64}}
\and E.~Maiorano\orcid{0000-0003-2593-4355}\inst{\ref{aff9}}
\and O.~Mansutti\orcid{0000-0001-5758-4658}\inst{\ref{aff5}}
\and O.~Marggraf\orcid{0000-0001-7242-3852}\inst{\ref{aff65}}
\and K.~Markovic\orcid{0000-0001-6764-073X}\inst{\ref{aff50}}
\and M.~Martinelli\orcid{0000-0002-6943-7732}\inst{\ref{aff27},\ref{aff66}}
\and N.~Martinet\orcid{0000-0003-2786-7790}\inst{\ref{aff19}}
\and F.~Marulli\orcid{0000-0002-8850-0303}\inst{\ref{aff67},\ref{aff9},\ref{aff10}}
\and R.~Massey\orcid{0000-0002-6085-3780}\inst{\ref{aff68}}
\and E.~Medinaceli\orcid{0000-0002-4040-7783}\inst{\ref{aff9}}
\and S.~Mei\orcid{0000-0002-2849-559X}\inst{\ref{aff69}}
\and M.~Melchior\inst{\ref{aff70}}
\and Y.~Mellier\inst{\ref{aff71},\ref{aff72}}
\and M.~Meneghetti\orcid{0000-0003-1225-7084}\inst{\ref{aff9},\ref{aff10}}
\and E.~Merlin\orcid{0000-0001-6870-8900}\inst{\ref{aff27}}
\and G.~Meylan\inst{\ref{aff73}}
\and M.~Moresco\orcid{0000-0002-7616-7136}\inst{\ref{aff67},\ref{aff9}}
\and L.~Moscardini\orcid{0000-0002-3473-6716}\inst{\ref{aff67},\ref{aff9},\ref{aff10}}
\and C.~Neissner\orcid{0000-0001-8524-4968}\inst{\ref{aff74},\ref{aff24}}
\and S.-M.~Niemi\inst{\ref{aff75}}
\and C.~Padilla\orcid{0000-0001-7951-0166}\inst{\ref{aff74}}
\and S.~Paltani\orcid{0000-0002-8108-9179}\inst{\ref{aff39}}
\and F.~Pasian\orcid{0000-0002-4869-3227}\inst{\ref{aff5}}
\and K.~Pedersen\inst{\ref{aff76}}
\and V.~Pettorino\inst{\ref{aff75}}
\and S.~Pires\orcid{0000-0002-0249-2104}\inst{\ref{aff42}}
\and G.~Polenta\orcid{0000-0003-4067-9196}\inst{\ref{aff43}}
\and M.~Poncet\inst{\ref{aff77}}
\and L.~A.~Popa\inst{\ref{aff78}}
\and F.~Raison\orcid{0000-0002-7819-6918}\inst{\ref{aff47}}
\and A.~Renzi\orcid{0000-0001-9856-1970}\inst{\ref{aff79},\ref{aff41}}
\and J.~Rhodes\orcid{0000-0002-4485-8549}\inst{\ref{aff50}}
\and G.~Riccio\inst{\ref{aff15}}
\and E.~Romelli\orcid{0000-0003-3069-9222}\inst{\ref{aff5}}
\and M.~Roncarelli\orcid{0000-0001-9587-7822}\inst{\ref{aff9}}
\and E.~Rossetti\orcid{0000-0003-0238-4047}\inst{\ref{aff8}}
\and R.~Saglia\orcid{0000-0003-0378-7032}\inst{\ref{aff44},\ref{aff47}}
\and Z.~Sakr\orcid{0000-0002-4823-3757}\inst{\ref{aff80},\ref{aff56},\ref{aff81}}
\and A.~G.~S\'anchez\orcid{0000-0003-1198-831X}\inst{\ref{aff47}}
\and D.~Sapone\orcid{0000-0001-7089-4503}\inst{\ref{aff82}}
\and B.~Sartoris\orcid{0000-0003-1337-5269}\inst{\ref{aff44},\ref{aff5}}
\and P.~Schneider\orcid{0000-0001-8561-2679}\inst{\ref{aff65}}
\and T.~Schrabback\orcid{0000-0002-6987-7834}\inst{\ref{aff83}}
\and A.~Secroun\orcid{0000-0003-0505-3710}\inst{\ref{aff45}}
\and G.~Seidel\orcid{0000-0003-2907-353X}\inst{\ref{aff57}}
\and S.~Serrano\orcid{0000-0002-0211-2861}\inst{\ref{aff84},\ref{aff85},\ref{aff86}}
\and C.~Sirignano\orcid{0000-0002-0995-7146}\inst{\ref{aff79},\ref{aff41}}
\and G.~Sirri\orcid{0000-0003-2626-2853}\inst{\ref{aff10}}
\and L.~Stanco\orcid{0000-0002-9706-5104}\inst{\ref{aff41}}
\and J.~Steinwagner\orcid{0000-0001-7443-1047}\inst{\ref{aff47}}
\and P.~Tallada-Cresp\'{i}\orcid{0000-0002-1336-8328}\inst{\ref{aff23},\ref{aff24}}
\and A.~N.~Taylor\inst{\ref{aff30}}
\and I.~Tereno\inst{\ref{aff37},\ref{aff87}}
\and R.~Toledo-Moreo\orcid{0000-0002-2997-4859}\inst{\ref{aff88}}
\and F.~Torradeflot\orcid{0000-0003-1160-1517}\inst{\ref{aff24},\ref{aff23}}
\and I.~Tutusaus\orcid{0000-0002-3199-0399}\inst{\ref{aff56}}
\and L.~Valenziano\orcid{0000-0002-1170-0104}\inst{\ref{aff9},\ref{aff89}}
\and T.~Vassallo\orcid{0000-0001-6512-6358}\inst{\ref{aff44},\ref{aff5}}
\and G.~Verdoes~Kleijn\orcid{0000-0001-5803-2580}\inst{\ref{aff90}}
\and A.~Veropalumbo\orcid{0000-0003-2387-1194}\inst{\ref{aff3},\ref{aff13},\ref{aff91}}
\and Y.~Wang\orcid{0000-0002-4749-2984}\inst{\ref{aff92}}
\and J.~Weller\orcid{0000-0002-8282-2010}\inst{\ref{aff44},\ref{aff47}}
\and G.~Zamorani\orcid{0000-0002-2318-301X}\inst{\ref{aff9}}
\and E.~Zucca\orcid{0000-0002-5845-8132}\inst{\ref{aff9}}
\and C.~Burigana\orcid{0000-0002-3005-5796}\inst{\ref{aff93},\ref{aff89}}
\and M.~Calabrese\orcid{0000-0002-2637-2422}\inst{\ref{aff94},\ref{aff22}}
\and A.~Pezzotta\orcid{0000-0003-0726-2268}\inst{\ref{aff47}}
\and V.~Scottez\inst{\ref{aff71},\ref{aff95}}
\and A.~Spurio~Mancini\orcid{0000-0001-5698-0990}\inst{\ref{aff96}}
\and M.~Viel\orcid{0000-0002-2642-5707}\inst{\ref{aff4},\ref{aff5},\ref{aff7},\ref{aff6},\ref{aff97}}}
                                                                                   
\institute{Jodrell Bank Centre for Astrophysics, Department of Physics and Astronomy, University of Manchester, Oxford Road, Manchester M13 9PL, UK\label{aff1}
\and
School of Mathematics and Physics, University of Surrey, Guildford, Surrey, GU2 7XH, UK\label{aff2}
\and
INAF-Osservatorio Astronomico di Brera, Via Brera 28, 20122 Milano, Italy\label{aff3}
\and
IFPU, Institute for Fundamental Physics of the Universe, via Beirut 2, 34151 Trieste, Italy\label{aff4}
\and
INAF-Osservatorio Astronomico di Trieste, Via G. B. Tiepolo 11, 34143 Trieste, Italy\label{aff5}
\and
INFN, Sezione di Trieste, Via Valerio 2, 34127 Trieste TS, Italy\label{aff6}
\and
SISSA, International School for Advanced Studies, Via Bonomea 265, 34136 Trieste TS, Italy\label{aff7}
\and
Dipartimento di Fisica e Astronomia, Universit\`a di Bologna, Via Gobetti 93/2, 40129 Bologna, Italy\label{aff8}
\and
INAF-Osservatorio di Astrofisica e Scienza dello Spazio di Bologna, Via Piero Gobetti 93/3, 40129 Bologna, Italy\label{aff9}
\and
INFN-Sezione di Bologna, Viale Berti Pichat 6/2, 40127 Bologna, Italy\label{aff10}
\and
INAF-Osservatorio Astrofisico di Torino, Via Osservatorio 20, 10025 Pino Torinese (TO), Italy\label{aff11}
\and
Dipartimento di Fisica, Universit\`a di Genova, Via Dodecaneso 33, 16146, Genova, Italy\label{aff12}
\and
INFN-Sezione di Genova, Via Dodecaneso 33, 16146, Genova, Italy\label{aff13}
\and
Department of Physics "E. Pancini", University Federico II, Via Cinthia 6, 80126, Napoli, Italy\label{aff14}
\and
INAF-Osservatorio Astronomico di Capodimonte, Via Moiariello 16, 80131 Napoli, Italy\label{aff15}
\and
INFN section of Naples, Via Cinthia 6, 80126, Napoli, Italy\label{aff16}
\and
Instituto de Astrof\'isica e Ci\^encias do Espa\c{c}o, Universidade do Porto, CAUP, Rua das Estrelas, PT4150-762 Porto, Portugal\label{aff17}
\and
Faculdade de Ci\^encias da Universidade do Porto, Rua do Campo de Alegre, 4150-007 Porto, Portugal\label{aff18}
\and
Aix-Marseille Universit\'e, CNRS, CNES, LAM, Marseille, France\label{aff19}
\and
Dipartimento di Fisica, Universit\`a degli Studi di Torino, Via P. Giuria 1, 10125 Torino, Italy\label{aff20}
\and
INFN-Sezione di Torino, Via P. Giuria 1, 10125 Torino, Italy\label{aff21}
\and
INAF-IASF Milano, Via Alfonso Corti 12, 20133 Milano, Italy\label{aff22}
\and
Centro de Investigaciones Energ\'eticas, Medioambientales y Tecnol\'ogicas (CIEMAT), Avenida Complutense 40, 28040 Madrid, Spain\label{aff23}
\and
Port d'Informaci\'{o} Cient\'{i}fica, Campus UAB, C. Albareda s/n, 08193 Bellaterra (Barcelona), Spain\label{aff24}
\and
Institute for Theoretical Particle Physics and Cosmology (TTK), RWTH Aachen University, 52056 Aachen, Germany\label{aff25}
\and
Institute of Cosmology and Gravitation, University of Portsmouth, Portsmouth PO1 3FX, UK\label{aff26}
\and
INAF-Osservatorio Astronomico di Roma, Via Frascati 33, 00078 Monteporzio Catone, Italy\label{aff27}
\and
Dipartimento di Fisica e Astronomia "Augusto Righi" - Alma Mater Studiorum Universit\`a di Bologna, Viale Berti Pichat 6/2, 40127 Bologna, Italy\label{aff28}
\and
Instituto de Astrof\'isica de Canarias, Calle V\'ia L\'actea s/n, 38204, San Crist\'obal de La Laguna, Tenerife, Spain\label{aff29}
\and
Institute for Astronomy, University of Edinburgh, Royal Observatory, Blackford Hill, Edinburgh EH9 3HJ, UK\label{aff30}
\and
European Space Agency/ESRIN, Largo Galileo Galilei 1, 00044 Frascati, Roma, Italy\label{aff31}
\and
ESAC/ESA, Camino Bajo del Castillo, s/n., Urb. Villafranca del Castillo, 28692 Villanueva de la Ca\~nada, Madrid, Spain\label{aff32}
\and
Universit\'e Claude Bernard Lyon 1, CNRS/IN2P3, IP2I Lyon, UMR 5822, Villeurbanne, F-69100, France\label{aff33}
\and
Institut de Ci\`{e}ncies del Cosmos (ICCUB), Universitat de Barcelona (IEEC-UB), Mart\'{i} i Franqu\`{e}s 1, 08028 Barcelona, Spain\label{aff34}
\and
Instituci\'o Catalana de Recerca i Estudis Avan\c{c}ats (ICREA), Passeig de Llu\'{\i}s Companys 23, 08010 Barcelona, Spain\label{aff35}
\and
UCB Lyon 1, CNRS/IN2P3, IUF, IP2I Lyon, 4 rue Enrico Fermi, 69622 Villeurbanne, France\label{aff36}
\and
Departamento de F\'isica, Faculdade de Ci\^encias, Universidade de Lisboa, Edif\'icio C8, Campo Grande, PT1749-016 Lisboa, Portugal\label{aff37}
\and
Instituto de Astrof\'isica e Ci\^encias do Espa\c{c}o, Faculdade de Ci\^encias, Universidade de Lisboa, Campo Grande, 1749-016 Lisboa, Portugal\label{aff38}
\and
Department of Astronomy, University of Geneva, ch. d'Ecogia 16, 1290 Versoix, Switzerland\label{aff39}
\and
INAF-Istituto di Astrofisica e Planetologia Spaziali, via del Fosso del Cavaliere, 100, 00100 Roma, Italy\label{aff40}
\and
INFN-Padova, Via Marzolo 8, 35131 Padova, Italy\label{aff41}
\and
Universit\'e Paris-Saclay, Universit\'e Paris Cit\'e, CEA, CNRS, AIM, 91191, Gif-sur-Yvette, France\label{aff42}
\and
Space Science Data Center, Italian Space Agency, via del Politecnico snc, 00133 Roma, Italy\label{aff43}
\and
Universit\"ats-Sternwarte M\"unchen, Fakult\"at f\"ur Physik, Ludwig-Maximilians-Universit\"at M\"unchen, Scheinerstrasse 1, 81679 M\"unchen, Germany\label{aff44}
\and
Aix-Marseille Universit\'e, CNRS/IN2P3, CPPM, Marseille, France\label{aff45}
\and
INAF-Osservatorio Astronomico di Padova, Via dell'Osservatorio 5, 35122 Padova, Italy\label{aff46}
\and
Max Planck Institute for Extraterrestrial Physics, Giessenbachstr. 1, 85748 Garching, Germany\label{aff47}
\and
Dipartimento di Fisica "Aldo Pontremoli", Universit\`a degli Studi di Milano, Via Celoria 16, 20133 Milano, Italy\label{aff48}
\and
Institute of Theoretical Astrophysics, University of Oslo, P.O. Box 1029 Blindern, 0315 Oslo, Norway\label{aff49}
\and
Jet Propulsion Laboratory, California Institute of Technology, 4800 Oak Grove Drive, Pasadena, CA, 91109, USA\label{aff50}
\and
Department of Physics, Lancaster University, Lancaster, LA1 4YB, UK\label{aff51}
\and
Felix Hormuth Engineering, Goethestr. 17, 69181 Leimen, Germany\label{aff52}
\and
Technical University of Denmark, Elektrovej 327, 2800 Kgs. Lyngby, Denmark\label{aff53}
\and
Cosmic Dawn Center (DAWN), Denmark\label{aff54}
\and
Universit\'e Paris-Saclay, CNRS/IN2P3, IJCLab, 91405 Orsay, France\label{aff55}
\and
Institut de Recherche en Astrophysique et Plan\'etologie (IRAP), Universit\'e de Toulouse, CNRS, UPS, CNES, 14 Av. Edouard Belin, 31400 Toulouse, France\label{aff56}
\and
Max-Planck-Institut f\"ur Astronomie, K\"onigstuhl 17, 69117 Heidelberg, Germany\label{aff57}
\and
NASA Goddard Space Flight Center, Greenbelt, MD 20771, USA\label{aff58}
\and
Department of Physics and Helsinki Institute of Physics, Gustaf H\"allstr\"omin katu 2, 00014 University of Helsinki, Finland\label{aff59}
\and
Universit\'e de Gen\`eve, D\'epartement de Physique Th\'eorique and Centre for Astroparticle Physics, 24 quai Ernest-Ansermet, CH-1211 Gen\`eve 4, Switzerland\label{aff60}
\and
Department of Physics, P.O. Box 64, 00014 University of Helsinki, Finland\label{aff61}
\and
Helsinki Institute of Physics, Gustaf H{\"a}llstr{\"o}min katu 2, University of Helsinki, Helsinki, Finland\label{aff62}
\and
NOVA optical infrared instrumentation group at ASTRON, Oude Hoogeveensedijk 4, 7991PD, Dwingeloo, The Netherlands\label{aff63}
\and
Centre de Calcul de l'IN2P3/CNRS, 21 avenue Pierre de Coubertin 69627 Villeurbanne Cedex, France\label{aff64}
\and
Universit\"at Bonn, Argelander-Institut f\"ur Astronomie, Auf dem H\"ugel 71, 53121 Bonn, Germany\label{aff65}
\and
INFN-Sezione di Roma, Piazzale Aldo Moro, 2 - c/o Dipartimento di Fisica, Edificio G. Marconi, 00185 Roma, Italy\label{aff66}
\and
Dipartimento di Fisica e Astronomia "Augusto Righi" - Alma Mater Studiorum Universit\`a di Bologna, via Piero Gobetti 93/2, 40129 Bologna, Italy\label{aff67}
\and
Department of Physics, Institute for Computational Cosmology, Durham University, South Road, Durham, DH1 3LE, UK\label{aff68}
\and
Universit\'e Paris Cit\'e, CNRS, Astroparticule et Cosmologie, 75013 Paris, France\label{aff69}
\and
University of Applied Sciences and Arts of Northwestern Switzerland, School of Engineering, 5210 Windisch, Switzerland\label{aff70}
\and
Institut d'Astrophysique de Paris, 98bis Boulevard Arago, 75014, Paris, France\label{aff71}
\and
Institut d'Astrophysique de Paris, UMR 7095, CNRS, and Sorbonne Universit\'e, 98 bis boulevard Arago, 75014 Paris, France\label{aff72}
\and
Institute of Physics, Laboratory of Astrophysics, Ecole Polytechnique F\'ed\'erale de Lausanne (EPFL), Observatoire de Sauverny, 1290 Versoix, Switzerland\label{aff73}
\and
Institut de F\'{i}sica d'Altes Energies (IFAE), The Barcelona Institute of Science and Technology, Campus UAB, 08193 Bellaterra (Barcelona), Spain\label{aff74}
\and
European Space Agency/ESTEC, Keplerlaan 1, 2201 AZ Noordwijk, The Netherlands\label{aff75}
\and
DARK, Niels Bohr Institute, University of Copenhagen, Jagtvej 155, 2200 Copenhagen, Denmark\label{aff76}
\and
Centre National d'Etudes Spatiales -- Centre spatial de Toulouse, 18 avenue Edouard Belin, 31401 Toulouse Cedex 9, France\label{aff77}
\and
Institute of Space Science, Str. Atomistilor, nr. 409 M\u{a}gurele, Ilfov, 077125, Romania\label{aff78}
\and
Dipartimento di Fisica e Astronomia "G. Galilei", Universit\`a di Padova, Via Marzolo 8, 35131 Padova, Italy\label{aff79}
\and
Institut f\"ur Theoretische Physik, University of Heidelberg, Philosophenweg 16, 69120 Heidelberg, Germany\label{aff80}
\and
Universit\'e St Joseph; Faculty of Sciences, Beirut, Lebanon\label{aff81}
\and
Departamento de F\'isica, FCFM, Universidad de Chile, Blanco Encalada 2008, Santiago, Chile\label{aff82}
\and
Universit\"at Innsbruck, Institut f\"ur Astro- und Teilchenphysik, Technikerstr. 25/8, 6020 Innsbruck, Austria\label{aff83}
\and
Institut d'Estudis Espacials de Catalunya (IEEC),  Edifici RDIT, Campus UPC, 08860 Castelldefels, Barcelona, Spain\label{aff84}
\and
Satlantis, University Science Park, Sede Bld 48940, Leioa-Bilbao, Spain\label{aff85}
\and
Institute of Space Sciences (ICE, CSIC), Campus UAB, Carrer de Can Magrans, s/n, 08193 Barcelona, Spain\label{aff86}
\and
Instituto de Astrof\'isica e Ci\^encias do Espa\c{c}o, Faculdade de Ci\^encias, Universidade de Lisboa, Tapada da Ajuda, 1349-018 Lisboa, Portugal\label{aff87}
\and
Universidad Polit\'ecnica de Cartagena, Departamento de Electr\'onica y Tecnolog\'ia de Computadoras,  Plaza del Hospital 1, 30202 Cartagena, Spain\label{aff88}
\and
INFN-Bologna, Via Irnerio 46, 40126 Bologna, Italy\label{aff89}
\and
Kapteyn Astronomical Institute, University of Groningen, PO Box 800, 9700 AV Groningen, The Netherlands\label{aff90}
\and
Dipartimento di Fisica, Universit\`a degli studi di Genova, and INFN-Sezione di Genova, via Dodecaneso 33, 16146, Genova, Italy\label{aff91}
\and
Infrared Processing and Analysis Center, California Institute of Technology, Pasadena, CA 91125, USA\label{aff92}
\and
INAF, Istituto di Radioastronomia, Via Piero Gobetti 101, 40129 Bologna, Italy\label{aff93}
\and
Astronomical Observatory of the Autonomous Region of the Aosta Valley (OAVdA), Loc. Lignan 39, I-11020, Nus (Aosta Valley), Italy\label{aff94}
\and
ICL, Junia, Universit\'e Catholique de Lille, LITL, 59000 Lille, France\label{aff95}
\and
Department of Physics, Royal Holloway, University of London, TW20 0EX, UK\label{aff96}
\and
ICSC - Centro Nazionale di Ricerca in High Performance Computing, Big Data e Quantum Computing, Via Magnanelli 2, Bologna, Italy\label{aff97}}

%
%
\abstract{

The tomographic approach to analysing the 3$\times$2pt signal involves dividing the observed galaxy sample into a configuration of redshift bins. We present a simulation-based method to explore the optimum tomographic binning strategy for \textit{Euclid}, focussing on the expected configuration of its first major data release (DR1). To do so, we 1) simulated a \textit{Euclid}-like observation and generated mock shear catalogues from multiple realisations of the 3$\times$2pt fields on the sky; and 2) measured the 3$\times$2pt Pseudo-$C_{\ell}$ power spectra for a given tomographic configuration and derived the constraints they place on the standard dark energy equation-of-state parameters, $(w_{0},w_{a})$. For a simulation including Gaussian-distributed photometric redshift uncertainties and shape noise under a $\Lambda$CDM cosmology, we find that bins that are equipopulated with galaxies yield the best constraints on $(w_{0},w_{a})$ for an analysis of the full 3$\times$2pt signal or the angular clustering component only. For the cosmic shear component, the optimum $(w_{0},w_{a})$ constraints can be achieved by bins equally spaced in fiducial comoving distance. However, the advantage with respect to alternative binning choices is only of a few per cent in the size of the $1\,\sigma\,(w_{0},w_{a})$ contour and we conclude that the cosmic shear is relatively insensitive to the binning methodology. We find that the information gain extracted on $(w_{0},w_{a})$ for any 3$\times$2pt component starts to become saturated beyond roughly seven or eight bins. Any marginal gains resulting from a greater number of bins are likely to be limited by additional uncertainties present in a real measurement and the increasing demand for accuracy of the covariance matrix. Finally, we considered a $5\%$ contamination from catastrophic photometric redshift outliers and found that if these errors are not mitigated in the analysis, the bias induced in the 3$\times$2pt signal for ten equipopulated bins results in dark energy constraints that are inconsistent with the fiducial $\Lambda$CDM cosmology at {$\gtrsim3\,\sigma$}.

    }
%
%
    \keywords{Cosmology: observations -- dark energy -- cosmological parameters -- Methods: statistical -- Gravitational lensing: weak}
%
%
   \titlerunning{Optimising 3$\times$2pt power spectrum tomography for \textit{Euclid} DR1}
   \authorrunning{J.~H.~W.~Wong et al.}
   
   \maketitle
%
%
%
%

\section{Introduction}
\label{sec:introduction}

The weak gravitational lensing of galaxies by the large-scale structure in the Universe, known as cosmic shear, has emerged as a key tool for cosmology given its ability to probe the evolution of matter through cosmic time. This property has allowed for weak lensing analyses (see e.g. \citealt{Kilbinger15} for a review) to be used to constrain the nature of dark energy, proposed as the origin of the observed accelerated expansion of the Universe (\citealt{Riess98}, \citealt{Schmidt98}, \citealt{Perlmutter99}).

Surveys such as \textit{Euclid} (\citealt{Euclid}, \citealt{EuclidSkyOverview}), the forthcoming \textit{Nancy Grace Roman} Space Telescope \citep{Roman}, and the Vera C. Rubin Observatory Legacy Survey of Space and Time (\citealt{LSST}), will provide samples of over 1 billion galaxies and are poised to offer an improvement (by at
least an order of magnitude) in the precision of cosmic shear measurements over current experiments. These data will be used to investigate the extent to which dark energy can be described as a time-evolving fluid contributing to the mass-energy budget of the Universe. In particular, \textit{Euclid}  will target constraints on a $w_{0}w_{a}$CDM cosmology, where the dark energy equation of state parameter, $w$, is assumed to take the functional form $w=w_{0}+(1-a)w_{a}$ (\citealt{Euclid}), where $a$ is the scale factor.

An intrinsic step with respect to the weak lensing method is the analysis of the 3D radial distribution of matter in the Universe, traced by the galaxies observed in the survey. While a cosmic shear study can employ the `3D weak lensing' analysis technique (\citealt{Heavens03}, \citealt{Kitching14}, \citealt{SpurioMancini18}) where the shear field is expanded radially in spherical Bessel functions, the common method (i.e., tomography) involves subdividing the observed galaxy population into binned samples at different distances. This technique effectively compresses the 3D cosmological information into a number of slices, through which a 2D two-point analysis of the projected galaxy images is used to recover the radial shear field \citep{Hu99}. 

Comparatively, 3D weak lensing measurements (\citealt{Castro05}, \citealt{Kitching07}) naturally offer more statistical power for cosmological parameter constraints. However, the information gain is dependent on the chosen parameter(s) and it has been demonstrated (e.g. \citealt{Hu02}; \citealt{Jain_Taylor03}; \citealt{Huterer02}; \citealt{Schrabback10}; \citealt{Benjamin13}; \citealt{Taylor18_b}) that the tomographic weak lensing approach recovers the majority of the cosmological information required to make measurements of dark energy with the benefit of a significant reduction in computational cost. Indeed, the current generation of surveys such as the Dark Energy Survey (DES; \citealt{DES_Y3}), the Kilo Degree Survey (KiDS; \citealt{Hildebrandt17}), and the Subaru Hyper Suprime-Cam (HSC; \citealt{Hikage19}) have consistently utilised weak lensing tomography to obtain constraints on parameters such as the matter density, $\Omega_{\mathrm{m}}$, the amplitude of matter fluctuations on scales of $8\,h^{-1}\,\mathrm{Mpc}$, $\sigma_{8}$, and the dark energy equation of state parameter, $w$. These analyses provide complementary measurements to constraints derived from alternative probes such as Type Ia supernovae (e.g. \citealt{Brout22}), the cosmic microwave background (CMB, e.g. \citealt{Planck18}), and baryon acoustic oscillations (BAO, e.g. \citealt{Shadab17}). 

Notably, the DES collaboration (\citealt{DES-w0wa}) have constrained the dark energy $(w_{0}, w_{a})$ parameters using a tomographic analysis of the 3$\times$2pt signal (the joint data vector of cosmic shear, angular galaxy clustering, and their cross-correlation, see Sect.~\ref{subsec:cosmic-shear-basics}). When combining the 3$\times$2pt signal with \textit{Planck} temperature and $E$-mode polarisation observables (\citealt{Planck18}), BAO results from eBOSS (\citealt{Ahumada20}) and the Pantheon SNe Ia sample (\citealt{Scolnic18}), they found $(w_{0}, w_{a})=(-0.95\pm0.08, -0.4^{+0.4}_{-0.3}),$ which is consistent with a standard Lambda cold dark matter ($\Lambda$CDM) model at $1\,\sigma$. 

Comparatively, studies with the Pantheon+ analysis of a larger sample of distinct SNe Ia (see \citealt{Brout22} and references therein) have found $(w_{0}, w_{a})=(-0.841^{+0.066}_{-0.061}, -0.65^{+0.28}_{-0.32})$ when combined with \textit{Planck} and BAO, which is moderately consistent with the cosmological constant at $2\,\sigma$. In addition, the Dark Energy Spectroscopic Instrument (\citealt{DESI16}, \citealt{DESI24}) has found support for $w_{0}>-1, w_{a}<0$ when combining BAO and \textit{Planck} CMB data, which is inconsistent with the $\Lambda$CDM cosmology at $\gtrsim2.5\sigma$. One of the targets of \textit{Euclid} is to provide a $\sim2\%$ and $\sim10\%$ measurement on $w_{0}$ and $w_{a}$, respectively \citep{Euclid}, which will help in further investigating the nature of dark energy and the validity of the standard $\Lambda$CDM model. 

The structure of this paper is as follows. In Sect.~\ref{sec:motivations}, we discuss previous attempts to investigate tomographic binning strategies for weak lensing surveys and highlight the need to determine an appropriate strategy for \textit{Euclid}. In Sects.~\ref{sec:background-theory} and~\ref{sec:methods-catalogues}, we review the analytical estimators used to measure the 3$\times$2pt signal and present our method to simulate mock Stage IV-like galaxy catalogues. In Sects.~\ref{sec:inference-routine} and~\ref{sec:results_tomography}, we discuss the inference method that we use to derive constraints on the $(w_{0}, w_{a})$ parameters from the 3$\times$2pt signal, exploring these constraints for different tomographic binning strategies applied to our mock catalogues. In Sect.~\ref{sec:catastrophic-photozs}, we explore the effects of catastrophic photometric redshift uncertainties. In Sect.~\ref{sec:discussion_conclusions}, we discuss the overall conclusions of our study to determine an optimum tomographic binning method. Finally, in the appendix, we review the validation testing we  conducted for our simulation method and discuss the finer details of the modelling and computational choices that we implemented in this work.

\section{Motivations}
\label{sec:motivations}

To construct a tomographic measurement, two analysis choices must be made: the number of bins to split the observed galaxy sample into and how to determine the bin boundaries along the radial (line-of-sight) direction. Each of these choices are governed by both scientific and practical considerations. 

With regard to choosing the number of tomographic bins, it is generally expected that a greater number of bins will increase the information captured in a shear analysis, but at a higher computational cost. However, there will be a point at which the information gain starts to saturate, since adding more sample bins increases the Poisson shot noise per bin, which will dominate over the underlying cosmological shear signal of interest (see Sect.~\ref{sec:background-theory} for more details). Additionally, the cosmological volume enclosed by a single bin reduces as a larger number of bins is used, resulting in a comparatively lower gain in information, as the lensing signal over a given bin displays less of a change. 

Several studies have been conducted to explore this issue and determine the `optimum' tomography for a Stage IV-like weak lensing survey. For example, \cite{Taylor18_a} used a principal component analysis (PCA) to compare the suitability of a range of tomographic binning strategies with respect to the 3D cosmic shear method. For a \textit{Euclid}-like weak lensing signal, they found that approximately ten tomographic bins with equal populations of galaxies can sufficiently capture the majority ($\sim97\%$) of the cosmological information of the 3D approach, but the tomographic method starts to fail at higher redshifts due to information loss and computational cost. Comparatively, a tomographic set-up with bins equally spaced in redshift does not capture as much information as the equipopulated case for a small number of bins; however, it can extract $\sim99\%$ of the information with respect to the 3D method for a large number of bins $(\sim50)$. Regardless of binning choice, the computational advantage of tomography is highlighted with respect to a 3D analysis. 

This result is consistent with \cite{Sipp21}, who explored different tomographic strategies to assess the statistical properties of a predicted \textit{Euclid}-like cosmic shear signal. For a range of science metrics, including minimising the uncertainties on individual cosmological parameter constraints and the optimisation of the \textit{Euclid} dark energy figure of merit \citep{Blanchard-EP7}, they found that the optimum tomographic configuration is very close to the equipopulated scenario, with a small number of bins (four to five) sufficient for the cosmological analysis.

In contrast, \cite{Kitching19} used a self organising map technique (SOM; \citealt{Kohonen}) to define bin boundaries in photometric colour space, while optimising the signal-to-noise ratio (S/N) on the dark energy equation of state parameter, $w_{0}$, measured from the cosmic shear power spectrum. For a small number of bins (< 5), the best-performing binning methodology identified by the SOM was found to be a close approximation of the case where the bins are defined to have equal width in redshift. 

This issue is further complicated if a `3$\times$2pt' analysis of the data is targeted (see Sect.~\ref{subsec:cosmic-shear-basics}), where the shear signal is combined with the tomographic galaxy number density (angular clustering) and galaxy-galaxy lensing (cross-correlation of weak lensing and angular clustering) signals into a joint data vector. The full 3$\times$2pt data vector will offer more constraining power than a shear-only analysis (see e.g. \citealt{Tutusaus20}). However, the choice of an appropriate tomographic binning configuration may also depend on which cosmological parameter is being targeted and which part of the data vector is most sensitive to the corresponding science case. 

In particular, by considering the photometric redshift estimation techniques for \textit{Euclid}, \citet{Pocino-EP12} found that for a galaxy clustering and galaxy-galaxy lensing analysis, tomographic bins equally spaced in redshift yield a higher dark energy figure of merit (FOM) than equipopulated bins. Furthermore, a relatively large number of bins (13) maximises the amount of cosmological information extracted. 

Additionally, \cite{Zuntz21} have coordinated a tomography challenge in which a range of tomographic binning configurations, guided by machine learning approaches on a test sample, were proposed by the community to target a variety of metrics such as the 3$\times$2pt power spectra S/N and the dark energy FOM. The findings suggest that the optimum tomographic binning set-up used for the analysis is strongly dependent on both the chosen 3$\times$2pt observable and the specific cosmological parameter FOM that is targeted. However, a source galaxy distribution that is split into redshift bins that are equally spaced in fiducial comoving distance was found to be a good general choice.

It is clear that these studies reach a variety of different conclusions and, in addition, they are not all performed for the same target survey set-up. This makes it challenging to use them for choosing an optimum binning scheme for \textit{Euclid}. Hence, the aim of this work is to present an alternative method to explore a range of 3$\times$2pt tomographic configurations and investigate an optimum tomographic binning approach for the specific survey characteristics expected for the analysis of \textit{Euclid}'s first data release (DR1). 

We construct a cosmological simulation pipeline that generates full \textit{Euclid}-like cosmic shear catalogues realised on the sky from a given fiducial cosmology, with realistic observational uncertainties such as errors from photometric redshift (`photo-$z$') estimation and shape noise injected into the catalogue observables. From our simulated catalogues we measure the 3$\times$2pt signal and compare constraints on the dark energy equation of state for each tomographic set-up chosen, which will help to further identify the optimum configuration for the cosmological analysis of the 3$\times$2pt observables in the \textit{Euclid} survey. Our simulation pipeline includes the following stages:
\begin{enumerate}
    \item The sampling of an observed galaxy population from a predicted \textit{Euclid}-like galaxy redshift distribution and spatial number density;
    \item The calculation of a theoretical 3$\times$2pt signal over a given redshift range and fiducial cosmology;
    \item Simulations of the correlated cosmic shear and galaxy density fields on the sky as a set of 2D maps that approximates the 3D evolution of the 3$\times$2pt signal over time;
    \item The sampling of galaxy angular positions and weak lensing observables from the density and shear fields, respectively;    \item The injection of observational uncertainties from photo-$z$ estimation and galaxy shape noise into the individual galaxies to generate a mock catalogue mimicking a realistic survey.
\end{enumerate}
We provide a full description of this simulation method and each stage in Sect.~\ref{sec:methods-catalogues}. We note that the previously discussed tomography studies typically focus on using numerical methods to explore a theoretical shear signal. In comparison, we highlight that our method offers the advantage of imitating a realistic survey observation and measuring the corresponding 3$\times$2pt signal from an underlying cosmology, which provides a more direct and realistic framework with which to investigate the cosmological impact of different tomographic strategies.

\section{Background and theory}
\label{sec:background-theory}

\subsection{Cosmic shear and the 3$\times$2pt signal}
\label{subsec:cosmic-shear-basics}

Gravitational lensing occurs when the foreground potential of a massive object distorts the light observed from a background source. In the regime of weak lensing (see e.g. \citealt{Bartelmann_Schneider_01} for a detailed review), where the distortion effects are small (i.e. at the per cent level), this phenomenon leads to subtle changes to a background galaxy's size and shape, which can be described using the convergence, $\kappa$ (spin-0 quantity), and shear, $\gamma$ (spin-2 quantity), respectively. 

For a population of multiple background galaxies, weak lensing leads to correlations in the distortions of individual galaxy shapes. Using tomography, the galaxy sample is split into a set of redshift bins and the cosmic shear signal is then measured using the angular power spectrum under the Limber approximation (\citealt{Limber53}, \citealt{LoVerde08}) as $C_{ij}(\ell)$, where $i, j$ denote the pairs of redshift bins in the sample. Explicitly, for a flat Universe, the tomographic shear power spectrum,  $C_{ij}^{\gamma\gamma}(\ell)$, is given by

\begin{equation}
    \label{eq:harmonic-shear}
    C_{ij}^{\gamma\gamma}(\ell) = \int_{0}^{\chi_{\mathrm{H}}} \mathrm{d}\chi \, \frac{q_{i}^{\kappa}(\chi)\,q_{j}^{\kappa}(\chi)}{\chi^{2}}\,P_{\delta}\left(\frac{\ell+1/2}{\chi},\,\chi\right) \,,
\end{equation}
where we take the matter density parameter at the present day, $\Omega_{\mathrm{m}}=0.3$, and the Hubble constant, $H_{0}=70\,\mathrm{km}\,\mathrm{s}^{-1}\,\mathrm{Mpc}^{-1}$, throughout this work. Here, the comoving radial coordinate, $\chi$, is integrated up to $\chi_{\mathrm{H}}$, which is the comoving distance to the cosmic horizon. Additionally, $a$ is the scale factor, $P_{\delta}$ is the matter power spectrum, and $q^{\kappa}_{i}(\chi)$ is the lensing efficiency, 
\begin{equation}
    \label{eq:shear-kernel}
    q_{i}^{\kappa}(\chi) = \frac{3}{2}\,\Omega_{\mathrm{m}}\left(\frac{H_{0}}{c}\right)^{2}\,\frac{\chi}{a(\chi)}\int_{\chi}^{\chi_{\mathrm{H}}} \mathrm{d} \chi' \, \frac{n_{i}(z)\, \mathrm{d} z/\mathrm{d}\chi'}{\Bar{n}_{i}} \, \frac{\chi'-\chi}{\chi'} ,
\end{equation}
which describes the redshift distribution of the background galaxies, $n_{i}(z)$, weighted by the angular number density,

\begin{equation}
    \Bar{n}_{i}=\int_0^\infty \mathrm{d} z \, n_{i}(z) \,,
\end{equation}
where the redshift $z=z(\chi)$, and $i$ denotes a given tomographic bin. A full 3$\times$2pt analysis also uses the galaxy angular positions on the sky, $\delta_{g}(\bm{\theta})$, and the corresponding galaxy clustering radial kernel, $q_{i}^{\delta_{g}}$,

\begin{equation}
    q_{i}^{\delta_{g}}(k, \chi) = b_{i}\left(k, z\right)\,n_{i}(z)\,\frac{\mathrm{d} z}{\mathrm{d} \chi} ,
\end{equation}
where $b_{i}(k, z)$ is the galaxy bias in tomographic bin $i$, which describes the spatial relation between the underlying matter field and the observed distribution of galaxies as a function of redshift and scale, $k$, where $k=(\ell+1/2)/\chi$.

In addition to weak lensing, the 3$\times$2pt signal consists of the galaxy clustering angular power spectrum, 


\begin{equation}
    \label{eq:harmonic-clustering}
    C_{ij}^{\delta_{g}\delta_{g}}(\ell)=\int_{0}^{\chi_{\mathrm{H}}} \mathrm{d}\chi\, \frac{q_{i}^{\delta_{g}}\left(\frac{\ell+1/2}{\chi},\, \chi\right)\,q_{j}^{\delta_{g}}\left(\frac{\ell+1/2}{\chi},\, \chi\right)}{\chi^{2}}\,P_{\delta}\left(\frac{\ell+1/2}{\chi}, \, \chi\right) \, ,
\end{equation}
and the cross correlation of the clustering and shear kernels, the galaxy-galaxy lensing power spectrum, $C_{ij}^{\delta_{g}\gamma}(\ell)$,

\begin{multline}
    \label{eq:harmonic-galgal}
    C_{ij}^{\delta_{g}\gamma}(\ell)=\int_{0}^{\chi_{\mathrm{H}}}\mathrm{d}\chi\,\frac{q_{i}^{\delta_{g}}\left(\frac{\ell+1/2}{\chi},\, \chi\right)\,q^{\kappa}_{j}(\chi)}{\chi^{2}}\,P_{\delta}\left(\frac{\ell+1/2}{\chi}, \, \chi\right) \, .
\end{multline}
We note that these expressions for the 3$\times$2pt signal do not include the effects of intrinsic alignments, which we did not consider in this work.

\subsection{Estimators for the 3$\times$2pt signal}
\label{subsec:estimators}

\subsubsection{Cosmic shear estimators}
\label{subsubsec:estimators-cosmic-shear}

The shear, $\gamma(\bm{\theta})$, which describes how weak lensing `stretches' and `rotates' an image, is a spin-2 field. It can be expressed as a complex quantity,

\begin{equation}
    \gamma(\bm{\theta})=\gamma_{1}(\bm{\theta})+\mathrm{i}\,\gamma_{2}(\bm{\theta}) \, .
\end{equation}
On the full sky, the shear signal can be expressed in spherical harmonic space via the $\prescript{}{\pm2}Y_{\ell, m}(\bm{\theta})$ spin-2 harmonics defined on the sphere,

\begin{equation}
    \gamma_{1}(\bm{\theta})\pm\mathrm{i}\,\gamma_{2}(\bm{\theta}) = \sum_{\ell, m}\,\prescript{}{\pm2}{\Tilde{\gamma}_{\ell, m}}\prescript{}{\pm2}{Y_{\ell,m}}(\bm{\theta}) \, ,
\end{equation}
where the $\Tilde{\gamma}_{\ell, m}$ coefficients can be expressed using the gradient $E$- and curl $B$-modes analogous to the $Q$ and $U$ Stokes parameters for an electromagnetic field (\citealt{Crittenden02}, \citealt{Kamionkowski98}, \citealt{Stebbins96}, \citealt{Kaiser92}),

\begin{equation}
    \prescript{}{\pm2}{\Tilde{\gamma}_{\ell,m}} = E_{\ell, m} \pm \mathrm{i}B_{\ell, m} = \int \left(\gamma_{1}\pm \mathrm{i}\gamma_{2}\right)\prescript{}{\pm2}{Y^{*}_{\ell, m}}(\bm{\theta})\,\mathrm{d} \bm{\theta} \, ,
\end{equation}
and the coefficients for each mode are given by

\begin{equation}
    E_{\ell, m} = \frac{1}{2}\int \left[\gamma(\bm{\theta})\prescript{}{+2}{Y^{*}_{\ell, m}}(\bm{\theta}) + \gamma^{*}(\bm{\theta})\prescript{}{-2}{Y^{*}_{\ell, m}}(\bm{\theta})\right] \mathrm{d}\bm{\theta} \, ,
\end{equation}

\begin{equation}
    B_{\ell, m} = -\frac{\mathrm{i}}{2}\int \left[\gamma(\bm{\theta})\prescript{}{+2}{Y^{*}_{\ell, m}}(\bm{\theta}) - \gamma^{*}(\bm{\theta})\prescript{}{-2}{Y^{*}_{\ell, m}}(\bm{\theta})\right] \mathrm{d}\bm{\theta} \, ,
\end{equation}
where each of these integrals are done over the full sphere. From the harmonic coefficients, we define the $E$-mode, $B$-mode, and $EB$ angular power spectra as

\begin{equation}
    \left<E_{\ell, m}\,E^{*}_{\ell', m'}\right>=\delta_{\ell\ell'}\,\delta_{mm'}\,C^{EE}(\ell) \, ,
\end{equation}

\begin{equation}
    \left<B_{\ell, m}\,B^{*}_{\ell', m'}\right>=\delta_{\ell\ell'}\,\delta_{mm'}\,C^{BB}(\ell) \, ,
\end{equation}

\begin{equation}
    \left<E_{\ell, m}\,B^{*}_{\ell', m'}\right>=\delta_{\ell\ell'}\,\delta_{mm'}\,C^{EB}(\ell) \, ,
\end{equation}
where $\delta_{XY}$ is the Kronecker delta and the triangular brackets denote an ensemble average over many realisations. Since we are only observing one Universe and are limited by the number of $m$ modes available, each $C(\ell)$ value is distributed among $2\ell+1$ degrees of freedom. Consequently, we can only measure an estimate of the true power spectrum, $\hat{C}(\ell)$, which, for the given tomographic bins $i, j$, we define as

\begin{equation}
    \hat{C}^{\alpha\beta}_{ij}(\ell)=\frac{1}{2\ell + 1} \sum_{m=-\ell}^{+\ell}\alpha_{\ell,m}^{(i)}\,\beta^{(j)\,*}_{\ell,m} \, ,
\end{equation}
where the indices $\alpha, \beta$ denote the modes for the shear power spectra, $\left[\hat{C}^{EE}_{ij}(\ell), \hat{C}^{BB}_{ij}(\ell), \hat{C}^{EB}_{ij}(\ell)\right]$. In the weak lensing regime, assuming the Born approximation, it has been shown that to leading order, the shear field contains {negligible} power arising from the $B$-mode curl component
(\citealt{Hilbert09}, \citealt{Krause_Hirata_10}), and the coefficient $B_{\ell, m}$ is taken as zero (e.g. \citealt{Giocoli16}). 

Therefore, the shear signal is typically fully characterised by the $E$-mode, which we use to estimate the theoretical shear power spectrum in this work (i.e. $C_{ij}^{\gamma\gamma}(\ell)=\left<\hat{C}_{ij}^{EE}(\ell)\right>$). The $B$-mode power spectrum can be used as an important null test for systematics. Finally, observed shear power spectra will typically contain a noise component. For this work, we consider the shape noise, associated with the intrinsic elliptical shape of galaxies (see also Sect.~\ref{subsubsec:methods-shape-noise}). This effect is random and uncorrelated between tomographic bins, so the final measured shear signal, $\hat{C}_{ij}^{EE, \mathrm{obs}}(\ell)$, contains a contribution from the shape noise in the autocorrelation power spectra,

\begin{equation}
    \label{eq:shear_powerspectrum_noise}
    \hat{C}_{ij}^{EE, \mathrm{obs}}(\ell) = \hat{C}_{ij}^{EE}(\ell) + N_{i}^{\mathrm{\epsilon}}(\ell)\,\delta_{ij}\, ,
\end{equation}
where $N_{i}^{\mathrm{\epsilon}}(\ell)$ is the power spectrum of the shape noise in the autocorrelation tomographic power spectra.

\subsubsection{Galaxy clustering estimators}
\label{subsubsec:estimators-galaxy-clustering}

The angular galaxy clustering signal is estimated using the normalised galaxy number count on a pixelised sky, which traces the underlying matter distribution. For the tomographic bin, $i$, we formulate the real-space overdensity estimator,

\begin{equation}
    \label{eq:gal_overdensity}
    \hat{\delta}_{g}^{(i)}(\bm{\theta})=\frac{n_{i}^{\mathrm{p}}(\bm{\theta})-\Bar{n}_{i}^{\mathrm{p}}}{\Bar{n}_{i}^{\mathrm{p}}} \, ,
\end{equation}
where $n_{i}^{\mathrm{p}}(\bm{\theta})$ is the per-pixel integer galaxy count in the redshift bin and $\Bar{n}_{i}^{\mathrm{p}}$ is the average number of galaxies observed per pixel in the same bin. 

Analogously to the shear field, the measured galaxy overdensity field can be represented in spherical harmonic space via the overdensity coefficients, $d_{\ell,m}$,

\begin{equation}
    \hat{\delta}_{g}^{(i)}(\bm{\theta})=\sum_{\ell, m} d_{\ell, m}^{(i)} Y_{\ell, m}(\bm{\theta}) \, .
\end{equation}
From the $d_{\ell, m}$ coefficients, we can measure the angular power spectrum of the estimated density field,

\begin{equation}
    D_{ij}(\ell)=\frac{1}{2\ell + 1}\sum_{m=-\ell}^{+\ell}d_{\ell,m}^{(i)}d_{\ell,m}^{(j)\,*} \, .
\end{equation}
The measured power spectrum $D_{ij}(\ell)$ contains contributions from both the underlying galaxy clustering signal and the Poisson noise associated with the discretised number counts in Eq.~(\ref{eq:gal_overdensity}). Explicitly, the Poisson noise contributes to the auto-power spectra in a tomographic analysis, while the power spectrum of the Poisson noise in a tomographic bin $i$ is given by (e.g. \citealt{Loureiro19}, \citealt{Nicola20})

\begin{equation}
    \label{eq:Poisson_nl}
    N_{i}^{\mathrm{P}}(\ell)=\Omega_{\mathrm{p}}\,\frac{\Bar{w}}{\Bar{n}_{i}^{\mathrm{p}}} \, ,
\end{equation}
where $\Omega_{\mathrm{p}}$ is the angular size in steradians of the pixels that form the pixelised overdensity map, and $\Bar{w}$ is the average value of the mask over all pixels on the sky. For this work, we used a binary mask where the pixels were set to 1 inside the observed footprint and 0 outside the footprint. The Poisson noise and underlying galaxy clustering signal are uncorrelated, so the angular power spectrum of the estimated density field is simply the sum of the two components:

\begin{equation}
    \label{eq:cl-poisson}
    D_{ij}(\ell)=C_{ij}^{\delta_{g}\delta_{g}}(\ell)+N_{i}^{\mathrm{P}}(\ell)\,\delta_{ij} \,.
\end{equation}
 
\subsubsection{Galaxy-galaxy lensing estimators}
\label{subsubsec:estimators-galaxy-galaxy}

To measure the galaxy-galaxy lensing signal, we can estimate the angular power spectrum from the harmonic coefficients of the density and shear fields. By taking $B_{\ell, m}=0$, the galaxy-galaxy lensing power spectrum is characterised by the cross-correlation of the density field and the shear $E$-mode, 

\begin{equation}
     \hat{C}_{ij}^{\delta_{g}\gamma}(\ell) = \hat{C}_{ij}^{\delta_{g}E}(\ell) = \frac{1}{2\ell + 1} \sum_{m=-\ell}^{+\ell}d_{\ell,m}^{(i)}E_{\ell, m}^{(j)\,*} \, ,
\end{equation}
for the given tomographic bins $i, j$.

\subsection{Estimators on the cut sky}
\label{subsec:pseudo-cl-estimators}

The above framework is defined for the case of a full sky being observed. In reality there will be partial sky coverage due to unobserved regions, or bright stars, meaning that a survey mask, $W(\bm{\theta})$, is applied to the shear and overdensity fields, which leads to mode mixing in the derived power spectra. These `pseudo' power spectra measured on the cut sky, $\Tilde{C}_{\ell}$, can be related to the true underlying full-sky elements, ${C}_{\ell}$, by a mode-mixing matrix, $M_{\ell\ell'}$ (see e.g. \citealt{Peebles73}, \citealt{Brown05}, and \citealt{Hikage11}), which contains the coupled harmonic space information about the survey mask,

\begin{equation}
    \label{Eq:mixing_matrix}
    \left<\Tilde{C}_{\ell}^{(i,j)}\right>=\sum_{\ell'}M_{\ell\ell'}^{(i,j)}\,C_{\ell'}^{(i,j)} \, .
\end{equation}
Here, the angular brackets denote an expectation value, and the tomographic dependence of the power spectra are introduced using the superscripts $(i, j)$.  In addition, $C_{\ell}$ denotes the elements of a column vector representing $C(\ell)$ evaluated at integer $\ell$, which we will adopt as a shorthand description of the power spectrum for the remainder of this work.

For our analysis, we worked with the Pseudo-$C_{\ell}$ power spectrum of the signal measured directly from the catalogue data, $\Tilde{S}_{\ell}$. The expected value for this measured signal is calculated as the sum of the cosmological signal on the masked sky and the noise,

\begin{align}
    \label{eq:data_pcl}
    \nonumber\left<\Tilde{S}_{\ell}^{(i,j)}\right>&=\left<\Tilde{C}_{\ell}^{(i,j)}\right> + \left<N_{\ell}^{(i,j)}\right>
    \\
    &=\left(\sum_{\ell'}M_{\ell\ell'}^{(i,j)}\,C_{\ell'}^{(i,j)}\right) + \left<N_{\ell}^{(i,j)}\right> \, .
\end{align}
Here, the noise term $N_{\ell}^{(i,j)}$ arises due to shape noise for the cosmic shear field and Poisson noise for the galaxy clustering field, and is only non-zero for the auto-power spectra. A full description of the matrix $M_{\ell\ell'}$ for each 3$\times$2pt component is given in Appendix~\ref{sec:appendix-cut-sky-estimators}. Throughout our work we measure the 3$\times$2pt Pseudo-$C_{\ell}$ power spectra using \texttt{NaMaster} (\citealt{Alonso19}).

\subsection{Bandpowers}
\label{subsec:bandpowers}

Following the technique of \cite{Hivon02}, it is often convenient to bin the Pseudo-$C_{\ell}$ measured on the cut sky in $\ell$ space. This gives a reduction in the errors on the power spectrum estimator and allows the mixing matrix (Eq.~\ref{Eq:mixing_matrix}) to remain invertible in the case of large and complex sky cuts. For a set of bins, denoted by the index $b$, with bin boundaries $\left\{\ell_{\mathrm{low}}^{b}<\ell_{\mathrm{high}}^{b}<\ell_{\mathrm{low}}^{b+1}<\ell_{\mathrm{high}}^{b+1}< ...\right\}$ we take a binning operator, $P_{b\ell}$, defined as

\begin{equation}
    P_{b\ell}=
    \begin{cases}
    \,\frac{1}{2\pi}\,\frac{\ell(\ell+1)}{\ell_{\mathrm{low}}^{b+1}-\ell_{\mathrm{low}}^{b}} \, , & \mathrm{if} \hspace{0.25cm} 2\leq\ell_{\mathrm{low}}^{b}\leq\ell<\ell_{\mathrm{low}}^{b+1}\\
    \,0 \, ,& \mathrm{otherwise} \, .
    \end{cases}
\end{equation}

The elements of the binned Pseudo-$C_{\ell}$ power spectrum are termed `bandpowers' and calculated as

\begin{equation}
    \label{eq:bandpowers}
    \Tilde{\mathcal{C}}_{b}=\sum_{\ell} P_{b\ell}\Tilde{C}_{\ell} \, .
\end{equation}
We note that our approach to estimate angular power spectra is similar to the analysis that is expected to be applied to the \textit{Euclid} DR1 sample \citep{EP-tessore}.

\section{Methods: Simulating mock catalogues}

\label{sec:methods-catalogues}

In this section, we describe the modelling process and present the simulation pipeline with which we construct Stage IV-like mock cosmic shear catalogues\footnote{Our code for this simulation and analysis pipeline, Simulator for WEak lensing Power spectrum Tomography (\texttt{SWEPT}), is found at \url{https://github.com/j-hw-wong/SWEPT}}.

For a given fiducial cosmology and background galaxy redshift distribution, we first generated a theoretical prediction for the full-sky 3$\times$2pt power spectra, $\left[C_{ij}^{\gamma\gamma}(\ell), C_{ij}^{\delta_{g}\,\delta_{g}}(\ell), C_{ij}^{\delta_{g}\gamma}(\ell)\right]$, at multiple, finely spaced redshift intervals. From the 3$\times$2pt data vectors, we generated a realisation on the sky of the galaxy clustering, $\delta_{g}(\bm{\theta})$, and shear, $[\gamma_{1}(\bm{\theta}), \gamma_{2}(\bm{\theta})]$, fields at each point in redshift space covered by our chosen galaxy sample. In this way, we approximated the full 3D cosmological information with a set of 2D sky maps that represents the behaviour of the 3$\times$2pt observable fields as the background Universe evolves over time and redshift.

Finally, we performed a random sampling routine and Poisson sampled the galaxy clustering fields at each redshift to populate the sky with galaxies that trace the underlying matter distribution. We constructed a mock catalogue by then assigning the weak lensing observables $[\kappa(\bm{\theta}), \gamma_{1}(\bm{\theta}), \gamma_{2}(\bm{\theta})]$ from the field values of the cosmic shear maps at the positions of each pixel hosting a galaxy.

This simulation pipeline consisted of a number of distinct stages of calculation that were executed in series. We describe each step in detail below. In Fig.~\ref{fig:simulation_pipeline} we present a flowchart diagram showing the workflow of our simulation to produce the final mock catalogues.

\subsection{Generation of the n(z) redshift sample}
\label{subsec:methods-nz}

The first step is the construction of a galaxy population that follows a given $n(z)$ redshift distribution. Following \cite{Sipp21}, we take a redshift probability distribution, $p(z)$, that has the functional form typical of a Stage IV survey,

\begin{equation}
    \label{eq:pz}
    p(z) \propto \left(\frac{z}{z_{0}}\right)^{2}\exp\left[-\left(\frac{z}{z_{0}}\right)^{1.5}\right] \, ,
\end{equation}
where $z_{0}$ is the characteristic redshift constant chosen such that the median redshift of the sample satisfies $z_{\mathrm{median}}=\sqrt{2}\,z_{0}$. We expect that this chosen form for $p(z)$ will be a good approximation for the \textit{Euclid} DR1 galaxy sample. We note that the measured cosmological constraints would change if the real galaxy distribution is observed to be significantly different. 

Following the definition of the redshift probability distribution, we then set the simulated redshift range, $[z_{\mathrm{min}}, z_{\mathrm{max}}]$, and a precision, $\diff z$, to which galaxy redshifts are generated. A raw galaxy sample, $n(z)$, is then created by performing a sampling routine to draw a chosen total number of galaxies, $N_{\mathrm{gal}}$, that traces the $p(z)$ redshift probability distribution.  

We highlight that the functional form for $p(z)$, the total observed galaxy density, $N_{\mathrm{gal}}$, and the redshift range and precision, $[z_{\mathrm{min}}, z_{\mathrm{max}}, \diff z]$, are left as free choices in our simulation that can arbitrarily be changed to match the specifications of a chosen survey. 

\subsection{Calculation of theoretical 3$\times$2pt data vector for each galaxy redshift}

\label{subsec:methods-theory-3x2pt-data-vector}

Following the simulation of a galaxy population, $n(z)$, we aim to generate 2D realisations on the sky of the galaxy clustering and weak lensing fields for each redshift that we sample galaxies at. To do this, we first need a prediction for the 3$\times$2pt power spectra at each redshift. This set of power spectra is constructed by converting the raw $n(z)$ sample into a `binned' population and creating a table where every column contains the galaxy number count at each specific, discrete redshift that has been sampled in the $p(z)$.

This process allows us to retain the correlations in the 3$\times$2pt fields between the different, finely spaced redshift slices. In Fig.~\ref{fig:nz_diagram} we show the \textit{Euclid} DR1-like redshift distribution we consider for our work and demonstrate the effects of binning and normalising the population. This ultimately allows us to generate a set of 2D slices of the 3$\times$2pt signal that approximates the full 3D cosmological information.

In addition to the `binned' $n(z)$, we require a matter power spectrum, $P_{\delta}(k)$, which consists of both a linear and non-linear component, and a galaxy bias model, $b(k, z)$, which represents the scale- and redshift-dependent relation between the distribution of galaxies and the underlying matter field that they trace. For the purposes of this work we assume a linear model and take the galaxy bias as a global constant, $b$. While it is beyond the scope of this study, we note that our framework has the flexibility to incorporate both scale- and redshift-dependence in the bias model, and the effects of these choices could be investigated in a future study.

To construct the final theoretical 3$\times$2pt power spectra at each redshift, we use \texttt{CosmoSIS} \citep{Zuntz15}, a parameter estimation code which specialises in the joint modelling of cosmological power spectra and exploration of parameter constraints for cosmic shear and galaxy clustering studies. Explicitly, we run a \texttt{CosmoSIS} pipeline, which executes the following modules:

\begin{itemize}
    \item  Code for Anisotropies in the Microwave Background (\texttt{CAMB}, \citealt{Lewis00}), used to model the linear matter power spectrum, which describes the evolution of density perturbations in the early Universe.
    \item \texttt{halofit\_takahashi} (\citealt{Smith03}, \citealt{Takahashi12}), used to model the matter power spectrum in the non-linear regime by scaling the linear component using fitting functions derived from simulations.
    \item \texttt{constant\_bias,}  used to define the functional form of the galaxy bias for modelling the galaxy clustering and galaxy-galaxy lensing signals. We made the assumption that on the large scales  we focus on in this work, the galaxy bias can be sufficiently described using a constant.
    \item \texttt{load\_nz}, used to process and normalise the `binned' $n(z)$ galaxy population based on the user-defined redshift boundaries of each column. The resulting $n(z)$ table products can then be used to generate the 3$\times$2pt signal.
    \item \texttt{project\_2d}, used for calculation of the theoretical 3$\times$2pt data vector by evaluating the Limber approximation to project the 3D line-of-sight information into 2D angular power spectra.
\end{itemize}
For the background cosmology in the \texttt{CosmoSIS} pipeline and our overall simulation, we worked with a flat $w_{0}w_{a}$ Universe with $\Omega_{\mathrm{m}}=0.3$ and $H_{0}=70\,\mathrm{km}\,\mathrm{s}^{-1}\,\mathrm{Mpc}^{-1}$. For the matter power spectrum generated using the \texttt{CAMB} module in \texttt{CosmoSIS}, we take the latest best fit \textit{Planck} values (see \citealt{Planck18} and references therein):

\begin{itemize}
    \item optical depth to reionisation, $\tau=0.05$;
    \item scalar spectral index, $n_{\mathrm{s}}=0.96$;
    \item primordial amplitude of power, $A_{\mathrm{s}}=2.1\times10^{9}$.
\end{itemize}
For the constant galaxy bias model, we chose a value $b=0.4$, which is chosen to avoid numerical issues encountered when generating the shear and clustering fields on the sky (see Sect.~\ref{subsec:methods-cat-compilation} for a further discussion of this point).  Lastly, we chose to work with the fiducial values $(w_{0}, w_{a})=(-1,0)$. We emphasise that the values chosen for the cosmological parameters listed in this section are free variables that can arbitrarily be changed in our simulation pipeline.

\subsection{Generation of weak lensing and clustering maps}
\label{subsec:methods-3x2pt-realisations}

To simulate the cosmological 3$\times$2pt signal on the sky, we generated a set of full-sky maps of the observables contained in the 3$\times$2pt power spectra calculated at each redshift sampled from the $n(z)$ distribution.

For the galaxy clustering component, we require the observed galaxy overdensity field, $\delta_{g}(\bm{\theta})$, a scalar quantity, while for weak lensing, there are three observable quantities on the sky: the scalar convergence field, $\kappa(\bm{\theta})$, describing the lensing-induced magnification of a source; and the components of the spin-2 shear field, $\gamma_{1}(\bm{\theta})$ and $\gamma_{2}(\bm{\theta})$.

These fields can be generated on the sky using \texttt{HEALPix} \citep{Gorski05}, a pixelisation technique which offers the ability to represent data on the sphere. To generate the pixelised maps from the 3$\times$2pt data vector, we used the Full-sky Lognormal Astro-fields Simulation Kit (\texttt{FLASK}; \citealt{Xavier16}), which creates realisations of an arbitrary set of correlated fields on the sky from their given tomographic power spectra via a Cholesky decomposition. 

While \texttt{FLASK} has the capacity to generate both lognormal and Gaussian-distributed realisations of the chosen fields, for the purposes of this work, we chose to consider only Gaussian fields, on the premise that complete information of a Gaussian field is directly captured in the power spectrum.

\subsection{Construction of mock catalogue}
\label{subsec:methods-cat-compilation}

The \texttt{FLASK} routine yields a 2D \texttt{HEALPix} map of the galaxy clustering, convergence, and cosmic shear fields at each redshift, $z_{s}$, which is sampled in the $p(z)$ distribution. To construct a mock catalogue, we first place galaxies at angular positions on the sky by Poisson-sampling the galaxy density field, $\delta_{g}^{z_{s}}(\bm{\theta}),$ at each redshift, $z_{s}$. We convert this into an observed galaxy population by Poisson-sampling the cell value to generate a pixelised integer number count of galaxies on the sky at the redshift slice, $z_{s}$,

\begin{equation}
    n_{\mathrm{gal}}^{z_{s}}(\bm{\theta})=\mathrm{Poisson}\left\{\Bar{n}_{\mathrm{gal}}^{z_{s}}[1+\delta_{g}(\bm{\theta})]\right\} \, ,
\end{equation}
where $\Bar{n}_{\mathrm{gal}}^{z_{s}}$ is the average number of galaxies per pixel on the sky at redshift $z_{s}$, which is defined from the sampled galaxy distribution, $n(z)$, in Eq.~(\ref{eq:pz}). In principle, this may vary on the sky due to effects such as the variable depth of a survey, but we did not consider such effects in this work.

To execute a Poisson sampling of the density field, the minimum value of the field must satisfy $\delta_{g}(\bm{\theta})\geq-1$. Due to the nature of Gaussian realisations on the sky, there can be random pixels generated that have the unphysical value $\delta_{g}<-1$. To reduce the impact of this effect, we took a redshift-independent galaxy bias value of $b=0.4$, which reduces the frequency of these random occurrences. We note that such values of $b<1$, while less realistic, are a choice taken by other studies in the literature (e.g. \citealt{Tessore23}) that have investigated Gaussian realisations of the 3$\times$2pt fields on the sky. Quantitatively, we would expect that (to first order) this choice will not affect the cosmic shear measurements, since the bias term does not feature in the weak lensing kernel (Eqs.~\ref{eq:harmonic-shear},~\ref{eq:shear-kernel}). For the angular clustering and galaxy-galaxy lensing signals, the bias term acts to scale the amplitude of the power spectrum. Since we used a constant bias over all angular scales and redshift, we expect that the choice of $b$ would change the absolute values of constraints on $(w_{0},w_{a})$ measured from our simulations, but would not alter the relative measurements that compare the constraints between different tomographic binning schemes. 

In addition, we rejected the redshift range $z<0.3,$ in which the evolution of galaxy clustering is more likely to produce such unphysical pixels. However, we note that at this redshift range there is also the least amount of information contained in the cosmic shear signal, since there is comparatively little foreground structure to induce the weak lensing effect. 

We assigned an angular (RA, Dec) position to each galaxy by placing them at the centre of the pixel they are Poisson-sampled in. The final galaxy `observation' to create a mock shear catalogue is then a straightforward assignment process. Each galaxy is described with a data vector of its (RA, Dec, $z$) values. We then append to this observed data the value of the cosmic shear $[\kappa(\bm{\theta}), \gamma_{1}(\bm{\theta}), \gamma_{2}(\bm{\theta})]$ fields at the specific sky position and redshift that the galaxy is found at. By repeating this procedure for each galaxy in our sampled population, we can produce a final cosmic shear catalogue, which:

\begin{itemize}
    \item consists of a galaxy population that traces a survey-specified number density and predicted redshift probability distribution, $p(z)$;
    \item statistically represents on the sky the information contained in the theoretical 3$\times$2pt power spectra calculated from a chosen underlying cosmology. 
\end{itemize}

\subsection{Pipeline overview}
\label{subsec:methods-overview}

\begin{figure}
        \includegraphics[width=\columnwidth]{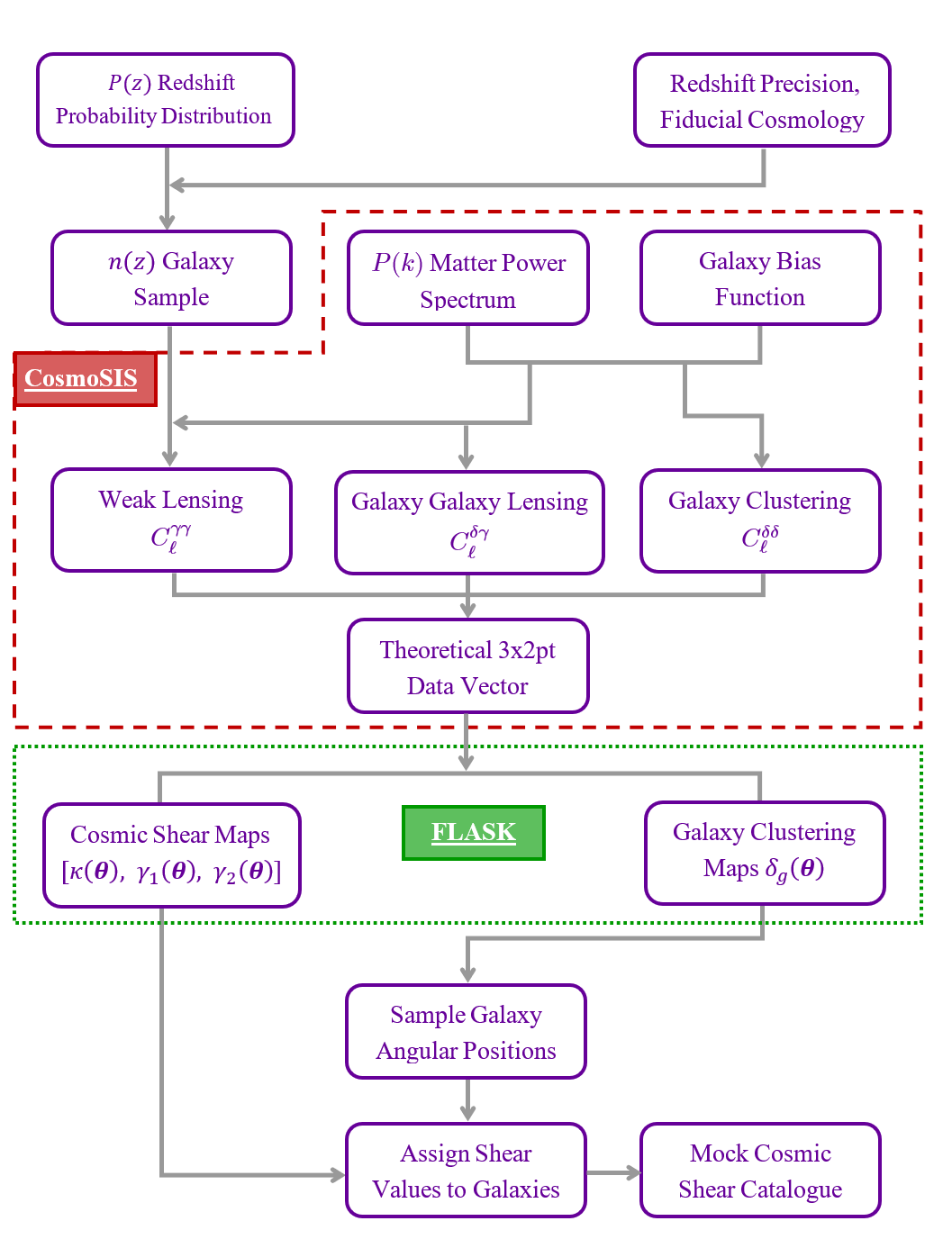}
    \caption{Flowchart describing the structure of our simulation to generate Stage IV-like mock galaxy catalogues from 2D realisations on the sky of the 3$\times$2pt signal that traces an underlying redshift distribution and fiducial cosmology. The red dashed and green dotted sections represent the simulation stages that are executed using \texttt{CosmoSIS} and \texttt{FLASK}, respectively.}
    \label{fig:simulation_pipeline}
\end{figure}

The overall workflow of our pipeline is presented in Fig.~\ref{fig:simulation_pipeline}. Our simulation-based approach offers an approximate, but relatively rapid technique for exploring the impact of different analysis choices on the precision of cosmological constraints that can be achieved with \textit{Euclid}. While a full $N$-body simulation to create a catalogue would provide the most complete astrophysical information, it is extremely computationally costly to generate and would typically yield only a few realisations. Correspondingly, introducing or removing any additional astrophysical or observational features in the simulation to test their impact on observables and cosmological measurements would be complex and inefficient.

As discussed in Sect.~\ref{sec:motivations}, previous studies on the choice of 3$\times$2pt tomography have typically focussed only on exploring a pre-defined signal, without generating a mock observation on the sky. 
Our method offers an alternative to such approaches. Since we are replicating a real analysis by measuring the 3$\times$2pt signal from the simulated data, we have a unique capability to introduce complex observational and astrophysical effects at the catalogue level, and rapidly explore their effects on measurements of cosmological parameters. We note that by using the map-based simulation approach, we can practically generate and work with many realisations of the Universe to further improve the statistical power of the analysis. To demonstrate our method, we focussed on two sources of uncertainty: photo-z errors and shape noise. We discuss them in the following sections and emphasise that a comprehensive suite of systematics can be explored in a follow-up investigation(s).

\subsection{Simulated noise and errors}
\label{subsec:methods-errors}

The basic execution of our pipeline contains intrinsic Poisson noise in the galaxy clustering signal, which has the noise power spectrum given in Eqs. (\ref{eq:Poisson_nl},~\ref{eq:cl-poisson}). In addition to the Poisson noise, we can also increase the realism of our simulation by introducing redshift uncertainties and shape noise in the galaxies, which are included as an optional feature in our simulation. 

\subsubsection{Redshift uncertainties}
\label{subsubsec:methods-photoz-uncertainties}

For a realistic survey, we require either spectroscopic or photometric analysis of individual galaxies to measure their redshift. While the spectroscopic measurement offers a very precise estimate of the galaxy's true redshift, it is unfeasible to yield a spectroscopic redshift for every galaxy in a survey. Comparatively, there are much larger uncertainties when deriving redshifts via the photometric method, which relies on fitting the galaxy's observed spectral energy distribution (see e.g. \citealt{Ilbert-EP11} for an overview of redshift measurement techniques for \textit{Euclid}).

When constructing the final survey catalogues in our pipeline (Sect.~\ref{subsec:methods-cat-compilation}) we can optionally introduce uncertainty in the galaxy redshifts associated with such measurement techniques. Explicitly, we expect that, to first order, the photometric uncertainty on a measured redshift will be Gaussian-distributed with standard deviation, $\sigma_{z}$, dependent on the redshift,

\begin{equation}
    \sigma_{z}=\sigma_{z}^{\mathrm{phot}}(1+z) \, ,
\end{equation}
where $\sigma_{z}^{\mathrm{phot}}$ is a constant.  To introduce this effect into our pipeline, for each galaxy at redshift $z$, we convert the `true' redshift of the galaxy in the catalogue to a `measured' redshift by drawing an estimated value from a Gaussian distribution with standard deviation $\sigma_{z}$ and mean $z$.

In addition to the Gaussian-distributed uncertainty in the photo-$z$s, there will be some catastrophic outliers due to limitations in the photometric fitting procedure. Explicitly, the measured SED of the galaxy can be fit to a specific template (see e.g. \citealt{Desprez-EP10} for an overview). However, confusion between given pairs of spectral lines can lead to a measured galaxy redshift that is catastrophically inaccurate. In the technique of \cite{Jouvel11}, we can model a `catastrophic' redshift estimate using the expression,

\begin{equation}
    \label{eq:catastrophic-photozs}
    z_{\mathrm{cata}}=(1+z)\,\frac{\lambda_{\mathrm{break-rf}}}{\lambda_{\mathrm{break-cata}}} -1\, ,
\end{equation}
where $\lambda_{\mathrm{break-rf}}$ is the true rest-frame wavelength of a break feature in an SED model, and $\lambda_{\mathrm{break-cata}}$ is the wavelength of the break feature that is predicted in the fitting procedure. As a first-order approximation, in the redshift range $z\leq2.5$, \cite{Jouvel11} suggest that the Lyman-$\alpha$ line, Lyman-break, Balmer-break and 4000-\r{A} break (D4000), and permutations between pairs of these four features, will contribute to the catastrophic redshift estimation. Within our simulation, we include the option to model this effect by converting the true redshifts of a chosen percentage of the galaxy population into a catastrophic measurement based on a given set of spectral line pairs.

\subsubsection{Shape noise}
\label{subsubsec:methods-shape-noise}

A further significant source of noise in a weak lensing analysis is the shape noise. Galaxies have an intrinsic ellipticity, $\epsilon^{\mathrm{int}}$, which acts as an irreducible source of confusion when estimating the weak lensing-induced shape distortion, $\gamma$. Under the assumption that galaxies have no preferred ellipticity on average, we can model the shape noise to be Gaussian-distributed with standard deviation $\sigma(\epsilon^{\mathrm{int}})$, known as the total ellipticity dispersion.

We introduce this as an optional effect in our simulation by drawing shape noise error quantities from a Gaussian distribution with standard deviation $\sigma(\epsilon^{\mathrm{int}})$. This error is then attributed to each galaxy by adding the component $\sigma(\epsilon^{\mathrm{int}})/\sqrt{2}$ to each of the $\gamma_{1}(\bm{\theta})$ and $\gamma_{2}(\bm{\theta})$ galaxy observable values in the catalogue.

Finally, when measuring tomographic shear power spectra from the catalogues, we model the uncertainty due to this shape noise using Eq.~(\ref{eq:shear_powerspectrum_noise}) for the cut sky signal, with the following expression for the shape noise contribution to the $E$-mode power spectrum,

\begin{equation}
    \Tilde{N}_{i}^{\epsilon}(\ell)=\frac{\left[\sigma(\epsilon^{\mathrm{int}})/\sqrt{2}\right]^{2}}{{1/{N}_{i}^{\mathrm{P}}}} \, ,
\end{equation}

\noindent where ${N}_{i}^{\mathrm{P}}$ is the {Poisson noise power spectrum} associated with the tomographic bin $i$ (Eq.~\ref{eq:Poisson_nl}). In this way, we ensure that the shape noise only contributes to the auto-correlation power spectra.

\section{Inference routine}
\label{sec:inference-routine}

\subsection{Grid-based Gaussian likelihood}
\label{subsec:grid-based-gaussian-likelihood}

The central aim of this work is to explore the 3$\times$2pt signal for different tomographic configurations, and identify the optimum tomographic binning strategy that provides the best constraints for the time evolving dark energy equation of state.  To derive parameter constraints, we perform a grid-based inference routine using a Gaussian likelihood in which we vary the $(w_{0}, w_{a})$ parameters in a $w_{0}w_{a}$CDM cosmology (see Sect.~\ref{sec:introduction}) while all other cosmological parameters are held at the fixed fiducial values presented in Sect.~\ref{subsec:methods-theory-3x2pt-data-vector}. An in-depth discussion on the validity of the Gaussian likelihood and the description of the grid-based framework adopted is given in \cite{Upham21}. Explicitly, the multivariate Gaussian likelihood is given by

\begin{equation}
    \mathcal{L}=\frac{1}{(2\pi)^{k/2}\,|\bm{\tens{C}}|^{1/2}}\,\exp\left[(\bm{D}-\bm{M})^{\mathrm{T}}\bm{\tens{C}}^{-1}(\bm{D}-\bm{M})\right], \, 
\end{equation}
where $\bm{D}$ is the (mock) observed data vector of length $k$, $\bm{M}$ is a given theory data vector, and $\bm{\tens{C}}$ is the covariance matrix for the data vector $\bm{D}$. The 2D posterior distribution, $p(\bm{\Theta} \, | \, \bm{D})$, of the model parameters, $\bm{\Theta}$, is then given by Bayes' theorem,

\begin{equation}
    p(\bm{\Theta} \, | \, \bm{D}) \propto \pi(\bm{\Theta}) \mathcal{L} \, (\bm{D} | \bm{\Theta}) \, ,
\end{equation}
where $\mathcal{L}(\bm{D} \, | \, \bm{\Theta})$ is the likelihood and $\pi(\bm{\Theta})$ is the prior knowledge. For this work, we assumed a flat prior in $(w_{0}, w_{a})$.

\subsection{Numerical covariance matrix}
\label{subsec:numerical-covariance-matrix}

A key ingredient of the Bayesian method is the covariance matrix, $\bm{\tens{C}}$. For our work, we use a numerical covariance matrix in which we estimate the errors using a large number of realisations. After measuring the bandpowers of the 3$\times$2pt Pseudo-$C_\ell$ data vector, $\Tilde{C}_{b}$ (Eq.~\ref{eq:bandpowers}) for each realisation, the elements of the covariance matrix are then constructed from the scatter of the realisations with respect to the mean,

\begin{equation}
    \mathrm{C}_{bb'}=\left<(\tilde{C}_{b}-\tilde{C}_{b,av})(\tilde{C}_{b'}-\tilde{C}_{b',av})\right>, \, 
\end{equation}
where $\tilde{C}_{b,av}$ is the mean recovered bandpower in a band, $b$, and the angled brackets denote an average over all realisations. Then, $\mathrm{C}_{bb'}$ are the elements of the covariance matrix, $\bm{\tens{C}}$.

{
While the numerical covariance matrix provides an estimate of the true errors intrinsic in our simulation method, it will also contain numerical noise due to the finite number of simulations used. By definition, a greater number of simulations leads to more suppression of this noise, until the limit of infinite realisations where the noise will average to zero and leave only the unbiased, `true' signal in the covariance matrix. This unbiased estimate of the covariance will contain information on both the cosmological 3$\times$2pt signal, and the effects of Poisson noise and shape noise in the measurement.

 To derive unbiased constraints using the numerical covariance matrix, we apply the Hartlap correction (\citealt{Hartlap07}) to the inverse of the covariance matrix, $\bm{\tens{C}}^{-1}$,

\begin{equation}
    \bm{\tens{C}}^{-1}\rightarrow\frac{N_{S}-N_{D}-2}{N_{S}-1}\,\bm{\tens{C}}^{-1}\,,
\end{equation}

\noindent where $N_{S}$ is the number of simulations used, and $N_{D}$ is the size of the data vector. For the Hartlap correction to be applicable, we require the condition $N_{S}>N_{D}+2$. The largest data vector considered in this work will be a full 3$\times$2pt analysis with ten tomographic bins and ten bandpowers, which gives $N_{D}=2100$. Hence, we generated 3000 realisations of our simulated catalogues to define the numerical covariance matrix.}

\section{Impact of tomography on dark energy constraints}
\label{sec:results_tomography}

\begin{figure}
        \includegraphics[width=\columnwidth]{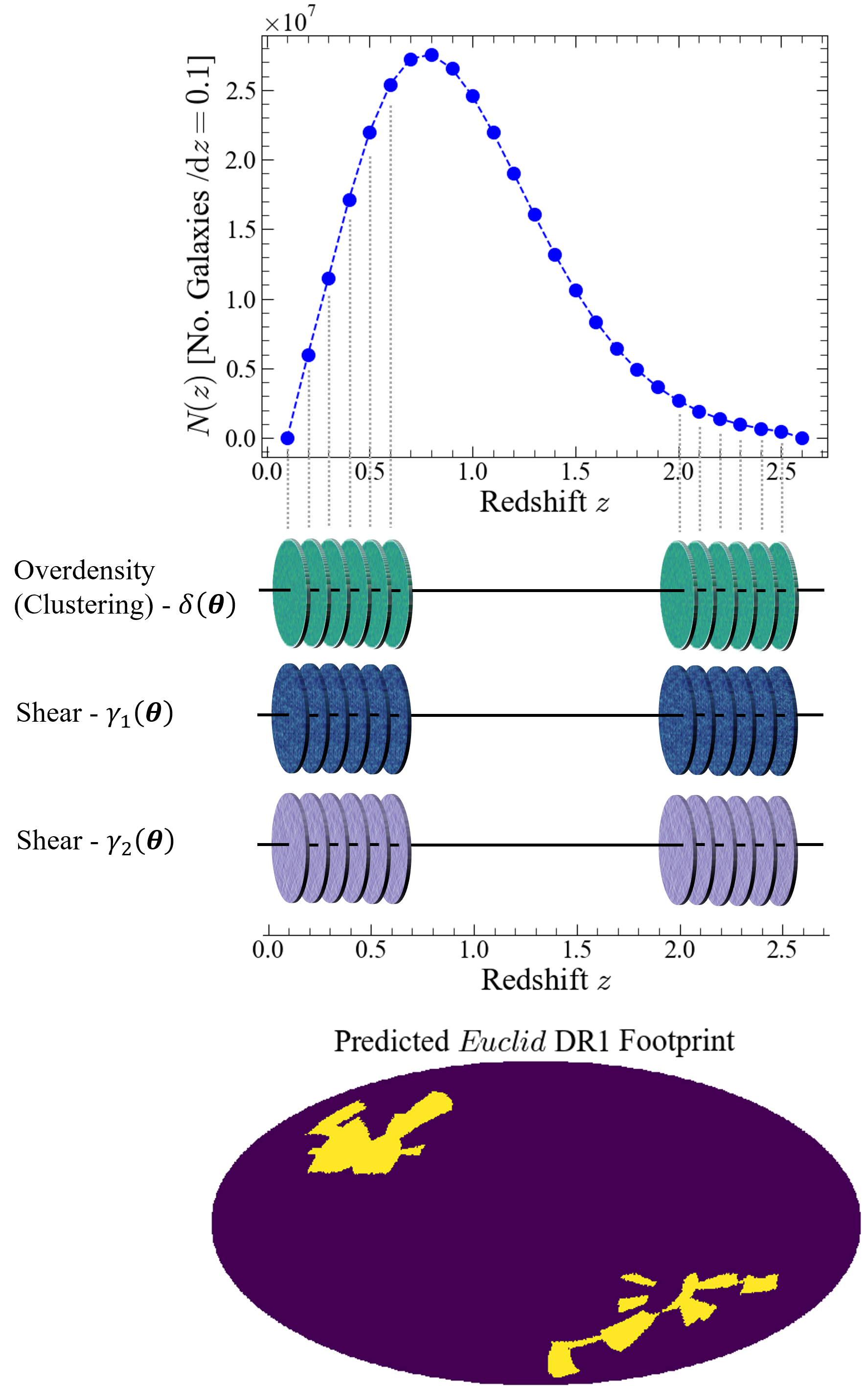}
    \caption{Illustration of the method used to simulate a mock weak lensing survey. 2D maps of the correlated 3$\times$2pt fields are generated at finely sampled points in redshift. These 2D maps are used to approximate the full 3D cosmological information of the 3$\times$2pt signal. From the overdensity fields, we Poisson sample a galaxy population that traces the underlying $n(z)$ distribution. We then assign the correlated weak lensing observables to each galaxy from the shear field values at the galaxy's angular position on the sky at a given redshift. We show an early approximation of the \textit{Euclid} DR1 footprint at the bottom of the figure which we have used to create our \textit{Euclid}-like simulations. The `observed' region is shown in yellow. Note: the actual \textit{Euclid} DR1 footprint will be significantly different to that shown here.}
    \label{fig:nz_diagram}
\end{figure}

We now present constraints on the $(w_{0}, w_{a})$ parameters measured from the {3000 realisations} of our simulated 3$\times$2pt catalogue data using the inference method described in Sect.~\ref{sec:inference-routine}, for a range of different tomographic binning configurations. For a \textit{Euclid}-like survey, we take a sample of $3\times10^{8}$ galaxies drawn in the range $0.3\leq z<2.7$ from the $p(z)$ defined in Eq.~(\ref{eq:pz}) using the constant $z_{0}=0.636$ to yield a sample with a median redshift $z_{\mathrm{med}}=0.9$, matching that predicted for \textit{Euclid} \citep{Euclid}. We place these galaxies within an approximation of the \textit{Euclid} DR1 footprint ($\sim2600$ deg$^{2}$, presented in Fig.~\ref{fig:nz_diagram}) to achieve the target galaxy number density, $30\,\mathrm{gal}/\mathrm{arcmin}^{2}$, on the sky. We use these survey characteristics for all results presented in this work. 

For generating the 3$\times$2pt fields on the sky, we use a \texttt{HEALPix} grid with resolution $N_{\mathrm{side}}=1024$. This allows us to generate catalogues from power spectra that are evaluated up to angular scales of $\ell=2000$. Figure~\ref{fig:nz_diagram} shows the $n(z)$ model that we use alongside examples of the generated 3$\times$2pt fields and the approximate \textit{Euclid} DR1 mask that we use. 

We sample galaxies to a redshift resolution $\diff z=0.1$, which we note is slightly more pessimistic than the target accuracy for the \textit{Euclid} photo-$z$ estimation for weak lensing science, $\sigma_{z}/(1+z)\leq0.05$. However, due to our simulation methodology, we find that this is the limiting precision that allows for a consistent recovery of the fiducial cosmology from the 2-point statistics measured from the catalogues. In Appendix~\ref{sec:results-validation} we present our validation methodology to investigate the accuracy of our simulation method and demonstrate the ability of our pipeline to self-consistently reproduce the underlying 3$\times$2pt signal predicted by the input cosmology. 

Following the procedure detailed in Sect.~\ref{subsec:estimators}, we measured the 3$\times$2pt Pseudo-$C_{\ell}$ bandpowers of the data in the catalogues to derive tomographic constraints on dark energy, meaning that the theoretical model for the full-sky power spectra needs to be convolved with the mixing matrices associated with the DR1 mask. In addition, a model for the noise components (see Eq.~\ref{eq:data_pcl}) is required for a direct comparison. We measure cosmic shear power spectra in the multipole range $100\leq\ell\leq1500$, and power spectra for the angular clustering and galaxy-galaxy lensing components in the range $100\leq\ell\leq600$, to isolate the scales where the assumption of the constant galaxy bias remains well-motivated. We evaluate all power spectra using ten bandpowers binned logarithmically in angular scale.

By fixing all other parameters in the likelihood analysis, we find that the inference process is extremely finely tuned, and even a $\ll1\%$ systematic effect in the 3$\times$2pt bandpowers (see also Appendices~\ref{subsec:validation-no-noise} and~\ref{subsec:validation-noise}) leads to a biased recovery of the fiducial parameter values, $(w_{0}, w_{a})=(-1, 0)$. Hence, to examine the effects of tomographic binning on the true cosmology, we choose to work with the fiducial data vector but use a numerical covariance matrix, which we derive from our simulated measurements (see Sect.~\ref{subsec:numerical-covariance-matrix}). This ensures that we will be probing the fluctuations about the true cosmology, but we do not expect that the size of the errorbars and the $(w_{0}, w_{a})$ contours will change, since we retain the realistic simulated noise that is observed in the 3$\times$2pt data from our catalogues.

To evaluate the performance of a given tomographic binning configuration, we plot the areas enclosed within the $(w_{0}, w_{a})$ posterior contours as a function of the number of tomographic bins used. We consider three binning cases: bins equipopulated with galaxies; bins equally spaced in redshift; and bins equally spaced in fiducial comoving distance. We consider constraints for both the `no-noise' simulation and the `realistic' set-up including Gaussian shape noise and Gaussian photo-$z$ estimation uncertainty. For the `noisy' simulation, we consider an intrinsic shape noise in the source galaxies that is described by a Gaussian parameterised by a constant, $\sigma\left(\epsilon^{\mathrm{int}}\right)=0.3$ (e.g. \citealt{Paykari-EP6}), and inject this uncertainty into the shear field values assigned to each galaxy in the catalogue using the method described in Sect.~\ref{subsubsec:methods-shape-noise}. 

To model the photo-$z$ uncertainty, we use Gaussian distributed errors with the redshift-dependent deviation discussed in Sect.~\ref{subsubsec:methods-photoz-uncertainties}, where we take the constant $\sigma_{z}^{\mathrm{phot}}=0.05$ to match the target uncertainty for the \textit{Euclid} analysis as defined in \cite{Euclid}. While we model such uncertainties in the redshift and galaxy shapes, in this section we consider only unbiased redshift and shear estimates, which we note would need to be controlled in a precision 3$\times$2pt analysis. While a comprehensive study of these issues is beyond the scope of this work, in Sect.~\ref{sec:catastrophic-photozs} we consider the effects of contamination by catastrophic redshift errors.

For both the no-noise and noisy cases, we investigate: a 1$\times$2pt cosmic shear only; a 1$\times$2pt angular clustering only; and lastly a full 3$\times$2pt analysis. For the noisy simulation measurements, we additionally plot the 1 and 2\,$\sigma$ posterior contours in the $(w_{0}, w_{a})$ plane for a selection of equipopulated binning constraints, which we find are similar in shape to the equivalent contours in the no-noise measurements.

\subsection{Signal-only simulations}

\label{subsec:results-nonoise}

\subsubsection{{1$\times$2pt cosmic shear only analysis, no-noise simulation}}
\label{subsubsec:1x2ptE_nonoise}

\begin{figure}
        \includegraphics[width=\columnwidth]{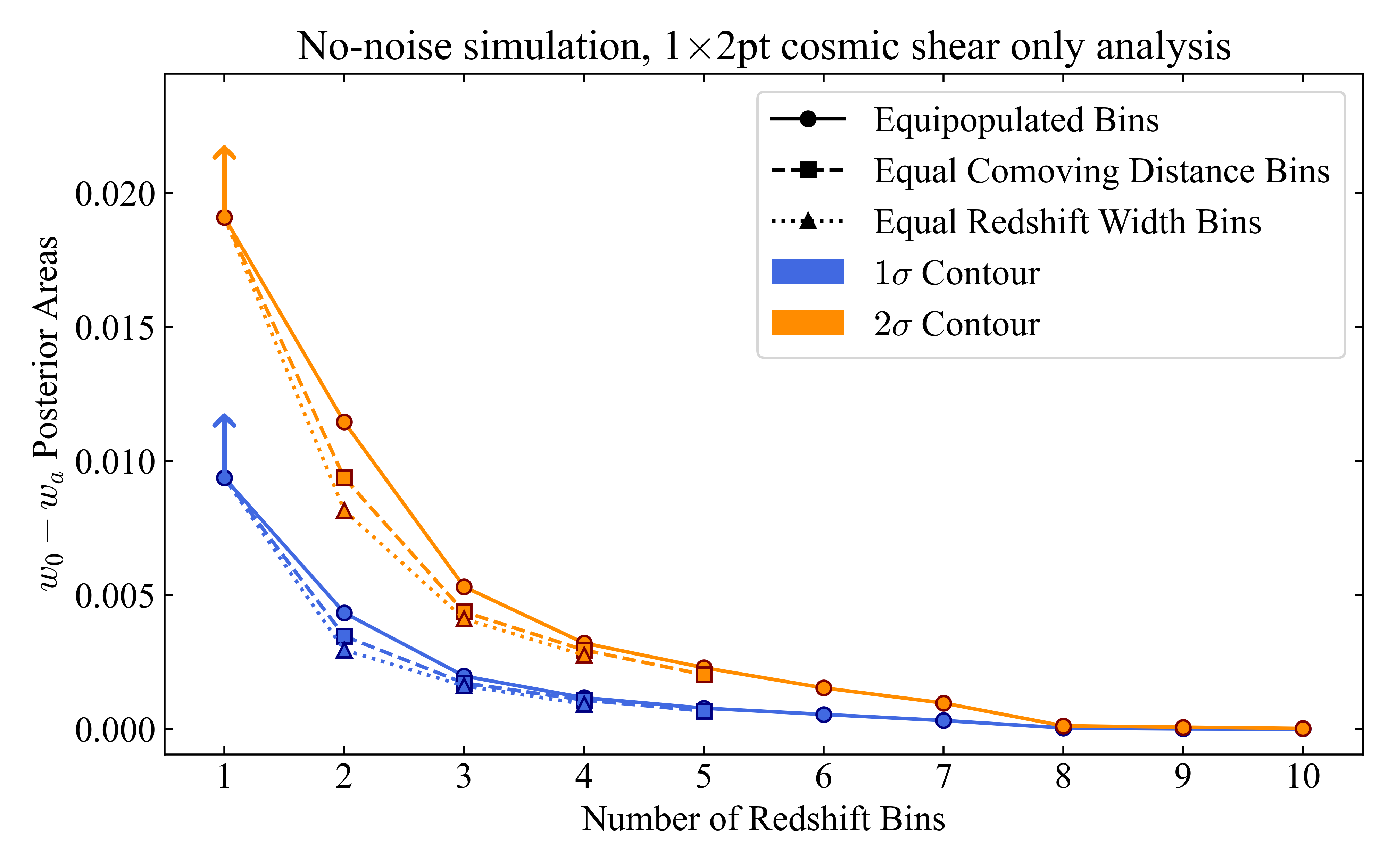}
    \caption{{Areas enclosed by the $1\,\sigma$ (Blue) and $2\,\sigma$ contours in the $(w_{0}, w_{a})$ plane, for different numbers of redshift bins used in a tomographic analysis of the cosmic shear signal measured from 3000 realisations of our simulation. We show in circular markers joined with solid lines the areas measured for the equipopulated binning choice; in square markers joined with dashed lines the equal comoving distance bins; and in triangular markers joined with dotted lines the equal redshift width binning choice. Since the cosmic shear component alone is relatively weakly constraining, the 1 bin measurement does not yield a closed contour in $(w_{0}, w_{a})$ within the ranges of the parameter grid. Hence, the data point for this case represents a lower bound of the true value, which we represent by using a vertical arrow.}}
    \label{fig:w0wa_NONOISE_1x2ptE}
\end{figure}

For a catalogue simulation in the absence of shape noise and photo-$z$ uncertainty, we measure the tomographic Pseudo-$C_{\ell}$ power spectra of the cosmic shear signal only, and plot in Fig.~\ref{fig:w0wa_NONOISE_1x2ptE} the area enclosed in the $(w_{0}, w_{a})$ plane by the 1 and 2\,$\sigma$ contours as a function of the number of redshift bins used for the analysis. We present these results for the three binning choices: equipopulated (solid line); equal redshift width (dotted line); and bins equally spaced in fiducial comoving distance (dashed line). 

We find that, comparatively, the equal redshift width bins cover the smallest area in $(w_{0}, w_{a})$ for a given number of redshift bins, indicating that this binning choice would yield the optimum constraints on the dark energy equation of state. We note that for the particular case of a single bin the shear signal is only weakly constraining and does not enclose a $2\,\sigma$ area within the prior boundaries of the parameter space. However, increasing the analysis to include just two bins provides sufficient constraining power to yield closed contours and place constraints on the dark energy parameters.

For a two-bin analysis, the difference in the contour area between each choice is at a considerable level of $\sim$15--25\%, but once three or four bins are considered for the analysis, the relative gain or loss between the binning choices reduces to the level of $<10$\%, on average. Throughout this work, we will define the percentage change in the $2\,\sigma$ contour area between any pair(s) of binning set-ups as the relative `information gain or loss' associated with the binning choices. Following the definition of the dark energy figure of merit (FOM; \citealt{Albrecht06}), whereby the FOM is taken as the inverse of the area of the $2\,\sigma$ $(w_{0}, w_{a})$ contour, our measured comparison between the areas for different binning choices similarly represents the level of improvement or degradation in the dark energy FOM value.

We note that for the equal redshift width and the equal comoving distance binning choices, the maximum number of bins that is used for the analysis is lower (four and five, respectively) than the maximum number we use for the equipopulated case. This is due to the fact that as a greater number of bins is used, there are fewer galaxies that trace the sky in a given bin, which leads to an undersampling of the observed field where pixels in the \texttt{HEALPix} map are left unfilled. For the case of a single realisation, this feature could be interpreted as a part of the mask itself and convolved with the mixing matrix formalism (see Sect.~\ref{subsec:pseudo-cl-estimators}). However, for the purposes of this work in which we consider multiple thousands of realisations, we highlight a crucial complication in that the numbers and locations of the unfilled pixels change per realisation, and then per tomographic bin, and per binning choice within each realisation. 

Hence, a new coupling matrix would need to be generated for each permutation and combination of these variations since the cosmological signal is different in each case. The simulation would be both computationally impractical, and would require a fundamentally different scientific analysis that would deserve a separate, independent investigation beyond this work. Regardless, a key point we emphasise is that as the number of redshift bins used for the tomographic analysis increases, the more similar the different binning choices become, until the limit of an infinite number of bins at which point the binning choices are identical. 

{
From Fig.~\ref{fig:w0wa_NONOISE_1x2ptE} we find that the area of the 1$\sigma$ $(w_{0}, w_{a})$ contours for the equipopulated binning choice starts to converge to a consistent value by six to seven bins. We find that only for no-noise measurements, there is additional uncertainty in the contour areas for relatively large numbers of tomographic bins, due to the redshift sampling precision, but we do not expect that this impacts any of the overall conclusions presented here and in the remainder of this work.}

In addition, we note that while the $(w_{0}, w_{a})$ constraints derived from the numerical covariance matrix are unbiased, there will be uncertainty in measurements of the contour areas due to excess numerical noise. Indeed, for a real measurement, any statistical gain in the errors on $(w_{0}, w_{a})$ that can be extracted by continuously introducing more bins will require increasing accuracy in estimating the covariance matrix.

\subsubsection{{1$\times$2pt angular clustering only analysis, no-noise simulation}}
\label{subsubsec:1x2ptN_nonoise}

\begin{figure}
        \includegraphics[width=\columnwidth]{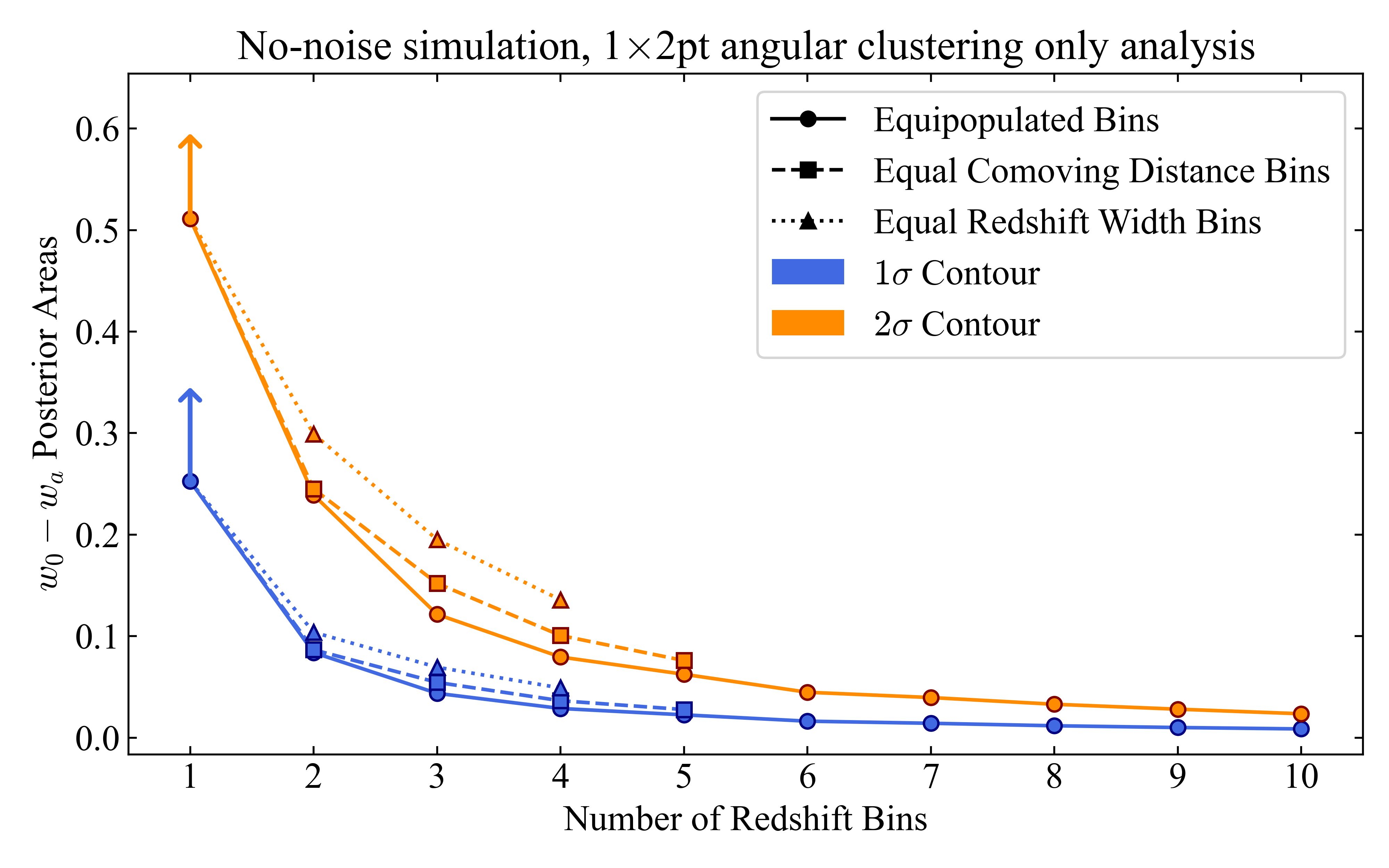}
    \caption{{Contour areas of the 1 and 2\,$\sigma$ constraints on $(w_{0}, w_{a})$, measured from the tomographic angular clustering component of our no-noise simulation. We plot the contour areas measured as a function of the number of redshift bins used in a tomographic analysis, considering the equipopulated, equal comoving distance, equally spaced in redshift binning strategies. The 1 bin measurement does not sufficiently constrain either parameter within the prior volume, hence we include a vertical arrow to denote that this data point is a lower bound.}}
    \label{fig:w0wa_NONOISE_1x2ptN}
\end{figure}

We present in this section the equivalent analysis of Sect.~\ref{subsubsec:1x2ptE_nonoise} for the clustering component of the 3$\times$2pt signal, showing in Fig.~\ref{fig:w0wa_NONOISE_1x2ptN} the areas of $(w_{0}, w_{a})$ contours derived for the three different binning choices as a function of the number of bins used for a tomographic Pseudo-$C_{\ell}$ power spectrum analysis.

Consistent with the cosmic shear measurements in Sect.~\ref{subsubsec:1x2ptE_nonoise}, we find that for a clustering-only analysis in the absence of noise, the use of two bins is required to sufficiently place firm constraints on the $(w_{0}, w_{a})$ parameters. Comparatively, our measurements demonstrate clearly that the equipopulated tomographic bins provide the optimum constraints on $(w_{0}, w_{a})$, which is in direct contrast with the results in Fig.~\ref{fig:w0wa_NONOISE_1x2ptE} that suggest that the equipopulated binning is the worst choice for the shear-only measurement.

Moreover, in the range of three to five tomographic bins, we find that there is a degradation in the clustering-only constraints on $(w_{0}, w_{a})$ of $\sim$20\% for the equal comoving distance bins, and $\sim$40\% for the equal redshift width bins, with respect to the equipopulated choice. This difference between the binning choices is a factor of $\sim$2 greater than the difference measured for the no-noise shear-only analysis.

Regarding the number of tomographic bins considered, it is clear that the rate of decrease of the area enclosed by the $(w_{0}, w_{a})$ contours converges to a constant value by six to seven bins, indicating that the information gain starts to saturate beyond this point. For the equipopulated binning choice we calculate that by using six bins the $2\,\sigma$ $(w_{0}, w_{a})$ contour decreases by $\sim$80\% compared to a two-bin analysis, and for every extra tomographic bin introduced beyond seven bins, there will only be a $\sim$1--2\% further improvement in the parameter constraints with respect to the two-bin case. 

\subsubsection{{Full 3$\times$2pt analysis, no-noise simulation}}
\label{subsubsec:3x2pt_nonoise}

\begin{figure}
        \includegraphics[width=\columnwidth]{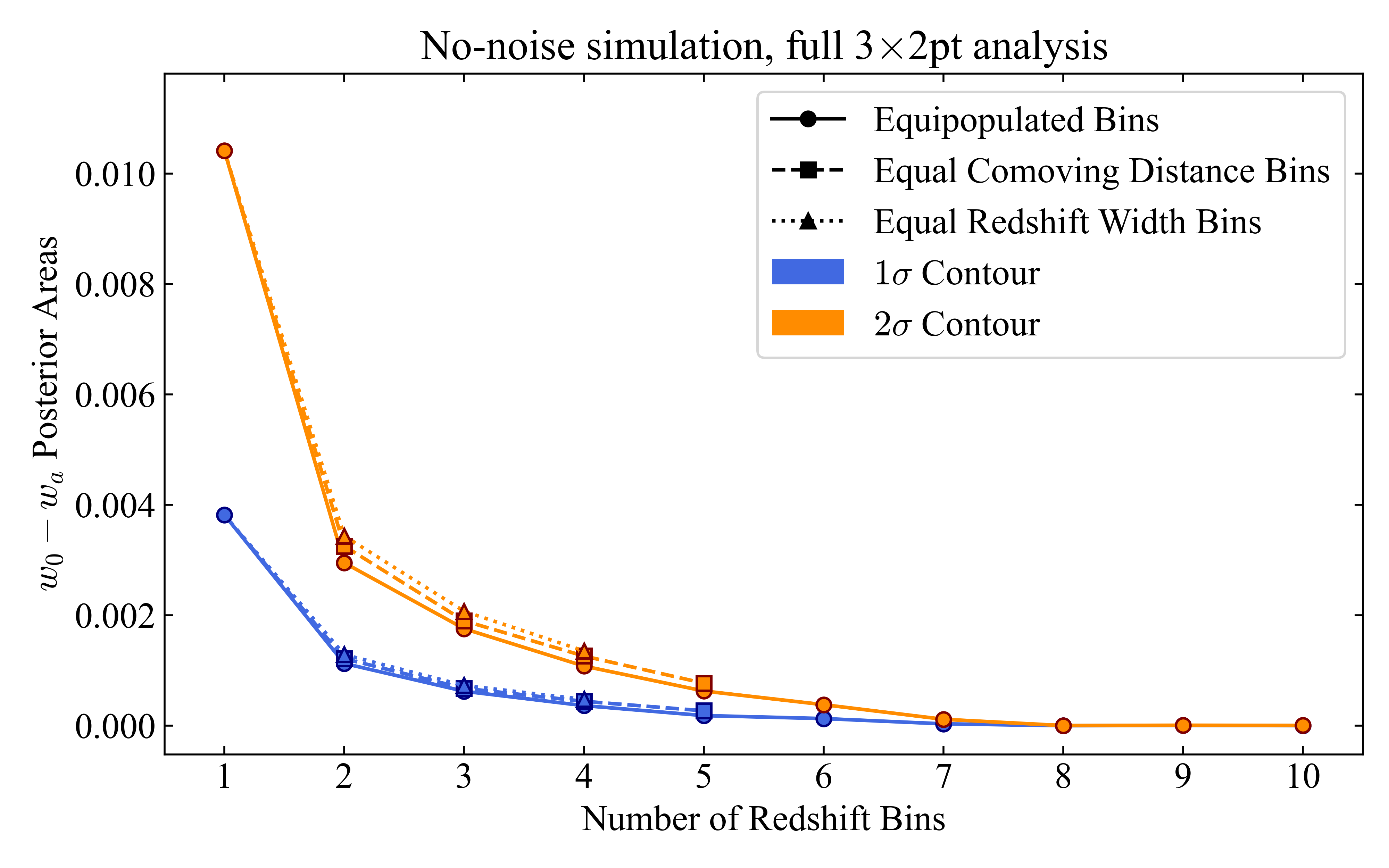}
    \caption{{Areas enclosed by the 1 and 2\,$\sigma$ contours of the posterior constraints on $(w_{0}, w_{a})$, derived from measurements of the full tomographic 3$\times$2pt signal from our `no-noise' mock catalogues. We plot the contour areas as a function of the number of redshift bins used in the tomographic analysis for the three different binning choices considered in this work.}}
    \label{fig:w0wa_NONOISE_3x2pt}
\end{figure}

For the no-noise simulations, we lastly considered a full 3$\times$2pt analysis. We plot the results of the measurements on $(w_{0}, w_{a})$ in Fig.~\ref{fig:w0wa_NONOISE_3x2pt}, demonstrating a comparison in the posterior constraints between the different binning choices. In contrast with measurements of the cosmic shear or angular clustering components
alone, we find that the full 3$\times$2pt data vector has sufficient constraining power to yield closed contours in $(w_{0}, w_{a})$ for a single redshift bin. This effectively demonstrates the joint power that the weak lensing and galaxy clustering signals have in investigating the effect of dark energy on the growth of structure in the Universe, since they probe different redshift-dependent information of the matter field along a line of sight (see also \citealt{Tutusaus20}).

By increasing the number of tomographic bins, we find that the equipopulated binning choice provides the smallest errors, followed by the equal comoving distance bins and the equal redshift width bins -- the same order of preference as for the clustering only analysis, which suggests that the tomographic behaviour of the clustering signal is the dominant component of the 3$\times$2pt data vector. The degradation in the $(w_{0}, w_{a})$ constraints with respect to the equipopulated choice for a given number of bins is $\lesssim$10\% for the equal comoving distance and $\sim$10--15\% for the equal redshift width choices, which is smaller in magnitude than for the clustering analysis. However, these magnitudes are consistent with the relative performances of the different binning choices in the shear-only analysis at small numbers of bins, which highlights the conclusion that the `optimum' binning choice is likely to be dependent on which observable is being targeted in the 3$\times$2pt measurement. 

{
We find that the contour areas, for the 3$\times$2pt case, converge by around six to eight bins. This is in good agreement with measurements of the shear- and clustering-only constraints (Figs.~\ref{fig:w0wa_NONOISE_1x2ptE},~\ref{fig:w0wa_NONOISE_1x2ptN}). Beyond this number of bins, we find that the information gain in constraints on dark energy, for any and all fields of the 3$\times$2pt signal, is relatively small in general. In this regime, we expect that for a real analysis the marginal statistical gains on the dark energy parameter constraints will likely be outweighed by the increased demand for accuracy of the systematics modelling, the likelihood method, and the estimation of the covariance matrix.}

The results for the noise-free `limiting-case' simulations are summarised in Table~\ref{tab:w0wa_summary}, alongside the equivalent conclusions for a realistic noisy set-up, which we discuss in detail next. 

\subsection{Realistic set-up including photo-$z$ and shape noise}
\label{subsec:results-noise}

We go on to present constraints on the $(w_{0}, w_{a})$ parameters derived from tomographic measurements of 3000 realisations of our catalogue simulation pipeline, in the presence of survey noise arising from Gaussian uncertainty in the photo-$z$ redshift estimation of galaxies and Gaussian shape noise. The characteristics of the noise are chosen to represent the level of uncertainty that is expected to be achieved in the \textit{Euclid} DR1 survey (see further details in Sect.~\ref{subsec:methods-errors}). As presented in the no-noise analysis in Sect.~\ref{subsec:results-nonoise} we examine each of the cosmic shear, angular clustering and full 3$\times$2pt signals, and measure the 1 and 2\,$\sigma$ constraints across different binning choices as a function of the number of tomographic bins used for the analysis.

\subsubsection{{1$\times$2pt cosmic shear only analysis, noisy simulation}}
\label{subsubsec:1x2ptE_noise}

\begin{figure*}
    \sidecaption
        \includegraphics[width=1.42\columnwidth]{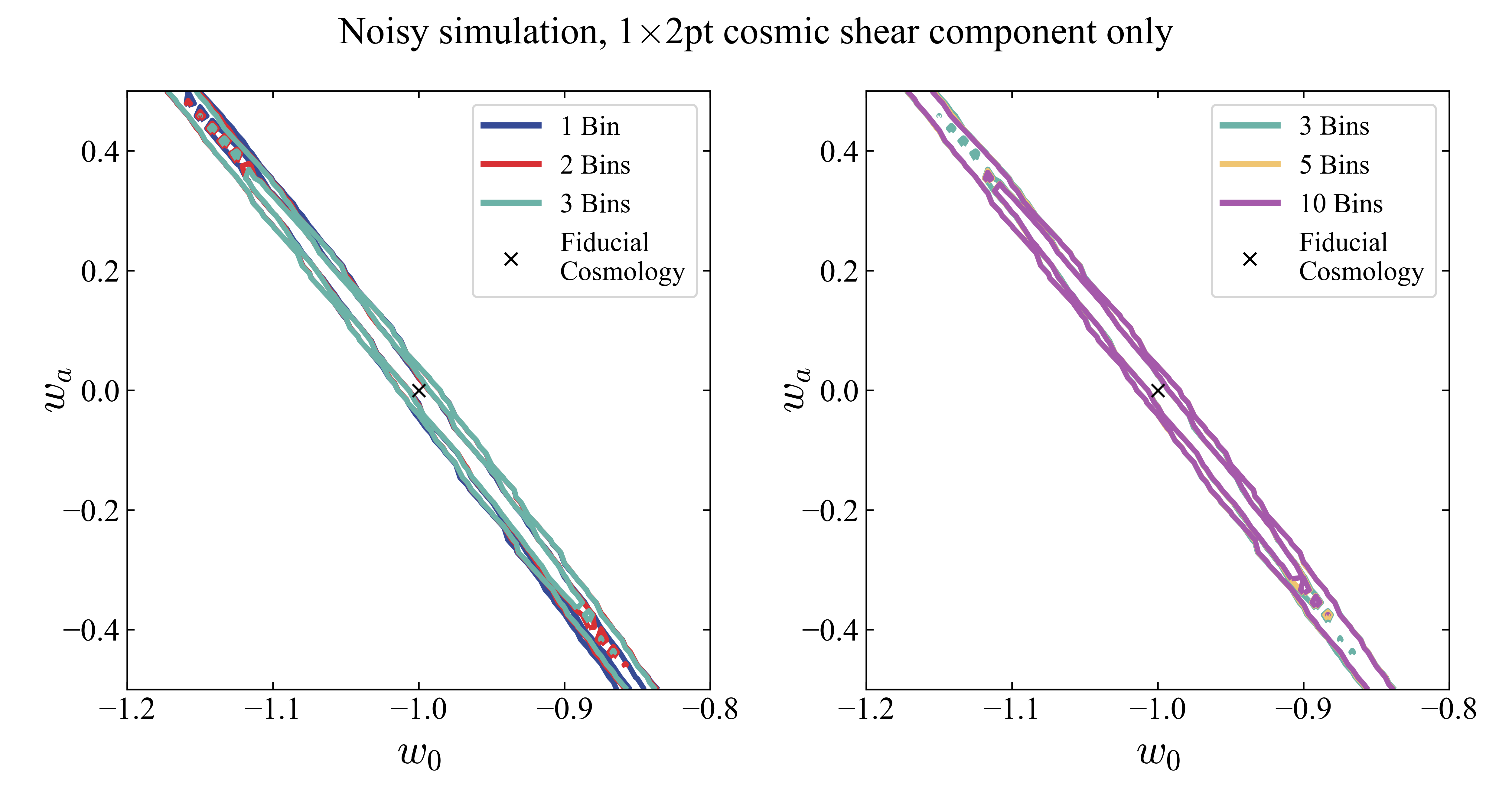}
    \caption{{ 1 and 2\,$\sigma$ constraints on the $(w_{0}, w_{a})$ parameters derived from measurements of the tomographic cosmic shear signal in our noisy simulation that includes contributions from Gaussian-distributed photo-$z$ errors and Gaussian-distributed shape noise. We present the constraints for different numbers of equipopulated bins but find that the signal has relatively weak constraining power that does not significantly improve as the number of bins increases. Hence, the contours corresponding to each different number of bins lie on top of each other.}}
    \label{fig:contours_equipop_noise_E}
\end{figure*}

In Fig.~\ref{fig:contours_equipop_noise_E} we plot the 2D posterior constraints on the time evolving dark energy parameters $(w_{0}, w_{a})$ from noisy tomographic cosmic shear measurements using one, two, three, five, and ten equipopulated bins. We find that in the presence of realistic noise, the cosmic shear is relatively weakly constraining and the constraints on both parameters do not change significantly when increasing the number of tomographic bins up to ten. Additionally, we note that within the prior boundaries of the grid-based likelihood approach, only the $1\,\sigma$ contour yields an enclosed area, while the $2\,\sigma$ contour coverage in the parameter space is extended such that it is computationally intractable to constrain. However, since we are sampling and fully characterising the parameter volume for the $1\,\sigma$ contour for all binning numbers considered, we do not believe that the true behaviour of the $2\,\sigma$ contours would yield different results or conclusions.

\begin{figure}
        \includegraphics[width=\columnwidth]{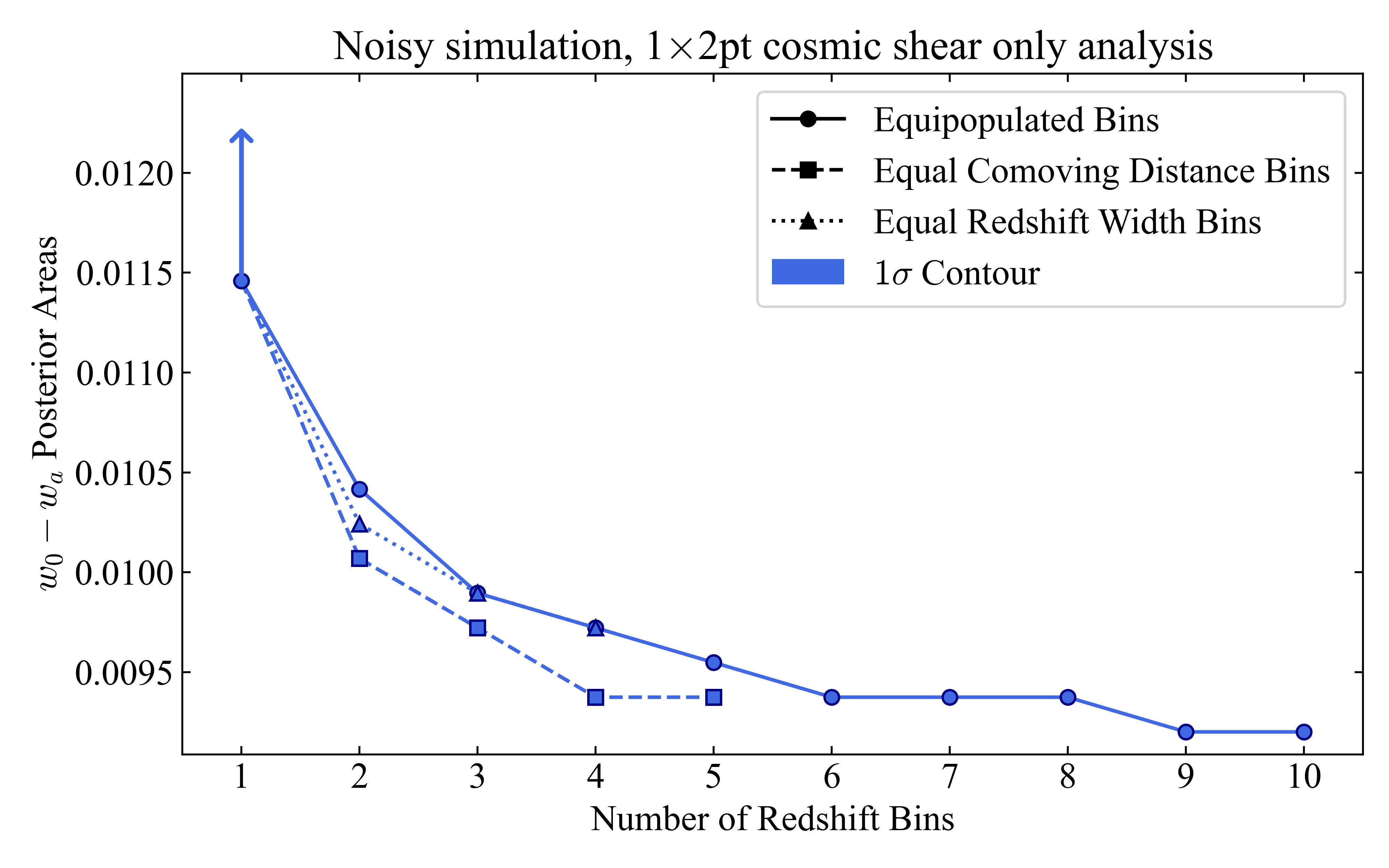}
    \caption{{The $1\,\sigma$ contour areas of constraints in $(w_{0}, w_{a})$ from measurements of the tomographic cosmic shear signal in our noisy simulation. We consider three different binning strategies and find that the 1 bin measurement does not have sufficient constraining power to generate closed contours within the prior ranges of the parameter grid. Hence, we plot the unbounded area measured within the prior ranges, and indicate that this is a lower bound of the true value by using a vertical arrow.}}
    \label{fig:w0wa_NOISE_1x2ptE}
\end{figure}

In Fig.~\ref{fig:w0wa_NOISE_1x2ptE}, we measured the area of the $1\,\sigma$ contours in the $(w_{0}, w_{a})$ plane for each binning choice. Here, we find that tomographic bins equally spaced in fiducial comoving distance produce the best constraints, followed by the equal redshift width bins and finally the equipopulated bins. This represents a different order of preference compared to the no-noise set of cosmic shear measurements (Fig.~\ref{fig:w0wa_NONOISE_1x2ptE}), for which equal redshift bins give the optimum constraints. However, the relative difference between the binning choices is at a minimal level of $< 5$\% and indicates that the gain that can be leveraged by choosing a specific binning choice for the cosmic shear is relatively small, regardless of the presence of realistic noise. 

In line with the 2D contour plot, we also find only a small reduction in the $(w_{0}, w_{a})$ area going from two to ten equipopulated bins of $\sim$10\%, which is a considerably lower gain than in the no-noise simulation, which demonstrates a $\gtrsim 99$\% gain between the same number of equipopulated bins. {The improvement in the $(w_{0}, w_{a})$ constraints shows a step-like behaviour between six to ten bins, which we attribute to the finite pixel size of the parameter grid. As the difference in the enclosed $(w_{0}, w_{a})$ contour areas is minimal in this range, we conclude that the convergence in the dark energy constraints for the noisy shear measurements is achieved at roughly seven to eight tomographic bins}

The simulated uncertainties from the shape noise in an individual galaxy's shear estimate, and the photo-$z$ uncertainties, which result in overlap in the redshift distributions of different tomographic bins, will both lead to a decrease in the constraining power. It is not necessarily straightforward to determine which one contributes more to the overall degradation here. One could explore simulations in which there is shape noise included but no photo-$z$ uncertainty, and vice versa, and conduct a similar analysis to investigate how the $(w_{0}, w_{a})$ constraints behave in each scenario. In particular, the former case would be more analogous to a spectroscopic weak lensing survey, in which the redshifts are precisely known and tomographic bins do not overlap. However, the fundamental survey characteristics such as the redshift range, survey footprint, and number density on the sky would change, and fundamentally require different tuning in our simulation parameters. It would then be difficult to make a direct comparison with the results presented here. 

Additionally, the reverse scenario, in which photo-$z$ uncertainties are included but shape noise is not present in the simulation, is unachievable in a real observation since observed galaxies have an intrinsic shape. As such, we consider that both of these intermediate simulation set-ups are beyond the scope of this work, but would nonetheless be interesting to explore in a future study.

\subsubsection{{1$\times$2pt angular clustering only analysis, noisy simulation}}
\label{subsubsec:1x2ptN_noise}

\begin{figure*}
    \sidecaption
        \includegraphics[width=1.42\columnwidth]{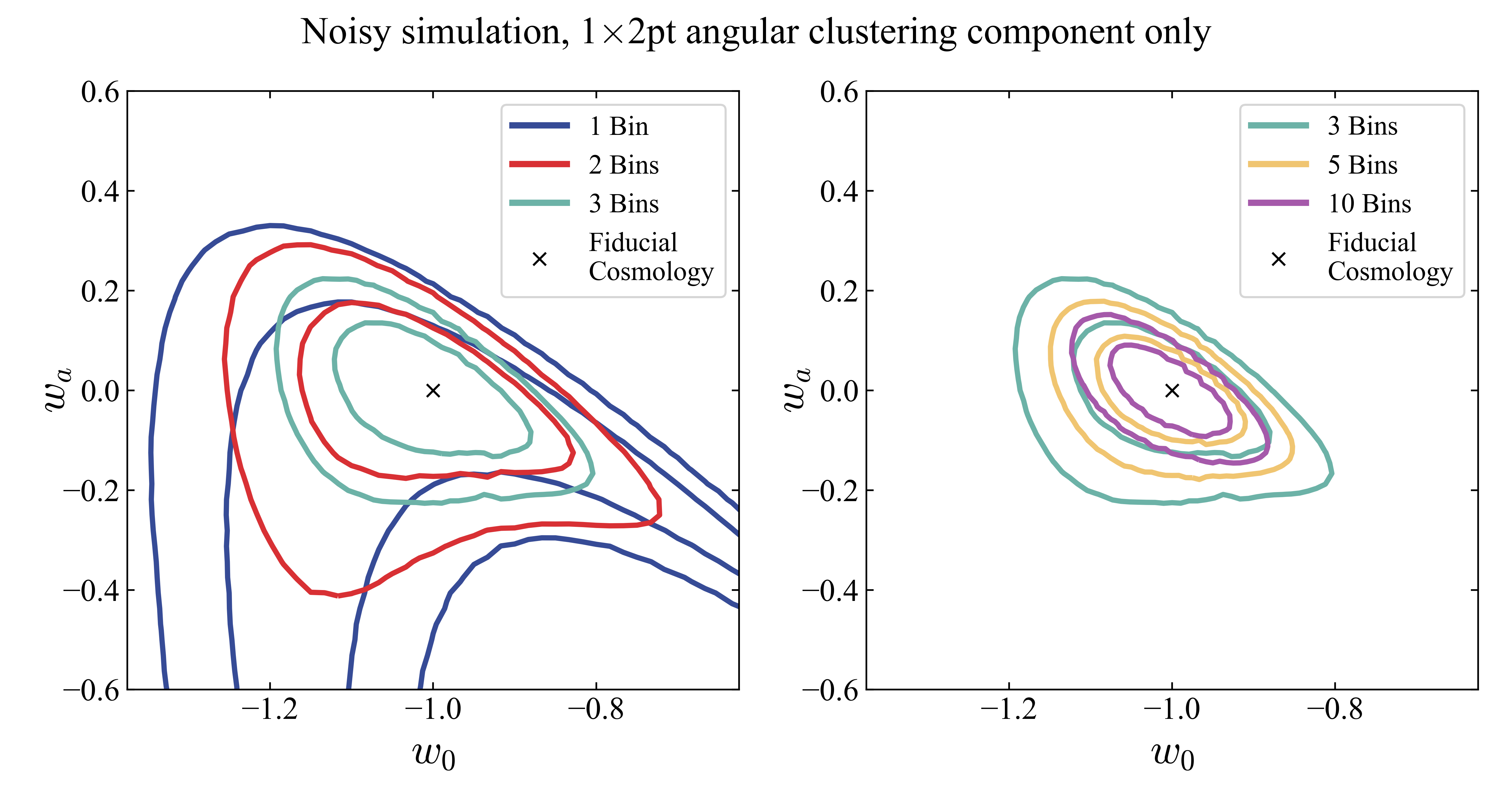}
    \caption{{ 1 and 2\,$\sigma$ posterior constraints on $(w_{0}, w_{a})$ measured from the tomographic angular clustering signal in our mock catalogues which include simulated noise from Gaussian photo-$z$ uncertainties and shape noise. We show the contours for different numbers of equipopulated bins used for the tomographic analysis.}}
    \label{fig:contours_equipop_noise_N}
\end{figure*}

\begin{figure}
        \includegraphics[width=\columnwidth]{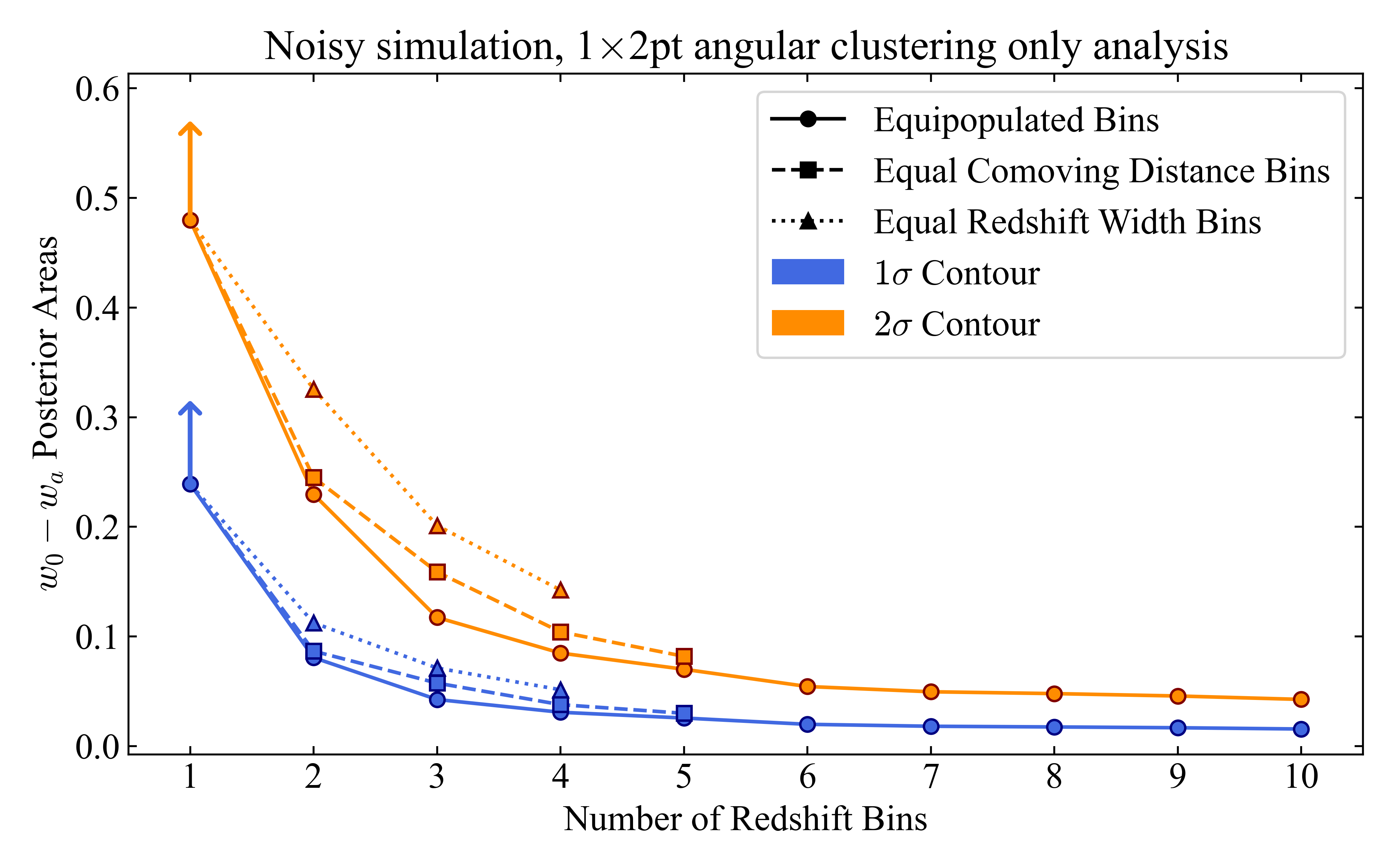}
    \caption{{ Contour areas of the 1 and 2\,$\sigma$ constraints on $(w_{0}, w_{a})$ for measurements of the tomographic angular clustering signal in our noisy simulation. We show the contour areas measured as a function of the number of redshift bins used, for the three different tomographic binning choices considered in this work.}}
    \label{fig:w0wa_NOISE_1x2ptN}
\end{figure}

In this section, we explore the tomographic constraints made on $(w_{0}, w_{a})$ by the clustering component of the 3$\times$2pt signal in the noisy simulation. In Fig.~\ref{fig:contours_equipop_noise_N} we show the 2D contours measured from varying the number of redshift bins used in the equipopulated binning choice, and in Fig.~\ref{fig:w0wa_NOISE_1x2ptN} we plot the areas enclosed by the 1 and 2\,$\sigma$ contours in the $(w_{0}, w_{a})$ plane as a function of the number of tomographic bins, for all three binning choices considered.
 
By inspecting Fig.~\ref{fig:contours_equipop_noise_N}, we find in particular that the shapes of the contours in the $(w_{0}, w_{a})$ plane for the clustering analysis are drastically different than the equivalent contour shapes derived from the cosmic shear analysis in Fig.~\ref{fig:contours_equipop_noise_E}. This property clearly highlights the power of the 3$\times$2pt combined probe to break degeneracies in the dark energy parameters and provide highly precise measurements of the equation of state. 

Collectively, the behaviour of $(w_{0}, w_{a})$ posterior areas measured as a function of the number of tomographic redshift bins is remarkably consistent with the no-noise simulation. In particular, the relative order of preference of the binning choices is identical across the noisy and no-noise analyses, whereby the equipopulated bins provide the optimum constraints, followed by the equal comoving distance bins and finally the equal redshift bins. 

Moreover, we find that simply using two bins is sufficient, for all binning choices in the noisy simulations, to make measurements of $(w_{0}, w_{a})$ which indicates that the presence of the photo-$z$ uncertainty and shape noise does not significantly degrade the constraining power of the cosmological signal. In an absolute sense, we find that compared to the no-noise measurements, the contour areas increase by $\sim$5--10\% on average for the equipopulated case, and $\sim$5\% for both the equal comoving distance and equal redshift binning choices. We note that for the equipopulated choice, the increase is only due to the degradation of the cosmological signal by the photo-$z$ uncertainties, as the shape noise does not affect the angular galaxy positions and the Poisson noise is the same in both the no-noise and noisy simulations. 

Most importantly, we highlight that, with and without the realistic noise included in the simulation, the improvement in the $(w_{0}, w_{a})$ constraints starts to saturate at six or seven tomographic bins, which is demonstrated by the convergence and flattening of the distributions corresponding to the 1 and 2\,$\sigma$ measurements in Fig.~\ref{fig:w0wa_NOISE_1x2ptN}. 

Consequently, we conclude that for an analysis of the angular clustering only, the optimum tomographic binning recommended by these results is the equipopulated choice. Here, the choice of about six or seven bins is  likely to be sufficient to achieve the optimum performance in the noisy case.

\subsubsection{{Full 3$\times$2pt analysis, noisy simulation}}
\label{subsubsec:3x2pt_noise}

\begin{figure*}
    \sidecaption
        \includegraphics[width=1.42\columnwidth]{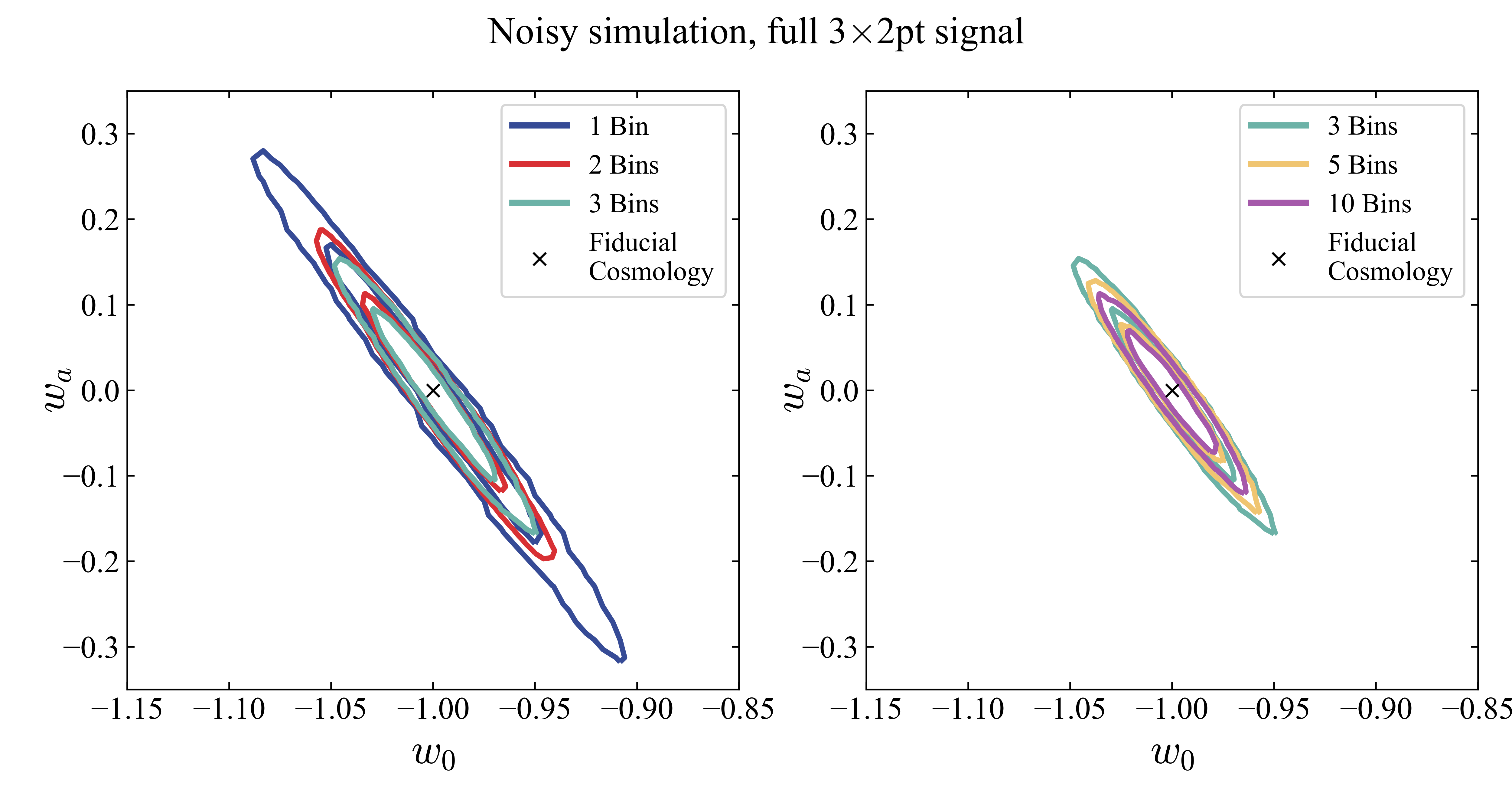}
    \caption{{ 1 and 2\,$\sigma$ constraints on $(w_{0}, w_{a})$ from tomographic Pseudo-$C_{\ell}$ measurements of the full 3$\times$2pt signal in our noisy simulation. We show constraints measured for different numbers of equipopulated bins used for the tomographic analysis. The cross marks the fiducial cosmology $(w_{0}, w_{a})=(-1,0)$ used for the simulation.}}
    \label{fig:contours_equipop_noise_3x2pt}
\end{figure*}

\begin{figure}
        \includegraphics[width=\columnwidth]{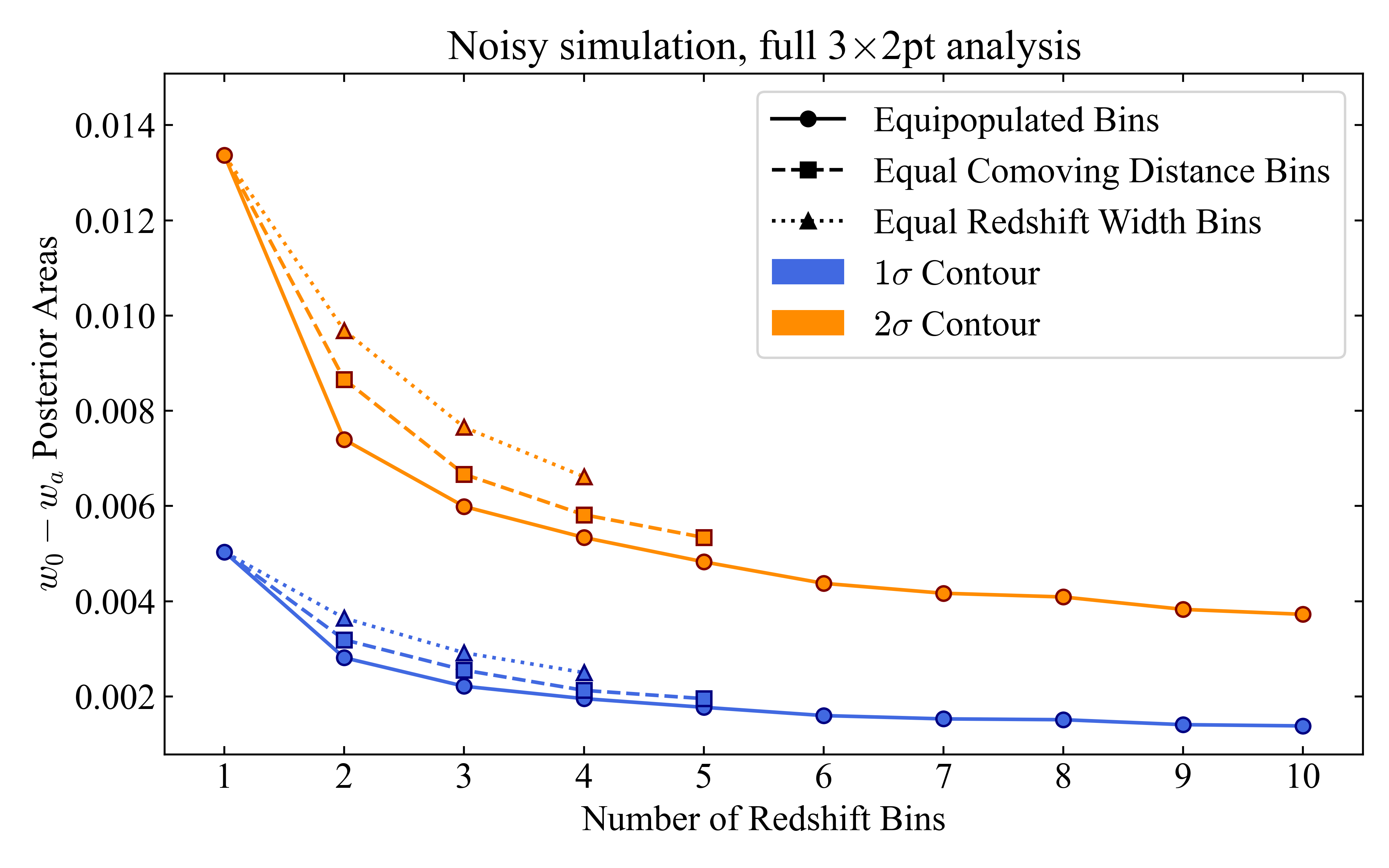}
    \caption{{ Contour areas enclosed by the 1 and 2\,$\sigma$ constraints on the dark energy $(w_{0}, w_{a})$ parameters, measured for the full 3$\times$2pt signal in the presence of Gaussian shape noise and photo-$z$ uncertainty. We vary the number of redshift bins used for each of the equipopulated, equally spaced in fiducial comoving distance, and equal redshift width binning strategies.}}
    \label{fig:w0wa_NOISE_3x2pt}
\end{figure}

Finally, we consider the effects of the tomographic binning choices on the $(w_{0}, w_{a})$ parameters for a full 3$\times$2pt analysis in the presence of noise, the target experimental and analytical set-up with which to compare to the aims of the \textit{Euclid} DR1 survey. In Fig.~\ref{fig:contours_equipop_noise_3x2pt}, we first plot the 1 and 2\,$\sigma$ $(w_{0}, w_{a})$ contours for tomographic measurements of the 3$\times$2pt signal using one, two, three, five, and ten equipopulated bins, and in Fig.~\ref{fig:w0wa_NOISE_3x2pt} we compare the areas enclosed by such contours across different binning strategies.

While a single tomographic bin used for analysis is only weakly constraining for measurements of the noisy cosmic shear and clustering signals only (Figs.~\ref{fig:contours_equipop_noise_E},~\ref{fig:contours_equipop_noise_N}), we find that by considering the joint information contained in the 3$\times$2pt signal, the single bin can fully constrain the $(w_{0}, w_{a})$ parameters and identify a time evolving nature of the dark energy equation of state. Consequently, this result suggests that the single-bin analysis could be used as an important first-order measurement to investigate whether the photo-$z$ uncertainties and shape noise are correctly accounted for in the modelling. 

With respect to the different binning choices in Fig.~\ref{fig:w0wa_NOISE_3x2pt}, the noisy 3$\times$2pt analysis is similarly consistent with the no-noise results and demonstrates that the equipopulated bins yield measurements on $(w_{0}, w_{a})$ that have the least error, followed by the equal comoving distance bins and then the equal redshift width bins. Since the same order of preference is found in the clustering-only investigation in Sect.~\ref{subsubsec:1x2ptN_noise}, and we find that the different binning choices only make per cent level changes to the $(w_{0}, w_{a})$ contour areas for the shear only analysis in Sect.~\ref{subsubsec:1x2ptE_noise}, we conclude that the performance of the different binning choices in the full 3$\times$2pt signal is likely to be dominated by the response of the angular clustering component to the tomographic binning. 

For the noisy measurements, the increase in the $(w_{0}, w_{a})$ areas between each binning choice in successive order of preference is more consistent with $\sim$10--15\% at low numbers of bins (two-four), compared to $\sim$5--10\% for the no-noise analysis, which demonstrates that in this range the specific binning choice becomes more important in deriving optimum constraints on dark energy with increasing realism in the simulation.

However, we find that beyond  six or seven tomographic redshift bins used for the analysis, the $(w_{0}, w_{a})$ areas corresponding to the equipopulated constraints start to plateau, demonstrating the same saturation in the information gain present in the noisy angular clustering measurements in Fig.~\ref{fig:w0wa_NOISE_3x2pt}. In Table~\ref{tab:w0wa_summary}, we summarise our findings on the behaviour of the $(w_{0}, w_{a})$ contours for the different tomographic measurements of the noisy cosmic shear-only, angular clustering-only, and full 3$\times$2pt signals. Also reported in Table~\ref{tab:w0wa_summary} are the limiting case no-noise results, for all 3$\times$2pt probes.

\begin{table*}
\renewcommand{\arraystretch}{1.5}
\begin{center}
\caption{{Optimum tomographic binning strategies evaluated for different science cases.}}
\label{tab:w0wa_summary}
\begin{tabularx}{\textwidth}{ >{\raggedright\arraybackslash}p{0.15\linewidth} >{\raggedright\arraybackslash}p{0.125\linewidth} >{\raggedright\arraybackslash}p{0.15\linewidth} >{\raggedright\arraybackslash}p{0.15\linewidth} p{0.31\linewidth}}

\hline
\hline
Signal & Simulation type& Optimum binning choice & Optimum No. bins & Properties\\
\hline
\hline
\multirow{4}{*}{Full 3$\times$2pt} &   Realistic noise & Equipopulated & Convergence by around six to seven    & Relative improvement between binning choices is up to $15\%$ -- represents motivation to use the equipopulated bins \\
                            & No-noise          & Equipopulated & Convergence by around six to eight  & Relative gain between binning choices is slightly less significant ($\sim$5--10\%) compared to the noisy simulation \\
  \hline
  \multirow{3}{*}{Cosmic shear} & Realistic noise   & Equal comoving distance   & Convergence by around seven or eight    & Minimal difference between different binning choices (percent level) \\
                                & No-noise          & Equal redshift width      & Convergence by around seven    & Improvement in $(w_{0}, w_{a})$ constraints by increasing from two to ten equipopulated bins is $\gtrsim99$\% for no-noise case compared to $\sim$10\% for noisy case. \\
  \hline
  \multirow{3}{*}{Angular clustering}   & Realistic noise   & Equipopulated & Convergence by around six to seven & Minimal difference in behaviour between no-noise and noisy simulations \\
                                        & No-noise          & Equipopulated & Convergence by around six to seven & Minimal difference in behaviour between no-noise and noisy simulations \\
 \hline
 
 \hline
 \hline
\end{tabularx}
\vspace{0.35cm}

\end{center}
\renewcommand{\arraystretch}{1}
\tablefoot{This table summarises the behaviour of the cosmic shear component, angular clustering component, and full 3$\times$2pt signal for different tomographic binning choices. The summary conclusions have been evaluated from the constraints on $(w_{0}, w_{a})$ measured across multiple binning choices and redshift bins in tomographic analyses of our simulated catalogues for both a no-noise and noisy set-up (see Sects.~\ref{subsec:results-nonoise} \& \ref{subsec:results-noise}).}
\end{table*}

\begin{figure}
        \includegraphics[width=\columnwidth]{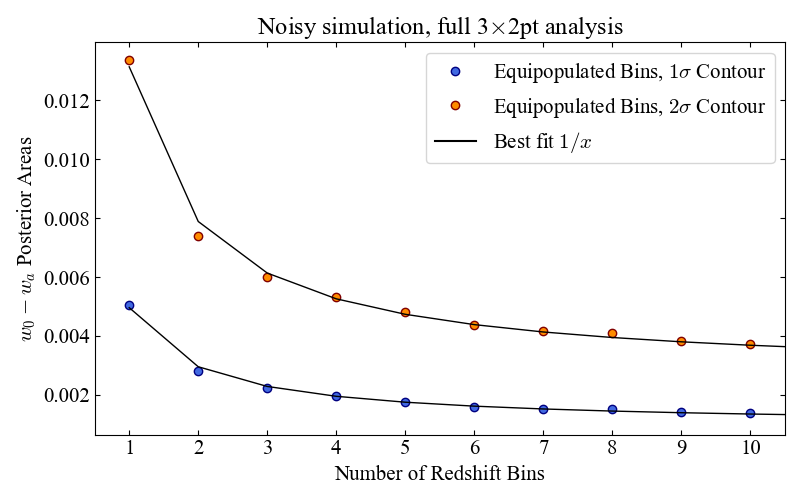}
    \caption{{Areas of the 1 and 2\,$\sigma$ constraints on $(w_{0}, w_{a})$, measured from our noisy catalogue simulations, plotted in circular markers as a function of the number of equipopulated bins used for the tomographic analysis. Black solid lines show the plot of the best fit functions of the form $y=a/x+b$ that we fit to these data sets, evaluated using the \texttt{SciPy} \texttt{curve\_fit} routine.}}
    \label{fig:w0_wa_fitting}
\end{figure}

Empirically, we find that the areas of the 1 and 2\,$\sigma$ contours in the $(w_{0}, w_{a})$ plane, measured as a function of the number of equipopulated tomographic redshift bins used, are well described by a $1/x$ function of the form $y=a/x + b$. We use the \texttt{SciPy} (\citealt{SciPy}) \texttt{curve\_fit} routine to derive {$(a,b)=(0.00401, 0.00094), (0.0105, 0.00263)$ for each of the 1 and 2\,$\sigma$ distributions, respectively.

We present these functions fit to the areas of the equipopulated constraints in Fig.~\ref{fig:w0_wa_fitting}, which demonstrates empirically how the rate of improvement in the $(w_{0}, w_{a})$ areas starts to decrease as the number of bins used in the tomographic analysis increases. Using the fitting functions as a baseline, we find that compared to a one-bin analysis, the areas of the $(w_{0}, w_{a})$ contours decrease by $69\%$ when using seven bins, and $72\%$ when using ten bins, which indicates only a further $3\%$ gain in information between the use of seven and ten bins.} 

By extending the fitting functions to explore the theoretical behaviour of tomographic analyses that use more than ten bins (which we find we cannot practically measure with our method due to the presence and randomness of untraced pixels on the sky), we can estimate the area of the $(w_{0}, w_{a})$ contours for the particular case of 13 equipopulated bins. We note this use of 13 bins was found in \cite{Pocino-EP12} to improve the dark energy FOM by 15\% for a 2$\times$2pt analysis of galaxy clustering and galaxy-galaxy lensing, based on photo-$z$ studies using the Euclid Collaboration Flagship Simulation (\citealt{EuclidSkyFlagship}). 

Using our extended fitting functions in Fig.~\ref{fig:w0_wa_fitting}, we find that the $(w_{0}, w_{a})$ contour areas for 13 bins decrease by $17\%$ and $7\%$ compared to analyses using seven and ten tomographic bins, respectively. Additionally, by measuring the total improvement on the $(w_{0}, w_{a})$ constraints with respect to the single-bin analysis, we find that this corresponds to only a further $5\%$ gain in the precision on the parameters when increasing from seven to 13 bins in the measurement, and a $2$\% level gain when going from ten to 13 bins. 

Overall, these results indicate that increasing the number of tomographic bins beyond ten leads to improvements in the $(w_{0}, w_{a})$ constraints that are less substantial than those found by \cite{Pocino-EP12}. However, we note that each of these results has been determined using different methods and with different assumed survey characteristics.

In particular, the latter study considers a 2$\times$2pt analysis and focuses on the configuration and cosmological information that is expected to be obtained from the complete \textit{Euclid} survey, rather than just the first data release (DR1) that our analysis has focussed on. Consequently, the underlying $n(z)$, spatial density of galaxies on the sky, and scale cuts taken for the 3$\times$2pt signal considered in \cite{Pocino-EP12} will be different than those considered in this study. Additionally, while we have considered an analytical method to simulate galaxy redshifts and their uncertainties, \cite{Pocino-EP12} adopted machine learning methods to estimate the photometric redshifts. 

We note for our analysis that there are still errors associated with the data points due to the fact we can only work with a finite number of realisations, which in an ideal case would be included in the derivation of the fitting function and its parameters. As such, the relative statistical gains and comparisons we present here should be considered only as a guide and not absolute results. In principle, a quantification of the errors could be achieved for example by repeating the entire analysis to measure the set of $(w_{0}, w_{a})$ contour areas over multiple batches of 3000 realisations, and then examining the deviation across all batches. However, this is unfortunately computationally expensive and impractical to achieve in this study. 

Nonetheless, we assert that the errors will be equivalent in magnitude for the different binning choices considered, and hence conclude that for a full 3$\times$2pt tomographic analysis of our simulated \textit{Euclid} DR1-like catalogues, in the presence of realistic noise, the optimum binning choice is the equipopulated case. 

While there perhaps could still be merit in using ten tomographic bins to extract a few more per cent precision in $(w_{0}, w_{a})$, we note that the use of more bins would increase the computational expense. Moreover, it is likely that any potential benefits are only realisable, within the framework of a theoretical study, such as the one presented here, where one has full control and knowledge of the uncertainties that are introduced. 

Comparatively, for a real Stage IV analysis, it is likely that this extra information probed by increasing the number of bins beyond about seven will be difficult to leverage, especially in light of the additional model and inference complexity that we do not consider here. Such effects include the shear bias in the shape measurements of galaxies (e.g. \citealt{Hirata03}, \citealt{Jansen24}, \citealt{EP-Congedo}), the effects of magnification in the galaxy sample and further high order terms (e.g. \citealt{Duncan22}, \citealt{Deshpande20}, \citealt{Deshpande-EP28}), {or the effects of marginalisation over nuisance parameters, such as the galaxy bias, or other cosmological parameters within a given prior range.}

We note additionally the impact of intrinsic alignments on 3$\times$2pt cosmology (e.g. \citealt{Troxel15}), which we do not model in this study. It has been proposed that the systematic effects of intrinsic alignments can be quantified by the use of tomography \citep{King02}, { and we highlight that such an investigation, combined with the effects of further systematics not considered in this study, can be examined in later work.}

Indeed, \cite{Pocino-EP12} consider marginalising over both intrinsic alignments and a constant galaxy bias in each tomographic bin for a 2$\times$2pt forecast of photometric galaxy clustering and galaxy-galaxy lensing with \textit{Euclid}. By targeting the dark energy figure of merit, they report that the marginalisation procedure motivates the use of fewer tomographic bins compared to conditional constraints, since less cosmological information can be extracted. 

In addition, they found that for the marginalised 2$\times$2pt constraints, there is negligible difference between the equal redshift width and equipopulated binning choices. Comparatively, there is a $\sim$30\% difference between these binning choices for conditional constraints for a clustering-only analysis using 13 tomographic bins. This result indicates that the choice of tomographic binning strategy for each 3$\times$2pt probe may have a complex dependence on different systematics that may be incorporated in the analysis. A detailed study covering different models for a range of systematics, in which their effect on tomography is considered both individually and collectively, is a study that is beyond the scope of this paper, but will be an important investigation for future work.

Lastly, we note that the true errors in the photo-$z$ estimates, the shear estimates and/or the measured power spectra may not be purely Gaussian (e.g. \citealt{Takada09}). We explore an example of this scenario in the following section (Sect.~\ref{sec:catastrophic-photozs}) where we consider the effects of catastrophic photo-$z$ uncertainties in the estimation of the galaxy redshifts and the subsequent impact on the inference on the dark energy parameters. 

\section{{Catastrophic \texorpdfstring{photo-$z$}{photo-z} uncertainties}}
\label{sec:catastrophic-photozs}

In this section, we demonstrate the flexibility of our simulation method to incorporate further astrophysical or survey-like effects by injecting catastrophic photo-$z$ uncertainties into our catalogue simulation and then propagating the resulting biased measurements into cosmological constraints on $(w_{0}, w_{a})$.  Photo-$z$ estimation codes can misidentify features in the spectral energy distribution (SED) of a galaxy, leading to a measured redshift that significantly fails to recover the true redshift of the galaxy. For the purposes of the 3$\times$2pt signal, the misestimated galaxy redshift will lead to an incorrect interpretation of the structure along the line of sight that has induced the shear in the galaxy, which will limit our ability to make accurate measurements of dark energy and its possible time-evolving behaviour. 

To achieve its science goals, \textit{Euclid} will target a $5\%$ level of contamination by catastrophic photo-$z$ errors (\citealt{Euclid}) in the real analysis. Hence, we introduced the effects of this systematic into our simulation by randomly selecting $5\%$ of our galaxy sample in the noisy set-up and distorting their estimated redshift measurement in the catalogue to a catastrophic estimation using Eq.~(\ref{eq:catastrophic-photozs}). The relation between the true redshift and the catastrophic redshift, $z_{\mathrm{cata}}$, modelled here is governed by the ratio between the rest-frame wavelength of a break feature in the SED and the wavelength that is predicted by the photo-$z$ estimation technique. As a first-order investigation, we considered catastrophic errors that result from the rest-frame Lyman-$\alpha$ line being confused with the Balmer and D4000 breaks, i.e. taking the pairs $[\lambda_{\mathrm{break-rf}}, \lambda_{\mathrm{break-cata}}]=[\mathrm{Ly}$-$\alpha,$ Balmer-break$],\, [\mathrm{Ly}$-$\alpha,$ D4000-break$]$ and splitting the $5\%$ contamination equally between the two pairs of misidentified features. We finally model the resulting catastrophic photo-$z$ estimate as a Gaussian-distributed value with mean $z_{\mathrm{cata}}$ and standard deviation $\sigma_{z, \mathrm{cata}}=0.1$.

Following this procedure, we measured the tomographic Pseudo-$C_{\ell}$ power spectra from these biased mock catalogues using the standard technique adopted in this work, as described in Sect.~\ref{subsec:estimators}. The presence of the catastrophic photo-$z$ errors will induce a change, $\Delta C_{\ell}$, in the measured power spectra compared to the standard noisy case.  To determine the impact of this bias on the dark energy $(w_{0}, w_{a})$ constraints, we: 1) added the measured $\Delta C_{\ell}$ onto the true, theoretical 3$\times$2pt data vector of the fiducial cosmology; and 2) performed the grid-based Gaussian likelihood analysis on this distorted 3$\times$2pt signal using an underlying tomographic redshift distribution $n(z)$ and a numerical covariance matrix that have both been derived from the simulations that include Gaussian noise. As such, the final constraints on $(w_{0}, w_{a})$ that we measure will be directly comparable to the Gaussian noise-like results of Sect.~\ref{subsubsec:3x2pt_noise}. Any difference will be due to the $5\%$ contamination by catastrophic photo-$z$s, where we do not retain knowledge of which individual galaxies are affected, and the use of a covariance matrix that does not include sufficient information on the catastrophic photo-$z$s. As such, we emphasise that we do not attempt to mitigate against the presence of the catastrophic outliers to derive our dark energy constraints, which we expect to be pessimistic. Comparatively, a real analysis could attempt to correct for these uncertainties in the assumed $n(z)$ distribution, or through nuisance parameters to marginalise over in the inference.

\begin{figure*}
    \centering
        \includegraphics[width=2\columnwidth]{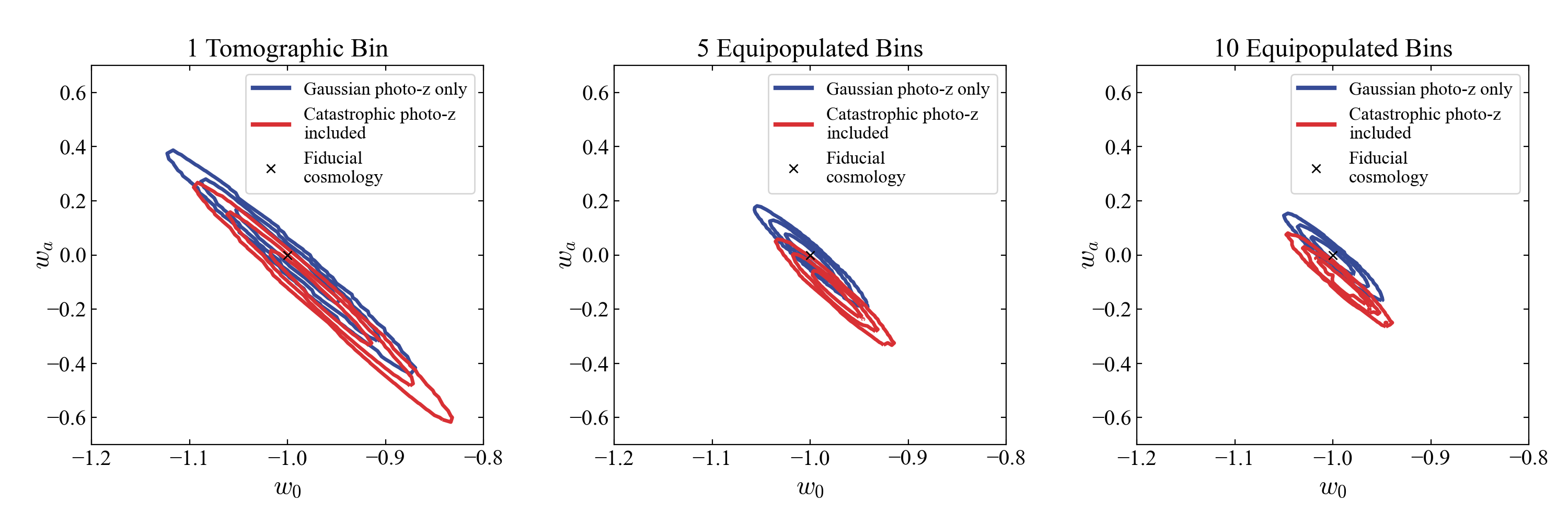}
    \caption{{Constraints on the dark energy $(w_{0}, w_{a})$ parameters measured from: simulations that include only Gaussian shape noise and Gaussian photo-$z$ uncertainties (blue) and simulations that include Gaussian shape noise, Gaussian photo-$z$ uncertainties, and a $5\%$ level of catastrophic photo-$z$ errors (outliers) that result from the Lyman-$\alpha$ line in the galaxy SED being misidentified as the Balmer and D4000 break features (red). From left to right, we plot the constraints derived from tomographic analyses of the full 3$\times$2pt signal using a single redshift bin, five equipopulated bins, and ten equipopulated bins.}}
    \label{fig:catzs}
\end{figure*}

In Fig.~\ref{fig:catzs}, we show the $1, 2,$ and $3\,\sigma$ constraints on $(w_{0}, w_{a})$ for three different tomographic analyses of the full 3$\times$2pt signal: a single redshift bin; five equipopulated bins, and ten equipopulated bins, comparing in blue the results from the Gaussian noise-like simulations in Sect.~\ref{subsubsec:3x2pt_noise} and in red the results from the catastrophic photo-$z$ analysis {averaged over 3000 realisations}. 

We find that the measurements across the three different tomographic binning choices all consistently yield the constraints $(w_{0}>-1,\,w_{a}<0)$. These parameters are strongly biased from the underlying $\Lambda$CDM cosmology and demonstrate that the presence of a $5\%$ level of contamination from catastrophic photo-$z$s, if entirely unaccounted for in the modelling, would lead us to infer significantly incorrect conclusions on a time-evolving behaviour of the dark energy equation of state. Indeed, such effects of catastrophic photo-$z$ uncertainties for weak lensing surveys have been extensively studied in the literature, e.g. ~\cite{Hearin10},~\cite{Ma06}, and ~\cite{Sun09}, for which our results act to support and consolidate.

Since the constraining power of the 3$\times$2pt signal increases with the number of redshift bins used for the tomographic analysis, the statistical disagreement with the true underlying cosmology also increases with the number of tomographic bins. Explicitly, for a single-bin analysis we find that if the catastrophic outliers are not modelled in the 3$\times$2pt data vector, the power spectrum measurements are in tension with $\Lambda$CDM at $\sim2\,\sigma$, {while for five and ten bins, the tension is at a $\gtrsim3\,\sigma$ level}. It is clear that the dark energy $(w_{0}, w_{a})$ parameters are highly sensitive to the catastrophic photo-$z$ uncertainties. Since we do not attempt to correct for these errors to derive the constraints presented here, these results aim to quantify the extent to which these uncertainties would need to be minimised or mitigated in the real analysis to determine the true nature of dark energy. 

In addition to the biased recovery of the true cosmology, we find that the presence of the catastrophic photo-$z$ errors leads to unusual structures in the 2D posterior. For our standard no-noise and noisy simulations (for which the photo-$z$ uncertainties are purely Gaussian), we expect that the 3$\times$2pt measurements will generate smooth, elliptical contours, as demonstrated in Fig.~\ref{fig:contours_equipop_noise_3x2pt}. In contrast, for the five- and ten-bin analyses we see that the contours are less elliptical in shape and have sharper features in the catastrophic photo-$z$ analysis. We attribute this behaviour to the $\Delta C_{\ell}$ change in the 3$\times$2pt power spectra that the catastrophic redshift errors induce. The $C_{\ell}$ bias is non-trivial and random across all redshifts, and hence the space of cosmological models in $(w_{0}, w_{a})$ that the resulting power spectra are compatible with is complex and possibly discontinuous.

We conclude that the introduction of the $5\%$ contamination by catastrophic photo-$z$ uncertainties can severely compromise the attempt to measure accurate cosmological constraints from the 3$\times$2pt signal if not adequately mitigated. In particular, the constraints yield a best fit model that is in tension with $\Lambda$CDM at $\gtrsim3\,\sigma$ for analyses using 5 or more tomographic redshift bins, and the posterior distributions in $(w_{0}, w_{a})$ demonstrate sharp features. 

Additionally, we note that catastrophic photo-$z$ uncertainties may be caused by further effects in addition to the confusion between the Lyman-$\alpha$ line, and the Balmer and D4000 breaks that we have considered throughout this work. For example, `uniform' catastrophic outliers may be caused when the light from a galaxy is contaminated by light from a nearby source on the sky which is at a different redshift (see e.g. \citealt{Hearin10}). A full investigation into the nature of the bias induced by further causes of catastrophic photo-$z$ errors is beyond the scope of this work. However, we emphasise that our simulation framework has the flexibility to incorporate a range of different distributions for the catastrophic photo-$z$ uncertainty that could be examined in a future study. 

These results emphasise that an accurate modelling of the catastrophic photo-$z$ errors would need to be accounted for in the theoretical model of the cosmological signal, the covariance matrix, and the likelihood distribution. Further methods to mitigate against catastrophic uncertainties include marginalisation over a nuisance parameter (\citealt{Bernstein10}) capturing catastrophic errors; or selectively removing galaxies from the redshift distribution (\citealt{Hearin10}), for example by monitoring the posterior likelihood distribution of galaxy redshift estimates (\citealt{Nishizawa10}).

\section{{Discussion and conclusions}}
\label{sec:discussion_conclusions}

Investigations attempting to determine an `optimum' binning configuration for a tomographic 3$\times$2pt analysis, mostly using machine learning or numerical methods (\citealt{Taylor18_a}, \citealt{Kitching19}, \citealt{Sipp21}, \citealt{Pocino-EP12}, \citealt{Zuntz21}, see Sect.~\ref{sec:motivations}), have returned inconsistent conclusions on both the binning choice and number of bins that ought to be targeted. In this work, we present an alternative method to explore the tomographic binning with the aim to mimic a real survey and subsequent cosmological analysis in a robust and comprehensive manner. We have developed a simulation pipeline that generates multiple realisations of mock galaxy catalogues for a Stage IV-like survey, which are constructed by generating a set of correlated 3$\times$2pt 2D maps on the sky, sampling a target redshift range, to model the full 3D cosmological information in the Universe. 

We simulated 3000 realisations of our mock catalogues and measured the 3$\times$2pt Pseudo-$C_{\ell}$ power spectra, evaluated over an approximation of the \textit{Euclid} DR1 footprint, for a range of tomographic configurations. We propagated these measurements into a grid-based Gaussian likelihood analysis to place constraints on the dark energy equation of state parameters, $(w_{0}, w_{a})$, and derive the optimum binning choice by comparing the size of the contours in $(w_{0}, w_{a})$ across all tomographic binning configurations considered. 

For a limiting case `no-noise' simulation, where we did not consider any effects of redshift uncertainties due to photo-$z$ estimation techniques or shape noise associated with the observed shear of a galaxy, we found that the optimum binning choice for a full 3$\times$2pt analysis is given by redshift bins that are equally populated with galaxies. For the angular galaxy clustering signal, the optimum binning choice is also the equipopulated case, which is likely to be the dominant component of the 3$\times$2pt data vector. However, for an analysis of the cosmic shear signal only, bins that are equally spaced in redshift yield the best constraints on $(w_{0}, w_{a}),$ which indicates that the optimum binning choice is dependent on which observable is targeted.

For a more realistic simulation in which we include Gaussian shape noise and Gaussian photo-$z$ uncertainties in the mock catalogues, we find that the equipopulated bins remain the best choice for measuring $(w_{0}, w_{a})$ for the full 3$\times$2pt and angular clustering only analyses. However, the cosmic shear signal itself gives the best constraints when bins that are equally spaced in fiducial comoving distance are used, where the fiducial cosmology we work with is the standard $\Lambda$CDM. 

We note that for a different underlying cosmology, the redshift bin boundaries for the equal comoving distance binning choice would change. However, we emphasise that as greater numbers of tomographic bins are used for the analysis, the derived constraints on $(w_{0}, w_{a})$ for each binning choice would converge together regardless of the fiducial cosmology. Hence, we do not expect that our results would change significantly in this range. 

Collectively, we find that the relative gain or loss between the different binning choices is least significant for the cosmic shear component, {in particular, at the $\sim$5$\%$ level when noise is included in the simulation}. This result suggests that the cosmic shear signal is less sensitive to the tomographic binning choice and there may not be significant benefit in targeting a given binning choice in the real analysis.

In comparison, we find that the difference in size of the $(w_{0}, w_{a})$ contour areas, between the different binning choices in successive order of preference, is $\sim$10\% on average for the full 3$\times$2pt and clustering-only measurements in the no-noise set-up. For the realistic noise simulation, this difference is slightly more amplified at $\sim$10--15\%. We conclude for these signals that the gain is considerable and it provides clear motivation to choose the equipopulated binning for a real Stage IV analysis.

With respect to the number of redshift bins chosen for tomography, we expect that the rate of information gain will decrease as the number of bins increases, until a point at which the area of the $(w_{0}, w_{a})$ contours will converge to a consistent value that does not continue to improve when arbitrarily adding more bins for the tomographic analysis. For each of the cosmic shear and clustering components individually, we find that this convergence is reached at around six to eight bins for the 1 and 2\,$\sigma$ contours across both the no-noise and noisy simulations. For each of these signals, the minimum number of bins required to fully constrain the dark energy parameters is two, and using this measurement as a baseline, we expect that for both components in the noisy set-up, there is only a further $\mathcal{O}(1\%)$ level improvement in the $(w_{0}, w_{a})$ constraints when increasing the number of tomographic bins successively from six to ten. 

{In terms of the full 3$\times$2pt analysis, we find a convergence in the $(w_{0}, w_{a})$ areas at the use of seven or eight tomographic redshift bins for the no-noise measurements. For the noisy simulation, the behaviour is similar and empirically. We find that the rate at which the $(w_{0}, w_{a})$ contour areas decrease (as a function of the number of equipopulated tomographic redshift bins) is well described by a $1/x$ function of the form $y=a/x+b$, where we derived $(a,b)=(0.00401, 0.00094), (0.0105, 0.00263)$ for the 1 and 2\,$\sigma$ trends, respectively.} Using this function to theoretically predict the contour size for the 13 bin analysis that has been proposed for \textit{Euclid} (e.g. \citealt{Pocino-EP12}), we measured the total gain in information with respect to the single redshift bin constraints, finding that there is only a $5\%$ further improvement in measurements of $(w_{0}, w_{a})$ when going from seven to 13 bins.

We note that for higher numbers of redshift bins, the 3$\times$2pt data vector becomes larger and the information gain on $(w_{0}, w_{a})$ becomes increasingly limited by the demand for accuracy in the covariance matrix. This is manifested as an increase in uncertainty due to excess random noise in the numerical covariance matrix used in this work; however, the trade-off would also be important to consider in a real analysis, particularly where an analytical covariance may be used. 

Furthermore, there will be additional uncertainties present in a real survey, such as from the shear calibration, redshift-dependent astrophysical systematics (e.g. intrinsic alignments), and the nature of non-Gaussianities in the likelihood or the errors. The accuracy to which these are accounted for will then limit whether the information gain can actually be realised and, hence, the extent to which arbitrarily increasing the number of redshift bins would be preferred. In principle, the flexibility of our simulation approach offers the capability to include the effects of these additional uncertainties in our mock catalogues and then propagate the resulting power spectra into cosmological constraints on $(w_{0}, w_{a})$. We note that such additional effects may lead to a convergence in the $(w_{0}, w_{a})$ areas at an even lower number of redshift bins or yield a different order of preference for the binning choice; however such a study is beyond the scope of this paper.

Nevertheless, for the 3$\times$2pt and angular clustering analyses such additional effects would have to be considerable since the relative difference between the different binning choices is reasonably significant at $\sim$10--15\%. Hence, we reassert that the best performing binning strategy, namely, the equipopulated case, is likely to be a good choice for these observables in the DR1 analysis for \textit{Euclid}. Furthermore, in the limit that the observation is dominated by shape noise and photo-$z$ uncertainty, we believe that any given binning choice is unlikely to result in significant degradation in the constraints achievable using the cosmic shear signal on $(w_{0}, w_{a})$. 

Additionally, we note that throughout this work we have considered a fixed scale range of $100\leq\ell\leq600$ for angular clustering and galaxy-galaxy lensing and $100\leq\ell\leq1500$ for cosmic shear when deriving dark energy constraints from tomographic Pseudo-$C_{\ell}$ power spectra. For the real analysis of the \textit{Euclid} sample, we expect that the upper limit is likely to be a conservative estimate for both probes. The effect of changing these angular scale cuts on conclusions of the optimum tomographic binning strategy could represent an informative extension of this study. However, such an investigation would require a careful consideration of a range of further modelling and analytical choices to adopt for the 3$\times$2pt measurements. In particular, as we probe increasingly smaller scales by including larger $\ell$ modes, the assumptions of the constant galaxy bias or even the Gaussianity of the underlying 3$\times$2pt fields are likely to break down. 

Furthermore, there is a considerable range of freedom in how a scale cut could be applied to each component of the 3$\times$2pt signal and it is not trivial to predict the degeneracy in the behaviour of the $(w_{0},w_{a})$ constraints when different combinations of the probes have different cuts applied. We further note that the practical and computational limitations of the simulation-based method will be exacerbated as smaller scales are introduced. Notably, the smaller scales will increase the presence of the unphysical pixels realised on the sky (those with $\delta_{g}<-1$ which mask out observable regions of the mock survey, see also Sect.~\ref{subsec:methods-3x2pt-realisations}); furthermore, it may increase the shot noise in the numerical covariance matrix if the 3$\times$2pt power spectra need to be resolved to a greater number of angular bandpowers to capture the small scale information. Hence, we have considered a thorough examination into the impact of the chosen angular scales on tomography as an important study, but this is beyond the scope of this work. However, as a first-order investigation, we have explored extending the assumption of the linear galaxy bias for the clustering and galaxy-galaxy lensing signals over the range $100\leq\ell\leq1500$ to match the cosmic shear measurements. Ultimately, we found that this does not change any of our overall conclusions.

Lastly, we have considered the effects of catastrophic photo-$z$ uncertainties for our dark energy constraints, where the errors associated with the redshift estimation of galaxies is no longer Gaussian. We injected a $5\%$ level of contamination by catastrophic photo-$z$ errors, randomly distributed across all redshifts, caused by the Lyman-$\alpha$ line being confused with the Balmer and D4000 breaks. 

If such a contamination is unaccounted for in the modelling, we find that a full 3$\times$2pt analysis using a single tomographic bin would return measurements of $(w_{0}, w_{a})$ that are in tension with the fiducial $\Lambda$CDM cosmology at $\sim2\,\sigma$. For a five- or ten-bin analysis with equipopulated bins, we find that the $(w_{0}, w_{a})$ constraints are in considerable tension with $\Lambda$CDM at $\gtrsim3\,\sigma$ and there are sharp features in the posterior distribution. This result reaffirms the necessity to minimise the presence of these catastrophic redshift uncertainties, and to develop techniques that model their effects in the cosmological signal, the covariance, and even the nature of the likelihood distribution. 

%
%

\begin{acknowledgements}

We thank colleagues in the \textit{Euclid} Consortium for useful comments and feedback. We thank the anonymous referee for their comments. We also thank Duncan Austin, Tom Harvey and Erik Rosenberg (JBCA) for their helpful discussions around modelling approaches for photo-$z$ outliers and the numerical covariance matrix. JHWW acknowledges support in the form of a PhD studentship from the UK Science and Technology Facilities Council (STFC) and the University of Manchester Centre for Doctoral Training. MLB and CAJD acknowledge funding from the STFC (grant number ST/X001229/1). 
\AckEC  
\end{acknowledgements}

%
%

\bibliography{Euclid}

%

\begin{appendix}

\section{3$\times$2pt pseudo-$C_{\ell}$ estimators}
\label{sec:appendix-cut-sky-estimators}

In this section, we summarise the expressions for the tomographic Pseudo-$C_{\ell}$ power spectra on the cut sky for a 3$\times$2pt analysis (as presented in Sect.~\ref{subsec:pseudo-cl-estimators}). For a spin-0 clustering field, the partial sky overdensity, $\Tilde{\delta}_{g}(\bm{\theta})$, is related to the full-sky field by the $W(\bm{\theta})$ window function,

\begin{equation}
    \Tilde{\delta}_{g}(\bm{\theta}) = W(\bm{\theta}) \, \delta_{g}(\bm{\theta}) \, ,
\end{equation}
and the spherical harmonic coefficients of the cut-sky field, $\Tilde{d}_{\ell, m}$, are related to the full-sky coefficients according to

\begin{equation}
    \label{eq:window_spin0}
    \Tilde{d}_{\ell, m} = \sum_{\ell ', m'} {}_{0}W_{\ell\ell'}^{mm'}\,d_{\ell',m'} \, ,
\end{equation}
where ${}_{0}W_{\ell\ell'}^{mm'}$ is the harmonic space window function characterised by the mask. For a general spin-$s$ field, the spin-weighted harmonic window function is given in terms of the spin-weighted ${}_{s}Y_{\ell,m}(\bm{\theta})$ polynomials and integrating over the full sphere:

\begin{equation}
    \label{eq:spin_window}
    {}_{s}W_{\ell\ell'}^{mm'}=\int\mathrm{d}\bm{\theta}\,{}_{s}Y_{\ell',m'}(\bm{\theta})\,W(\bm{\theta})\,{}_{s}Y_{\ell,m}^{*}(\bm{\theta}) \, .
\end{equation}
This framework can be extended to relate the observed cut-sky shear field components, $\Tilde{\gamma}_{1, 2}(\bm{\theta})$, to their full-sky counterparts, ${\gamma}_{1, 2}(\bm{\theta})$, via a spin-2 harmonic decomposition,

\begin{align}
    \nonumber\Tilde{\gamma}_{1}(\bm{\theta})\pm \mathrm{i}\Tilde{\gamma}_{2}(\bm{\theta})&=W(\bm{\theta})\left[\gamma_{1}(\bm{\theta})\pm \mathrm{i} \gamma_{2}(\bm{\theta})\right]\\
    &=\sum_{\ell,m}\left(\Tilde{E}_{\ell,m}\pm \mathrm{i}\Tilde{B}_{\ell,m}\right)\prescript{}{\pm2}{Y_{\ell,m}}(\bm{\theta}) \, .
\end{align}
Following \cite{Lewis01}, the cut sky harmonic coefficients $(\Tilde{E}_{\ell,m},\Tilde{B}_{\ell,m})$ are related to the spin-weighted harmonic window functions according to:

\begin{equation}
    \Tilde{E}_{\ell,m}=\sum_{\ell',m'}\left(E_{\ell',m'}W_{\ell\ell'mm'}^{+}+B_{\ell',m'}W_{\ell\ell'mm'}^{-}\right) \, ,
\end{equation}

\begin{equation}
    \Tilde{B}_{\ell,m}=\sum_{\ell',m'}\left(B_{\ell',m'}W_{\ell\ell'mm'}^{+}-E_{\ell',m'}W_{\ell\ell'mm'}^{-}\right) \, ,
\end{equation}
where we have defined the compound matrices, 

\begin{equation}
    W_{\ell\ell'mm'}^{+}=\frac{1}{2}\left(\,\prescript{}{2}{W_{\ell\ell'}^{mm'}}+\prescript{}{-2}{W_{\ell\ell'}^{mm'}}\right) \, ,
\end{equation}

\begin{equation}
    \label{eq:window_pm}
    W_{\ell\ell'mm'}^{-}=\frac{\mathrm{i}}{2}\left(\,\prescript{}{2}{W_{\ell\ell'}^{mm'}}-\prescript{}{-2}{W_{\ell\ell'}^{mm'}}\right) \, ,
\end{equation}
and where Eq.~(\ref{eq:spin_window}) is evaluated for $s=\pm2$. 

The cut-sky Pseudo-$C_{\ell}$ power spectra are related to the full-sky signal by a mixing matrix, $M_{\ell\ell'}$ (see e.g. the appendix in \citealt{Brown05} for a derivation and discussion). Explicitly, for each 3$\times$2pt component, the Pseudo-$C_{\ell}$ power spectra are given by the following matrix expressions.

\begin{itemize}
    \item Angular clustering,
    \begin{equation}
    \left<\Tilde{C}_{\ell}^{\delta_{g}(i)\,\delta_{g}(j)}\right>=\sum_{\ell'}W_{\ell\ell'}^{00}(i, j)\,C_{\ell'}^{\delta_{g}(i)\,\delta_{g}(j)} \, .
    \end{equation}

    \item Cosmic shear,

    \begin{align}
    \nonumber&\begin{pmatrix}
    \left<\Tilde{C}_{\ell}^{E(i)E(j)}\right>  \\
    \left<\Tilde{C}_{\ell}^{E(i)B(j)}\right>  \\
    \left<\Tilde{C}_{\ell}^{B(i)B(j)}\right>
    \end{pmatrix}
    = \sum_{\ell'} \left[\rule{0cm}{0.9cm}\right. \\
    &
    \nonumber\begin{pmatrix}
    W_{\ell\ell'}^{++}(i,j)     &   W_{\ell\ell'}^{-+}(i,j)+W_{\ell\ell'}^{+-}(i,j)     &   W_{\ell\ell'}^{--}(i,j)  \\
    -W_{\ell\ell'}^{+-}(i,j)    &   W_{\ell\ell'}^{++}(i,j)-W_{\ell\ell'}^{--}(i,j)     &   W_{\ell\ell'}^{-+}(i,j)  \\
    W_{\ell\ell'}^{--}(i,j)     &   -W_{\ell\ell'}^{-+}(i,j)-W_{\ell\ell'}^{+-}(i,j)    &   W_{\ell\ell'}^{++}(i,j)  \\
    \end{pmatrix},
    \\ &
    \begin{pmatrix}
    {C}_{\ell'}^{E(i)E(j)}  \\
    {C}_{\ell'}^{E(i)B(j)}  \\
    {C}_{\ell'}^{B(i)B(j)}
    \end{pmatrix} \hspace{-0.075cm} \left.\rule{0cm}{0.9cm}\right] \, .
    \end{align}

    \item Galaxy-galaxy lensing

    \begin{equation}
    \begin{pmatrix}
        \left<\Tilde{C}_{\ell}^{\delta_{g}(i)\,E(j)}\right> \\
        \left<\Tilde{C}_{\ell}^{\delta_{g}(i)\,B(j)}\right>
    \end{pmatrix}
    =
    \sum_{\ell'}
    \begin{pmatrix}
        W_{\ell\ell'}^{0+}(i,j)  & W_{\ell\ell'}^{0-}(i,j)   \\
        -W_{\ell\ell'}^{0-}(i,j) & W_{\ell\ell'}^{0-}(i,j)
    \end{pmatrix}
    \begin{pmatrix}
        {C}_{\ell'}^{\delta_{g}(i)\,E(j)} \\
        {C}_{\ell'}^{\delta_{g}(i)\,B(j)}
    \end{pmatrix} \, . 
    \end{equation}

\end{itemize}
Here, we have defined the tomographic mixing functions,

\begin{equation}
    \label{eq:window_mixing}
    W_{\ell\ell'}^{MN}(i,j)=\frac{1}{2\ell+1}\sum_{mm'}W_{\ell\ell'mm'}^{M, i}\left(W_{\ell\ell'mm'}^{N, j}\right)^{*} \, ,
\end{equation}
for $M,N=(0, +, -)$. Here, the $W_{\ell\ell'mm'}^{+/-}$ are as defined in Eq.~(\ref{eq:window_pm}), and $W_{\ell\ell'mm'}^{0}={}_{0}W_{\ell\ell'}^{mm'}$ from Eq.~(\ref{eq:window_spin0}).

\section{Results and validation}
\label{sec:results-validation}

In this section, we present the validation framework we used to investigate the accuracy of our simulation method to reproduce a target tomographic 3$\times$2pt signal for both a no-noise and noisy set-up. We measured the 3$\times$2pt Pseudo-$C_{\ell}$ power spectra over {3000 realisations} of our mock catalogues, and demonstrate that the sample generated by our pipeline can self-consistently reproduce the underlying power spectra predicted by the input cosmology to an encouraging accuracy. For our validation framework, we adopted the simulation set-up parameters that are defined at the start of Sect.~\ref{sec:results_tomography}, and measured the Pseudo-$C_{\ell}$ power spectra using ten log-spaced bandpowers, taking the range $100\leq\ell\leq600$ for angular galaxy clustering and galaxy-galaxy lensing, and $100\leq\ell\leq1500$ for cosmic shear.

\subsection{No-noise simulation}
\label{subsec:validation-no-noise}

\begin{figure*}
    \hspace{-0.25cm}
        \includegraphics[width=2.1\columnwidth]{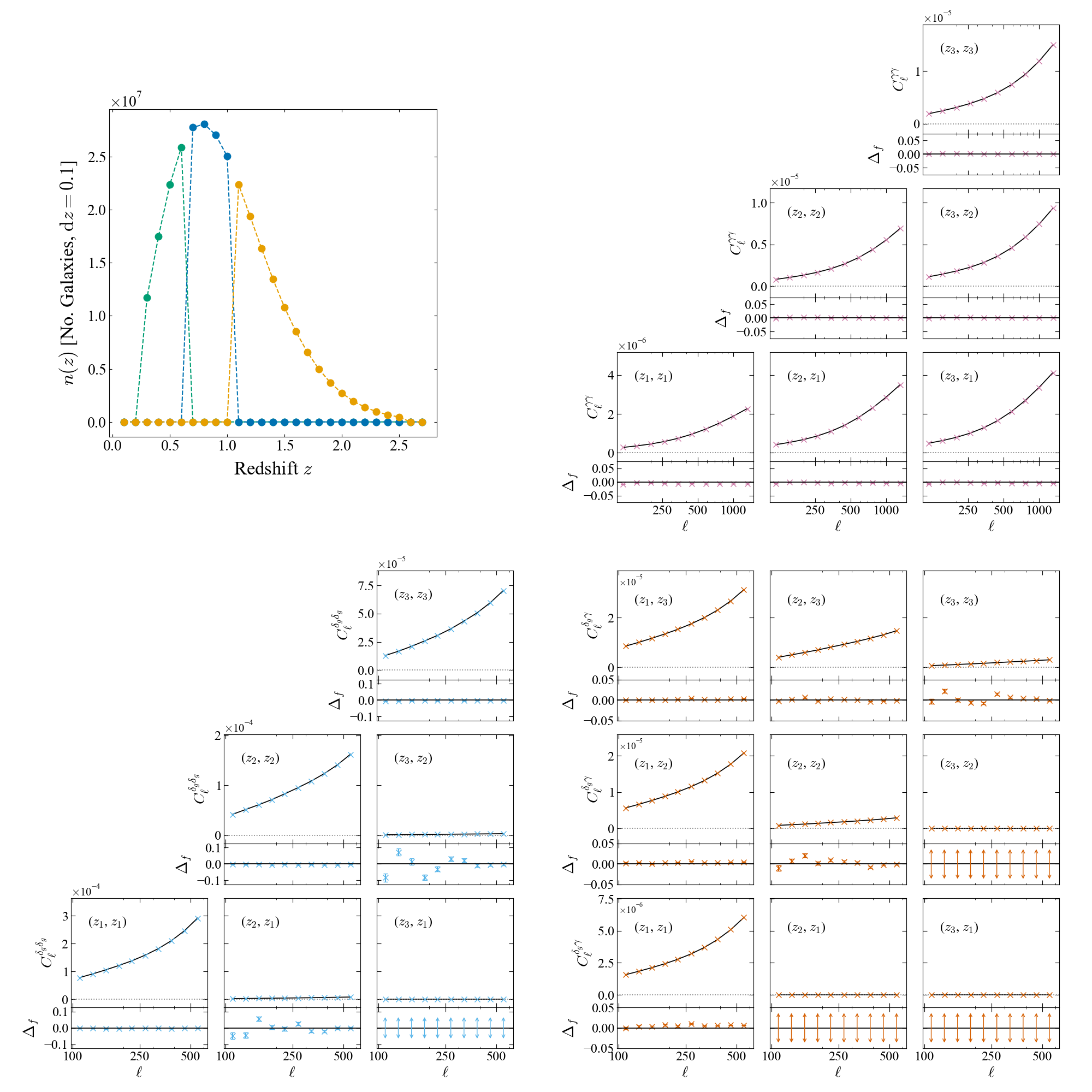}
    \caption{{\textbf{Top-left:}  Binned $n(z)$ for a tomographic 3$\times$2pt analysis of our mock catalogues using three equipopulated bins, with no contribution from shape noise or photo-$z$ uncertainty in the simulation. \textbf{Top-right, bottom-left, bottom-right:}  Measured 3$\times$2pt Pseudo-$C_{\ell}$ power spectra (cosmic shear, angular clustering, galaxy-galaxy lensing, respectively) measured over 3000 realisations for this chosen tomography, shown in crosses. We bin the power spectra into ten log-spaced bandpowers and measure the signal in the range $100\leq\ell\leq1500$ for cosmic shear, and $100\leq\ell\leq600$ for galaxy clustering and galaxy-galaxy lensing. The solid black lines show the input model power spectra. Beneath each individual power spectrum we plot the normalised residuals of the data, defined as the fractional quantity ($C_{\ell}^{\mathrm{Measured}}/C_{\ell}^{\mathrm{Theory}})-1$, which we denote as $\Delta_{f}$ on the $y$-axis label. The vertical double-headed arrows in the off-diagonal subplots indicate that the fractional residuals, and their errors, for these spectra are extremely large due to the division by an extremely small signal.}}
    \label{fig:3x2pt_cls_3bin_nonoise}
\end{figure*}

For our first-order validation demonstration, we present results for a `no-noise' simulation which includes no systematic contribution from redshift uncertainty or shape noise. In Fig.~\ref{fig:3x2pt_cls_3bin_nonoise} we present the binned $n(z)$ for this validation tomography and the corresponding measurements of the Pseudo-$C_{\ell}$ bandpowers of the data for each of the 3$\times$2pt components. Alongside, we plot the theoretical prediction, constructed using Eqs.~(\ref{eq:data_pcl}--\ref{eq:window_mixing}), and the normalised (fractional) residuals for each spectrum, which we denote as $\Delta_{f}$. We calculated the errorbars using the standard deviation of the data points measured from the 3000 realisations. 

In general, we find that for the `no-noise' set-up, the 3$\times$2pt measurements from our catalogues show very good agreement with the fiducial model. Collectively, the residuals of the 3$\times$2pt data vector demonstrate that our method can accurately reconstruct the underlying 3$\times$2pt signal to a sub-per cent level of accuracy, which we highlight as a very encouraging performance of our simulation approach.

Comparatively, the galaxy clustering power spectra show that the measured signal in the auto-correlation bins shows a minimal $(\sim0.5\%)$ under-recovery with respect to the theoretical prediction. This behaviour is not present in the shear or galaxy-galaxy lensing power spectrum, which both show a more random scatter about the model. Additionally, we note that for a no-noise simulation, in which source galaxy redshifts are estimated with complete precision, there is no overlap between the redshift distributions of different tomographic bins. Hence, the cross correlation power spectra of the galaxy clustering signal, and the galaxy-galaxy lensing cross spectra, in which a foreground (low redshift) shear field is correlated with a background clustering field, each have extremely low signal. The corresponding normalised residuals are therefore extremely large due to the small number statistics associated with the very small quantities.

\subsection{Realistic, `noisy' realisations with shape noise and \texorpdfstring{photo-$z$}{photo-z} uncertainty}
\label{subsec:validation-noise}

In this section, we present results of the 3$\times$2pt Pseudo-$C_{\ell}$ power spectra measured from our simulated mock catalogues that now include contributions from shape noise and photometric redshift estimation uncertainty. We modelled the uncertainties arising from each of these sources using Gaussian-distributed errors, taking $\sigma\left(\epsilon^{\mathrm{int}}\right)=0.3$ and $\sigma_{z}^{\mathrm{phot}}=0.05$ as defined in Sect.~\ref{sec:results_tomography}.

\begin{figure*}
    \hspace{-0.25cm}
        \includegraphics[width=2.1\columnwidth]{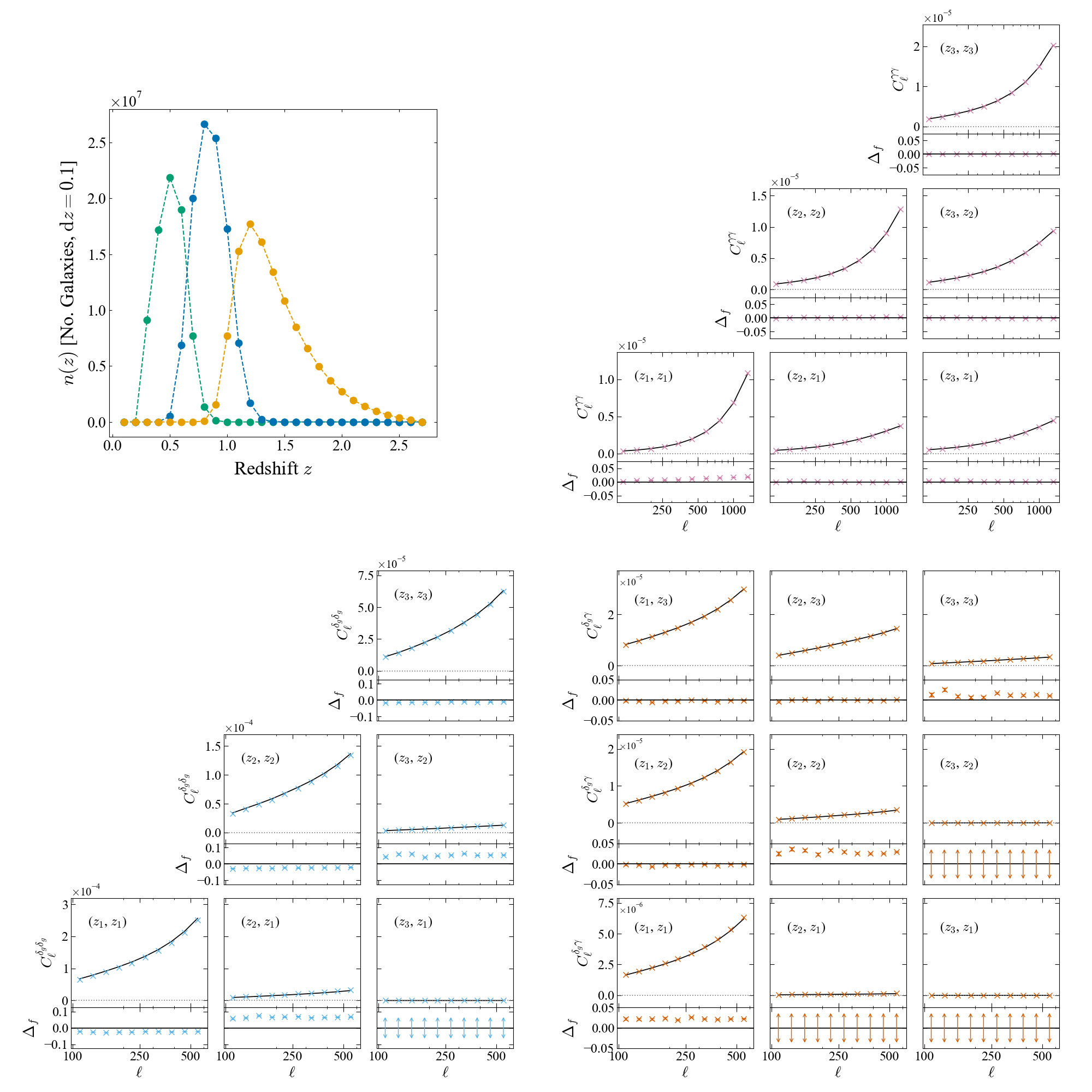}
    \caption{{\textbf{Top-left:}  Binned $n(z)$ for a three equipopulated bin tomographic set-up, measured from our mock catalogues that include realistic Gaussian shape noise and photo-$z$ uncertainties in the simulation. \textbf{Top-right, bottom-left, bottom-right:}  Cosmic shear, angular clustering, and galaxy-galaxy lensing components (respectively) of the 3$\times$2pt data vector. We show in crosses the Pseudo-$C_{\ell}$ power spectra measured over 3000 realisations of our simulation for this chosen tomography. We bin the multipoles into ten log-spaced bandpowers and measure the signal in the range $100\leq\ell\leq1500$ for cosmic shear, and $100\leq\ell\leq600$ for galaxy clustering and galaxy-galaxy lensing. We plot the theoretical prediction in black alongside the measured data, and underneath each individual power spectrum we plot the normalised residuals, which we define as the fractional quantity ($C_{\ell}^{\mathrm{Measured}}/C_{\ell}^{\mathrm{Theory}})-1$, denoted as $\Delta_{f}$ on the $y$-axis label.}}
    \label{fig:3x2pt_cls_3bin_noise}
\end{figure*}

In Fig.~\ref{fig:3x2pt_cls_3bin_noise}, we present the bandpowers of the 3$\times$2pt Pseudo-$C_{\ell}$ power spectra measured over 3000 simulations using the same validation framework of three bins equipopulated with galaxies as considered in the no-noise case in Appendix~\ref{subsec:validation-no-noise}. We show the normalised residuals under each power spectrum component in Fig.~\ref{fig:3x2pt_cls_3bin_noise}, and present the binned $n(z)$ in the upper left corner.

We find that the 3$\times$2pt power spectra measured from the simulated noisy catalogues are in good general agreement with the theoretical prediction. In the presence of Gaussian shape noise and photo-$z$ errors, the shear and galaxy-galaxy lensing signals both show a similar level of recovery of the fiducial model, at the $\sim$1\% level. This is consistent with that seen in the noise-free 3$\times$2pt simulations. However, for the galaxy clustering, the measured auto-correlation power spectra are consistently $\sim$2\% lower than the prediction from the theoretical model. This is significantly larger than the $\sim$0.5\% deficit found in the corresponding no-noise simulations. Furthermore, the cross-correlation power spectra between bins are in more considerable disagreement, with the recovered power spectra being 5--6\% in excess of the theoretical prediction. While the absolute values of these cross spectra are over a magnitude smaller than the auto-correlation signals, this 5--6\% level discrepancy in the cross-correlation clustering power spectra represents the limiting precision of our simulation approach. 

We attribute the above discrepancies to a small (but non-negligible) fundamental inconsistency between our simulation process and our model predictions. The former is an attempt to simulate the full 3D distribution and evolution of matter in the Universe, approximating it as a stack of 2D fields on the sky that have been evaluated using a tomographic projection over finely binned, discretised points in redshift. The resulting catalogues are then binned into much broader redshift bins for the cosmological power spectrum analysis. In contrast, the theoretical model directly predicts the signal in these much broader redshift bins, again using a tomographic projection but now performing the Limber integration over a much broader redshift kernel (see Eqs.~\ref{eq:harmonic-shear}--\ref{eq:harmonic-galgal}). Consequently, our simulation and modelling approaches contain different levels of approximation. This is the fundamental reason for the small discrepancies described above. Furthermore, one expects that the disagreement would be exacerbated in the presence of photometric redshift errors, which is what we have found.

\end{appendix}

\end{document}